\documentclass[natbib,letterpaper]{iopart}
\eqnobysec
\newcommand{\myapp}{Appendix}
\usepackage{amsfonts}
\usepackage{appendix}
\usepackage{fancyhdr}
\newcommand{\mytitle}[1]{\review{#1}}
\newcommand{\here}{\textcolor{red}{$\Box$}}

\usepackage{graphicx}

\usepackage[us,12hr]{datetime} 

\usepackage[usenames,dvipsnames]{xcolor}   

\usepackage{hyperref}
\hypersetup{
    colorlinks=true, 
    linktoc=all,     
    linkcolor=blue,  
    citecolor=red
}

\newcommand{\mycomment}[1]{}

\usepackage{iopams}

\begin{document}
\mytitle{Macroscopic Quantum Mechanics: Theory and Experimental Concepts of Optomechanics}
\author{Yanbei Chen}
\address{Theoretical Astrophysics 350-17, California Institute of Technology, Pasadena, California 91125, USA}

\begin{abstract}
Rapid experimental progress has recently allowed the use of light to prepare macroscopic mechanical objects into nearly pure quantum states.  This research field of {\it quantum optomechanics} opens new doors toward testing quantum mechanics, and possibly other laws of physics, in new regimes.  
In the first part of this paper, I will review a set of techniques of quantum measurement theory that are often used to analyze quantum optomechanical systems. Some of these techniques were originally designed to  analyze how a classical driving force passes through a quantum system, and can eventually be detected with optimal signal-to-noise ratio --- while others focus more on the quantum state evolution of a mechanical object under continuous monitoring. 
In the second part of this paper, I will review a set of experimental concepts that will demonstrate quantum mechanical behavior of macroscopic objects --- quantum entanglement, quantum teleportation, and the quantum Zeno effect.  Taking the interplay between gravity and quantum mechanics as an example, I will review a set of speculations on how quantum mechanics can be modified for macroscopic objects, and how these speculations --- and their generalizations --- might be tested by optomechanics.
\end{abstract}

\maketitle

\tableofcontents{}

\section{Introduction}

The macroscopic world is governed by classical physics,  a set of deterministic laws that govern particles in terms of their precise trajectories, fields in terms of their precise distribution of  amplitudes --- both in a space-time which has a geometry that is determined by the distribution and flow of energy and momentum due to particles and fields. 
%
%
%

Quantum mechanics is the law of motion for the microscopic world.  In quantum mechanics: each canonical degree of freedom in classical physics (e.g., a particle moving along one dimension, a single spatial mode of a field, etc.), which has one pair of canonical coordinate and momentum, gets elevated into a one-dimensional, complex-valued wave:
\begin{equation}
\label{class_quantum}
\{q,p\} \rightarrow \{\psi(q),\psi^*(q)\}
\end{equation}  
These waves evolve deterministically under quantum mechanics.  

By applying the short-wavelength approximation, quantum dynamics can  be reduced to classical dynamics --- in a way similar to the transition from wave optics to geometric optics.  
A more subtle process that often takes place during such a quantum-to-classical transition
is that for a quantum system that has, or interacts with, a large number of degrees of freedom (sometimes a ``heat bath''), any imprecision in initial data or subsequent observation (``coarse-graining'')  will lead to {\it decoherence}~\cite{Zurek:2003}, a loss of coherence between the waves that describe all these degrees of freedom, and the disappearance of their interference patterns.  

This article reviews a set of theoretical techniques and experimental concepts that will lead to answering the question of whether macroscopic objects --- even human-sized objects --- satisfy the same laws of quantum mechanics, if we consider those of their mechanical degrees of freedom~\footnote{For example the center of mass, or any mechanical eigenmode that has a long relaxation time.} that are well isolated from the environment, therefore well-protected from decoherence.    Studying this type of {\it Macroscopic Quantum Mechanics} has become possible with recent progress in {\it quantum optomechanics}:
physicists have been able to use light (sometimes microwaves) to prepare macroscopic mechanical objects into nearly pure quantum states~\cite{Schwab:2005,Aspelmeyer:2012}.  They will soon be able to let these mechanical objects evolve without much decoherence, and measure the final states, thereby making comparison with predictions of quantum mechanics.



\subsection{The Quantum Measurement Process}

Unfortunately for the human beings, the realization of the existence of the microscopic quantum world does not convert into an ability to directly perceive it.  
The initial condition and the final results of quantum experiments must still be communicated to and from the experimentalist in the form of classical information.  
Among other things, we can only {\it infer} the quantum state of a system, and the way we infer them is based on the {\it quantum measurement} process.


In a quantum measurement process, we  always try to obtain classical information in the form of a definitive answer on the value of a physical observable $O$ of a system $\mathcal{S}$ at a state $|\psi\rangle$, even though ``the value of $O$'' usually does not have a definition within quantum mechanics.   In such cases, we will inevitably be hit by {\it quantum uncertainty}: the answer we obtain {\it each time} can be any one of the operator $\hat O$'s eigenvalues --- only  as we perform the same measurement process many times on exact copies of the same state (destroying each one of them after each measurement) will we obtain a predictable answer, in the form of the {\it probability distribution} of obtaining all the eigenvalues of $\hat O$, which is given by the projection of the quantum state $|\psi\rangle$ onto eigenstates of $\hat O$.  
%
%
This repetition in measurements is a necessary consequence of the fact that a much larger amount of information is stored in the quantum description of a system than in its classical description [Cf.~Eq.~\eref{class_quantum}]. 

In this way, the validation of quantum mechanics as a {\it deterministic} law is quite involved, because both the {\it preparation} of the initial state and the {\it verification} of the final state require quantum measurements which can only be characterized probabilistically.  A preparation-verification experiment will therefore have to be carried out repeatedly for many times in order to gather statistics.  In particular, for preparation, only in special cases can we unconditionally prepare a quantum state --- and that preparation procedure depends heavily on the state we intend to prepare; in other cases, we will have to discard incidences which are not compatible with our intention --- and wait till our intended quantum state to appear.  For verification, the same final state can only be verified by measuring the statistical distributions of several observables, and then synthesizing those distributions.

Despite the above peculiarity in its mode of attack, quantum mechanics has so far proven tremendously successful, not only for microscopic systems (interactions between atoms, nuclei and fundamental particles), but also in determining properties of macroscopic objects that are governed by collective motions of microscopic objects (e.g., electrons' motion in solids).  Due to the faster time scales of the above physical processes, quantum nature of microscopic degrees of freedom are more easily visible.  For macroscopic quantum mechanics, on the other hand, we will have to pay more attention to the issue of quantum measurement.



\subsection{Two Approaches to Quantum Measurement}

Historically, physicists have studied quantum measurement theory, in particular, the theory on the evolution of a single system under continuous monitoring, because of two different reasons --- and these have lead to two different approaches. 

The {\it first approach} arose from the need of precision measurement, e.g., detecting weak classical forces like those due to gravitational waves, in presence of thermal and ultimately quantum fluctuations of the transducers~\cite{Braginsky:1968,Braginsky:1986,Braginsky:1992a}.    The aim is to {\it bypass} as much as possible the quantum fluctuations of the transducer (test masses) and the measuring device (light, microwave or electric field), and detect the classical signal at the output port with minimum additional noise.  
It was realized that, as the position of a macroscopic test mass is being monitored continuously, Heisenberg Uncertainty relation between position and momentum often imposes a Standard Quantum Limit (SQL) on the device's sensitivity~\cite{Braginsky:1968}.  It was later realized that the SQL can be circumvented by Quantum Non-Demolition (QND) devices~\cite{Caves:1980a,Braginsky:1996}, although the design and implementation of such devices requires careful consideration of quantum correlations that are built up between the measuring device and the test mass being measured.   
%
%

Further theoretical developments in this direction of research includes the characterization of the maximal amount of information extractable from a system by a measuring device, and the best observable of the system to measure in order to extract a certain type of information~\cite{Braunstein:1994,Tsang:2011,Tsang:2012,Tsang:2012c,Tsang:2012d}.  In the field of gravitational-wave detection, a program exploring laser interferometer configurations that can best achieve sub-SQL sensitivity is being pursued, both theoretically and experimentally~\cite{Danilishin:2012a,Chen:2009,Punturo:2010,Miao:2012c}.
As for the methodology employed by this approach, because one focuses on the detection of a classical signal, and because most of the scenario is a linear system or a system operating within a range of linearization, the Heisenberg picture is often employed, and for many such situations, the quantum-ness of the problem only shows up in the spectrum of fundamental field fluctuations. 

A {\it second approach} focuses on the quantum state of a system under continuous measurement, and therefore stays in the Schr\"odinger Picture.   This approach can be traced back to the study of non-equilibrium quantum statistical physics, where physicists, motivated by applications to condensed matter physics and chemistry, and later quantum optics,  were interested in studying the evolution of an ensemble of quantum systems in contact (but not in equilibrium) with a heat bath (which has an infinite number of degrees of freedom, and causes dissipation)~\cite{Feynman:1963,Keldysh:1964,Caldeira:1983,Weiss:1999}. These systems are also called {\it open quantum systems}.    The methodology initially was to derive equations of motion (Master Equations) for the density matrix of the ensemble of systems alone, while the bath degrees of freedom are traced out. 
%
%
%
%
%

The master equation was later adapted to the situation in which the system is  under continuous measurement, and Stochastic Master Equations (SMEs) were obtained to describe the evolution of such systems, conditioned upon the measurement result~\cite{Davies:1969,Davies:1970,Barchielli:1991,Caves:1986,Caves:1987a,Caves:1987b}. For a single system under an idealized (lossless) measurement, the random walk of its conditional state, which is pure, is described by a Stochastic Schr\"odinger Equation (SSE), and is also referred to as a ``quantum trajectory''~\cite{Carmichael:1991,Carmichael:1993}.   Quantum trajectories have also been used as a mathematical technique to analyze systems coupled to heat baths, when Master Equations are ``unravelled'' into averages over Stochastic Schr\"odinger Equations~\cite{Strunz:1999,Strunz:1999b}.

In this paper, being more interested in the state of a macroscopic mechanical object under continuous measurement, we shall mostly take the second approach.  Nevertheless, we shall often make connections to the first approach, mainly because for many optmechanical systems, their noise levels are often characterized in terms of their performance as measuring devices.  


\subsection{A Detailed Outline of This Paper}

In the {\it first part} of this paper (Secs.~\ref{sec:linear}, \ref{subsec:stochastic} and \ref{sec:unified}), we will use a straw-man linear optomechanical system to illustrate the key features of quantum measurement processes, and to introduce theoretical tools for treating such systems.
[This part is mainly pedagogical, the expert reader already familiar with quantum measurement theory is strongly encouraged to skip Secs.~\ref{sec:linear} and \ref{subsec:stochastic}, and quickly browse through Sec.~\ref{sec:unified}.]  More specifically:

In Sec.~\ref{sec:linear}, we will describe the simplest optomechanical system, and write down the its linear Heisenberg Equations of motion. We shall subsequently discuss the {\it Standard Quantum Limit} (SQL) in Secs.~\ref{subsec:spectrum} and \ref{subsec:SQL}, a sensitivity limitation for weak-force measurements, which arises when the noise due to the measuring device's stochastic back action to the system being measured significantly influences measurement sensitivity.  
Even in classical measurement processes, attachment of the measuring device may  modify the dynamics of the system being measured.  As we shall discuss in Sec.~\ref{rigidity}, in quantum measurement processes, the level of this modification can often be connected to the level of back-action noise.

In Sec.~\ref{subsec:stochastic}, we will discuss the Stochastic Schr\"odinger Equation (SSE) and the Stochastic Master Equation (SME), which describe the stochastic evolution of a system's pure state (SSE), or the density matrix of an ensemble (SME), when the system or ensemble is subject to a continuous measurement and the measurement result is recorded.  We shall refer to such a state as the {\it conditional state}, because it is conditioned on the measurement result.   For the simplest example of a harmonic oscillator:  the conditional expectation values of its position and momentum will always undergo a random walk, although the conditional covariance matrix will reach a constant shape after an initial transient period. As the measurement strength becomes stronger, the scale of the conditional covariance matrix will be set by the {\it measurement timescale}, and tends to be highly position squeezed. 
%
%
%
Although the SSE/SME approach is valid for nonlinear systems, it is only directly applicable to {\it Markovian} systems, i.e., those driven by quantum and/or classical noise with white spectra.  In order to incorporate fluctuations that are correlated in time (i.e., in non-Markovian systems), one must explicitly include the dynamics that generate those correlations.

In Sec.~\ref{sec:unified}, we will return to the Heisenberg Picture, and show that for linear systems, one can bypass the stochastic approach in Sec.~\ref{subsec:stochastic}, and directly obtain statistical characteristics involving the conditional state using tools of classical linear regression --- this will often lead to formulas that are more compact and therefore more informative about the physical nature of the process.  Unlike the SSE/SSE approach in Sec.~\ref{subsec:stochastic}, these formulas here are not limited to Markvoian systems.   This approach can be formulated in terms of path integrals, and will later be used to study non-Markovian linear systems driven by highly non-classical light.   

For more pedagogical treatments of quantum measurement theory, the reader is referred to the recent textbook of Wiseman and Milburn~\cite{Wiseman:2010}; for detailed discussions that emphasizes on force measurements, the reader is referred to the textbook of Braginsky and Khalili~\cite{Braginsky:1992a} and a recent review article by A.~Clerk et al.~\cite{Clerk:2010b}.

In the {\it second part} of the paper (Secs.~\ref{sec:exp_linear}, \ref{sec:further},  \ref{sec:nonlinear}, \ref{sec:test}),  we will outline a set of experimental concepts that can be performed to illustrate and test quantum mechanics --- with focus on the quantum state of the mechanical object.  Owing to the author's own background, many of the experimental concepts that are discussed in detail were originally designed for gravitational-wave detectors and prototype experiments; however,  they can be easily converted to mechanical oscillators in other regimes. Because these concepts apply to  a large set of experiments with a huge span in physical scales, we shall often describe optomechanical systems using dimensionless quantities --- for example, by comparing their various noise spectra  with the free-mass Standard Quantum Limit.   
More specifically:

In Sec.~\ref{sec:exp_linear}, we describe basic experimental strategies that can bring the linear optomechanical system into the quantum regime ({\it state preparation}, or sometimes referred to as ``{\it cooling}''), and to reconstruct the quantum state of the mechanical object with an error less than Heisenberg Uncertainty ({\it state verification}).  In this section, we establish a direct connection between the device's ability to beat the free-mass SQL and our ability to use it to perform macroscopic quantum mechanics experiments. 

In Sec.~\ref{sec:further}, we review experimental strategies that take further advantage of a {\it linear} quantum optomechanical system, and  illustrate features of quantum mechanics --- often involving the creation and detection of quantum entanglement. Many of these further strategies will be based on the concepts of the more basic strategies discussed in Sec.~\ref{sec:exp_linear}.
%
In Sec.~\ref{sec:nonlinear}, we shall study two scenarios of {\it non-linear optomechanical}  systems; both turns out to require the ``strong coupling condition'' that the momentum transfer from a single photon to the mechanical oscillator must be comparable to the quantum uncertainty of the oscillator.
  
In Sec.~\ref{sec:test}, we will look beyond demonstration of standard quantum mechanics and quantum measurement theory with macroscopic mechanical objects, and  consider examples of alternative theories for macroscopic quantum mechanics, and speculate on how a more general program on testing macroscopic quantum mechanics can be built.

Finally, Sec.~\ref{sec:conclusion}, summarizes the main conclusions of the paper, and comments on topics of research in optomechanics that are left out from this paper.  The Appendix contains several elementary facts about statistics and quantum mechanics, in order to make this paper more self contained.

\begin{figure}
\centerline{\includegraphics[width=3.125in]{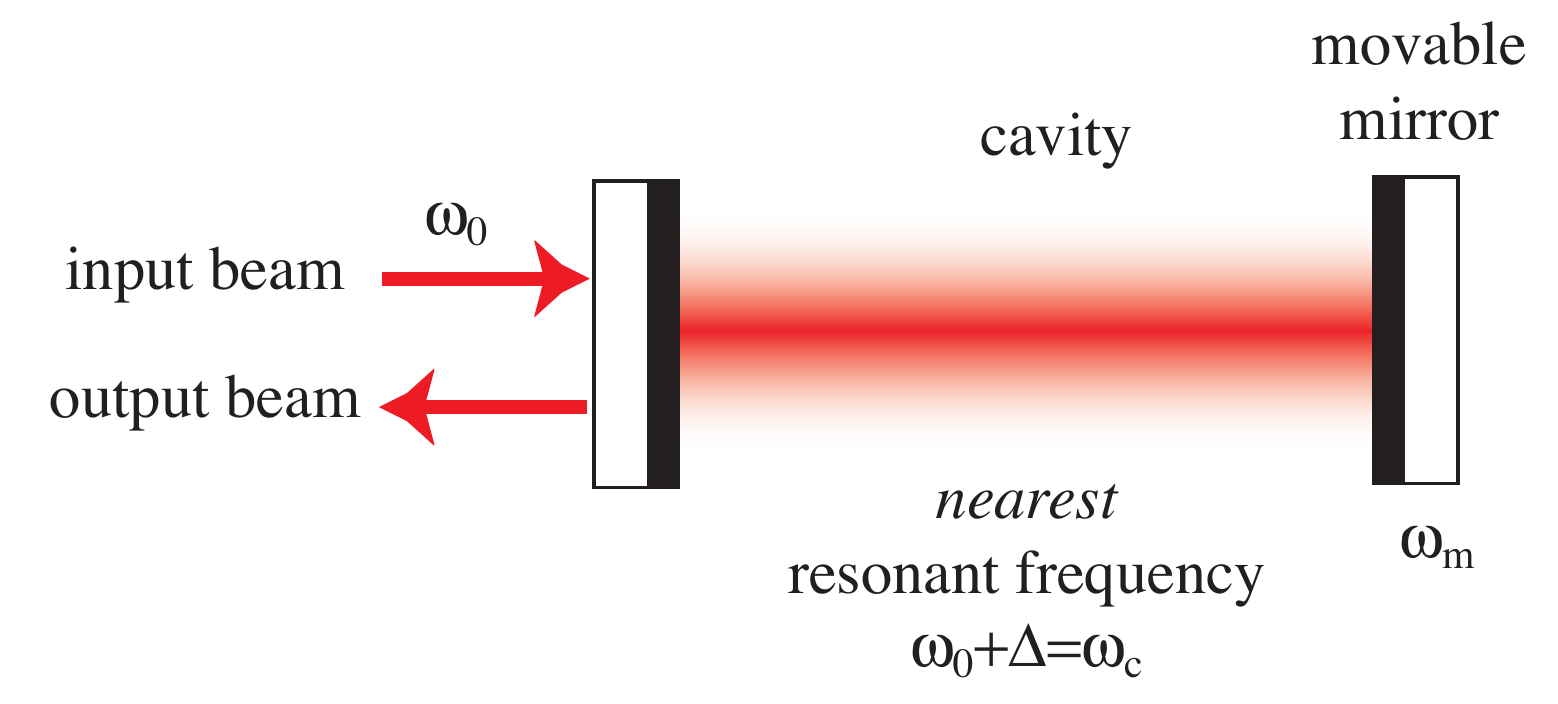}}
\caption{Simple optomechanical device used in Sec.~\ref{sec:linear} as a strawman system.  The mechanical object being measured is the movable end mirror of the cavity, which is driven near resonance. }
\end{figure}

\section{A strawman optomechanics system and linear quantum measurement theory}
\label{sec:linear}

In this section, we will describe a straw man optomechanical system, namely a simple harmonic oscillator, which is also the movable mirror of a high-finesse cavity; the cavity is driven from outside at a nearly resonant frequency; the out-going light  is sensed by homodyne detection (see Sec.~\ref{subsec:spectrum}) as a measure of the mirror's position.  Let us first write down the Hamiltonian~\cite{Law:1995}:
\begin{eqnarray}
\label{optoH} 
\hat H  &=&\hbar\omega_m \hat B^\dagger \hat B - F \hat x+ \hbar (\omega_0+\Delta) \hat A^\dagger \hat A  - \hbar g \hat x \hat A^\dagger \hat A 
\nonumber\\ &+&
 i   \hbar \sqrt{2\gamma}\int_0^{+\infty}\frac{d\omega}{2\pi} \left[\hat A^\dagger  c_\omega- \hat A  c_\omega^\dagger \right]  \nonumber\\
&+& \int_{0}^{+\infty} \frac{d\omega}{2\pi}\hbar\omega \hat c_{\omega}^\dagger \hat c_{\omega}\,.
 \end{eqnarray}
In this Hamiltonian, the mirror's mechanical resonant frequency is $\omega_m$ (with $\hat B$ and $\hat B^\dagger$ the annihilation and creation operators, $\hat x$ is position operator), the pumping frequency is $\omega_0$ ($\hat c_\omega$ and $\hat c^\dagger_\omega$ are the annihilation and creation operators of the external field), and cavity's nearest resonant frequency is $\omega_0+\Delta$ ($\hat A$ and $\hat A^\dagger$ are annihilation and creation operators for this cavity mode).   We have ignored all other modes of the cavity, which do not get excited in our case.  Note that the external field $(\hat c_\omega,\hat c_\omega^\dagger)$ contains an infinite number of degrees of freedom; we have chosen to use $\omega$ to represent the spatial mode with spatial angular frequency (wavenumber) of $\omega/c$. 
The mirror is under an external classical force $F$, and also coupled to the cavity mode via the coupling constant 
\begin{equation}
\label{eq:g}
g=\omega_c/L\,;
\end{equation}
this parametric coupling term accounts for the shift of the cavity's resonant frequency when the mirror is displaced, and the force acting on the mirror due to radiation pressure.  
The cavity mode is coupled to the external vacuum with a coupling constant of $\sqrt{2\gamma}$, and we shall later identify $\gamma$ as the decay rate of the cavity  mode.
The cavity mode is to be driven from outside at $\omega_0$, although this is not reflected in this Hamiltonian.




\subsection{Linearized Treatment: Hamiltonian and Input-Output Relation}
\label{subsec:linham}

In this section, let us assume that the level of pumping at $\omega_0$ is high, so that there is on average a large number of photons inside the cavity, and we can linearize the system's dynamics, using 
\begin{equation}
\hat A \rightarrow \bar A + \delta \hat A
\end{equation}
where $\bar A$ is the expectation value of $\hat A$ at a steady state close to the states we are going to consider.  Here by assuming $\bar A \in \mathbb{R}$ and $\bar A>0$, we have fixed the phase of light inside the cavity, the value of $\bar A$ will later be related to the power circulating in the cavity [See Eq.~\eref{eq:Ecavity}]. We then simply write $\delta \hat A$ as $\hat A$ itself. We shall also go into an interaction picture with $\hbar\omega_0 \hat A^\dagger \hat A$ removed, as well as the corresponding $\hbar\omega_0$ part of the external continuum field.  This results in:
\begin{eqnarray}
\label{linearH}
\hat H &=& \hbar \omega_m \hat B^\dagger \hat B  +  \hbar\Delta \hat A^\dagger \hat A 
- F \hat x 
-   \hbar G  \hat x (\hat A + \hat A^\dagger) \nonumber\\
&+& i \hbar \sqrt{2\gamma}  \int_{-\infty}^{+\infty}\frac{d\Omega}{2\pi} \left[ \hat A^\dagger \hat c_{\omega_0+\Omega}  -
\hat A  \hat c_{\omega_0+\Omega}^\dagger
\right] \nonumber\\ &+& \int_{-\infty}^{+\infty} \frac{d\Omega}{2\pi}\hbar\Omega \hat c_{\omega_0+\Omega}^\dagger \hat c_{\omega_0 +\Omega} .
\end{eqnarray} 
Here we have also restricted ourselves to field fluctuations around $\omega_0$ --- therefore the integration bounds $\pm\infty$ should be understood as numbers that are much larger than the frequency scale we care about, yet much lower than $\omega_0$.  The linear optomechanical coupling is 
\begin{equation}
\label{eq:G}
G=\bar A g.
\end{equation}
where  $\bar A$  is related to circulating power $I_c$ in the cavity by calculating the total energy $\mathcal{E}$ stored in the cavity: 
\begin{equation}
\label{eq:Ecavity}
\mathcal{E} ={2I_c L}/{c}  =\hbar \omega_0 \bar A^2
\end{equation}


Here one should be careful to note that the integrals involving $\Omega$, although related to frequencies at which $c$ oscillate, is actually a decomposition into spatial modes.  We can define (setting the speed of light to unity)
\begin{equation}
\hat c_z =\int_{-\infty}^{+\infty} \frac{d\Omega}{2\pi} \hat c_{\omega_0 +\Omega} e^{+i\Omega z}\,,\quad 
\end{equation}
this gives
\begin{equation}
\left[\hat c_z, \hat c_{z'}\right] = \left[\hat c_z^\dagger, \hat c_{z'}^\dagger\right] =0\,,\quad \left[\hat c_z, \hat c_{z'}^\dagger\right] =\delta(z-z')
\end{equation}
and terms involving $(\hat c_\omega,\hat c_\omega^\dagger)$ in the Hamiltonian will be written as (after having removed a constant):
\begin{equation}
i\hbar\sqrt{2\gamma} \left[\hat A^\dagger \hat c_{z=0}-\hat A \hat c^\dagger_{z=0}\right] + i\hbar\int_{-\infty}^{+\infty} \hat c_z^\dagger \left(\partial_z \hat  c_z\right) dz\,.
\end{equation}
This describes a wave propagating along the $+z$ axis, but coupled locally with $(\hat A,\hat A^\dagger)$ at the position of $z=0$.  This leads to the Heisenberg equations of 
\begin{eqnarray}
\label{eqdcz}
\frac{d \hat c_z}{dt} &=& \partial_z \hat c_z -\sqrt{2\gamma} \hat A \delta (z) \,, \\
\label{eqhatA}
 \frac{d\hat A}{dt}& =& \sqrt{2\gamma} \hat c_{z=0} +\Big(\!\!\mbox{\begin{tabular}{c}\small other \\ \small terms\end{tabular}}\!\!\Big)
\end{eqnarray}
One way to resolve this $\delta$-function is to first integrate Eq.~\eref{eqdcz} across $z=0$, which  leads to a jump of
\begin{equation}
\label{cinout}
\hat c_{z=0+} =\hat c_{z=0-} - \sqrt{2\gamma} \hat A
\end{equation}
and then use $\left[\hat c_{z=0-} +\hat c_{z=0+}\right]/2$ to replace the $\hat c_{z=0}$ in Eq.~\eref{eqhatA}:
\begin{equation}
\label{crep}
\hat c_{z=0} \rightarrow \frac{\hat c_{z=0-} +\hat c_{z=0+}}{2}\,.
\end{equation}
 The averaging here can be justified if we instead consider a distributed coupling between $c_z$ and $A$ near $z=0$, for example $\int  u(z) \hat c_z \hat  A^\dagger dz + \mbox{h.c.}$.  Note that the substitution~\eref{crep} will bring a damping to $\hat A$ --- if we use Eq.~\eref{cinout} to re-express Eq.~\eref{eqhatA} as having $\hat A$ driven by the incoming field $\hat c_{z=0-}$ alone:
\begin{equation}
\label{eqhatAnew}
 \frac{d\hat A}{dt} = -\gamma \hat A +\sqrt{2\gamma} \hat c_{z=0-} +\Big(\!\!\mbox{\begin{tabular}{c}\small other \\ \small terms\end{tabular}}\!\!\Big)\,.
\end{equation}
In the subsequent treatment, we will only need to use Eqs.~\eref{cinout} and \eref{eqhatAnew}. We shall use $\hat a$ to represent the Heisenberg operator of $\hat c_{z=0-}$, the {\it incoming field},  and $\hat b$ to represent the the Heisenberg operator of $\hat c_{z=0+}$, the {\it out-going field}.


From the Hamiltonian~\eref{linearH} and the above input-output formalism, we can easily obtain linear Heisenberg equations of motion in the frequency domain.  For the mirror, we have it moving under the influence of the classical force and the radiation-pressure force: 
\begin{eqnarray}
\label{eqx}
-i\Omega {\hat x}_\Omega &=& {\hat p_\Omega}/{M} \,, \\
\label{eqp}
-i\Omega  {\hat p}_\Omega &= & - M\omega_m^2 \hat x_\Omega + \underbrace{\hbar G[\hat A^\dagger_\Omega + \hat A_\Omega]}_{\hat F_{BA}}+ F_\Omega\,.
\end{eqnarray}
Here the Fourier transform of the radiation-pressure force is:
\begin{equation}
F_{\rm BA} =  \hbar G[\hat A^\dagger_\Omega + \hat A_\Omega] \,.
\label{eq:Fba}
\end{equation}
For the cavity mode, we have it driven not only by the incoming field, but also by the motion of the mirror:~\footnote{Note that $\hat A^\dagger_\Omega$ is the $\Omega$-Fourier component of the Heisenberg operator $\hat A^\dagger(t)$.  It is not the same as $(\hat A_\Omega)^\dagger$, which is the Hermitian conjugate of the $\Omega$-Fourier component of the Heisenberg operator $\hat A(t)$.}
\begin{eqnarray}
\label{eqA}
-i\Omega  {\hat A}_\Omega =& (- i\Delta-\gamma) \hat A_\Omega  + i G \hat x_\Omega + \sqrt{2\gamma} \hat a_{\omega_0+\Omega}\,,\\
\label{eqAd}
-i\Omega  {\hat A}^\dagger_\Omega =& (+i\Delta  -\gamma) \hat A^\dagger_\Omega  - i G \hat x_\Omega +\sqrt{2\gamma} \hat a_{\omega_0-\Omega}^\dagger\,.
\end{eqnarray}
Finally for the out-going field, it picks up a certain amount of the cavity mode,
\begin{eqnarray}
\hat b_{\omega_0+\Omega} &=& \hat a_{\omega_0+\Omega} - \sqrt{2\gamma}\hat A_\Omega\,, \quad  \\
\hat b^\dagger_{\omega_0-\Omega} &=&  \hat a^\dagger_{\omega_0-\Omega}  - \sqrt{2\gamma}\hat A^\dagger_\Omega\,.
\label{eqcoutd}
\end{eqnarray}
We can also organize the creation-annihilation operators into quadrature operators (following the Caves-Schumaker two-photon formalism~\cite{Caves:1985,Schumaker:1985}):
\begin{equation}
\label{eq:quadratures}
\hat a_{1\,\Omega} =  \frac{\hat a_{\omega_0 +\Omega} + \hat a^\dagger_{\omega_0 -\Omega}}{\sqrt{2}}, \; 
\hat a_{2\,\Omega} =  \frac{\hat a_{ \omega_0 +\Omega} - \hat a^\dagger_{\omega_0 -\Omega}}{\sqrt{2}i}.
\end{equation}
In this way, if the quadratures are superimposed with a classical carrier light $\propto\cos\omega_0 t$, then  $\hat a_1$ represents the amplitude quadrature of the field, while $\hat a_2$ the phase quadrature (both for ``in'' and ``out'' fields).  See \myapp~\ref{app:twomode} for more details. 

Equations~\eref{eqx}--\eref{eqcoutd} are analytically solvable, and characterizes all aspects of the system's evolution.  This will be the dynamical system that underlies most of our discussions in Secs.~\ref{sec:exp_linear} and \ref{sec:further}. In Sec.~\ref{subsec:elimination}, this system will be further simplified to eliminate the dynamics of the cavity's optical mode.

\subsection{Weak force measurement and the Standard Quantum Limit}
\label{subsec:spectrum}

Let us first look at the issue of measuring a classical force.  Starting from the simplest case, we assume cavity detuning $\Delta =0$, and obtain, from Eqs.~\eref{eqx}--\eref{eq:quadratures}, an input-output relation~\cite{Kimble:2001}:
\begin{eqnarray}
\label{c12compact}
\hat b_1 &=& e^{2i\beta} \hat a_1\,,  \\
\label{eqb2compact}
\hat b_2 &=& e^{2i\beta}\left [\hat a_2 -\mathcal{K} \hat a_1\right] +e^{i\beta} \left| \frac{2 \mathcal{K}}{S^F_{\rm SQL} }\right|^{1/2}F\,.
\end{eqnarray}
Here we have defined
\begin{equation}
e^{2i\beta} =\frac{\Omega-i\gamma}{\Omega+i\gamma}
\end{equation}
which is due to the storage of light in the cavity,  and
\begin{equation}
 \mathcal{K} =  \frac{2\Theta^3 \gamma}{(\Omega^2-\omega^2)(\Omega^2+\gamma^2)}\,,
 \end{equation}
 with
 \begin{equation}  
 \Theta^3 \equiv \frac{2G^2}{M} =\frac{4\omega_0 I_c}{(M L c)}\,.
 \end{equation}
Being proportional to $G^2$ and hence $I_c$ (circulating power in the cavity), $\mathcal{K}$ represents the strength of optomechanical coupling ($\Theta$ is a characteristic angular frequency).  We have also defined
\begin{equation}
S_{\rm SQL}^F = 2\hbar M |\Omega^2-\omega^2|
\end{equation}
which, as we shall explain below in Sec.~\ref{subsec:SQL}, is the the Standard Quantum Limit for force measurement.   In the case of vacuum input field, we have 
\begin{equation}
S_{\hat a_1} =S_{ \hat a_2}=1\,,\quad S_{\hat a_1 \hat a_2} =0\,,
\end{equation}
see \myapp~\ref{app:classicalrandom} for details on spectral density, and \myapp~\ref{app:twomode} for details on spectra of quadrature fields.

\begin{figure*}
\centerline{
\includegraphics[width=6.5in]{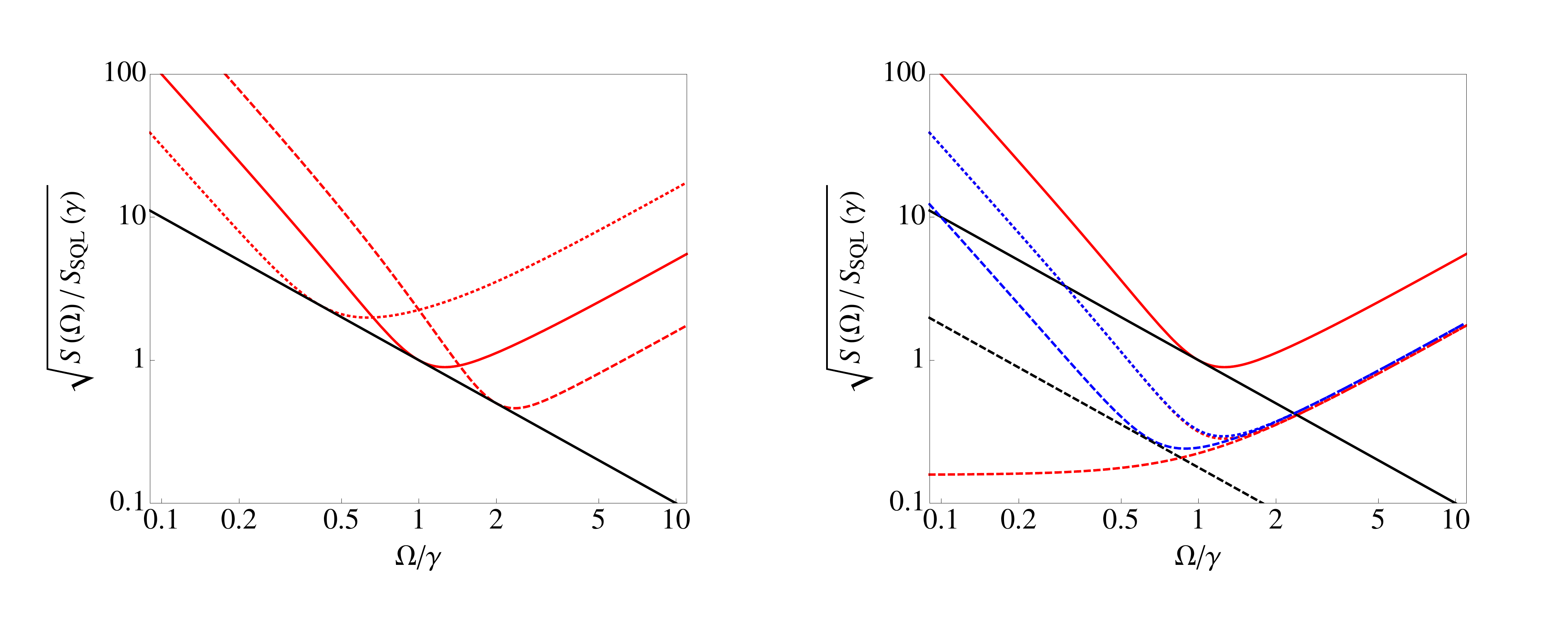}}
\caption{Left panel: noise spectrum of SQL-limited interferometer configurations, with phase-quadrature readout. The red solid curve depicts $\Theta=\gamma$; red dotted curve corresponds to 1/10 the power, and red dashed curve corresponds to 10 times the power. The black solid line is the free-mass Standard Quantum Limit. A typical GW detector may have $\Theta \approx \gamma \approx 2\pi\times 100\,$Hz. For $M=10\,$kg (this corresponds to the reduced mass of the four mirrors in Advanced LIGO~\cite{Harry:2010}), we have $\sqrt{S_x(2\pi\cdot 100\,{\rm Hz})}= 1.5 \times 10^{-20}\,\mathrm{m}/\sqrt{\rm Hz}$.  For an arm length of $4\,$km, the solid curve is achievable when the circulating optical power in the arms is $I_c =830$\,kW. [See Sec.~\ref{subsec:spectrum} for details.] 
   Right panel: solid curve is as in the left panel; dashed curves correspond to back-action-evading interferometer with variational readout scheme plus 10\,dB frequency-independent squeezing (red for lossless interferometer, blue for $1\%$ total optical loss); dotted curve has 10\,dB frequency-dependent input squeezing (red for lossless, and blue for $1\%$ total optical loss; in this case a 1\% loss does not significantly affect sensitivity).  The black dashed line indicates the limit of all variational readout schemes given $1\%$ optical loss and 10\,dB squeezing. [See Sec.~\ref{subsec:sqlbeating} for details.]  \label{fig:noise}}
\end{figure*}

Let us assume that a {\it homodyne detection} is made on the out-going light, measuring a quadrature field, which is a linear combination of $b_1$ and $b_2$, for example
\begin{equation}
\label{eq:bzeta}
\hat b_\zeta \equiv \hat b_1 \cos\zeta + \hat b_2 \sin\zeta\,.
\end{equation}
This can be achieved by detecting the beating amplitude between the out-going light and a local oscillator light:
\begin{equation}
\hat i(t) \propto  \left(\hat b_1 \cos\omega_0 t + \hat b_2 \sin\omega_0 t\right) D \cos (\omega_0 t -\zeta)\,.
\end{equation}
Here $D$ is the amplitude, and $\zeta$ is the phase of the local oscillator.  Note that quantum fluctuations of the local oscillator, ideally, should not enter the photocurrent $i(t)$.  Homodyne detection can be achieved in several ways: (i) using a highly transmissive mirror to combine a high fraction of the light to be detected and a low fraction of a nearly quantum-limited carrier field originally with high amplitude (this is somewhat similar to the so-called DC readout scheme of gravitational-wave detectors), (ii) use a 50/50 beamsplitter, interfere the light to be detected with a nearly quantum-limited carrier field, use two identical photodetectors to detect photocurrents $i_1(t)$ and $i_2(t)$  at the two output ports, then subtract from each other.  This is also referred to as {\it balanced homodyne detection}.

Rigorously speaking, a field quadrature is not always the optimal out-going observable to measure, especially when we are aiming at detecting a very weak displacement signal~\cite{Tsang:2012,Tsang:2012b}. However, these more optimal observables to detect will be nonlinear (i.e., in terms of field quadratures), and will likely depend on features of a specific waveform we aim at --- these would be undesirable because in practice we have many families of possible waveforms that need to be detected.   In addition, in gravitational-wave detection, the mirror also moves due to classical noise, with amplitude that is much higher than the SQL at low frequencies --- this will likely detract from the advantage of nonlinear detection schemes.  In any case, let us restrict ourselves to the linear regime, and consider homodyne detection.

In absence of any prior knowledge to $F$,  our best estimator for $F$ is simply to filter  $b_\zeta$ by inverse the coefficient in front of $F$.  The most conventional to measure would be $\zeta =\pi/2$ or $b_2$, because it is the one quadrature that contains all the signal.  

Within $b_2$ [Cf.~Eq.~\eref{eqb2compact}], the term containing $F$ is signal,
the term $e^{2i\beta}a_2$ is shot noise, the term $-e^{2i\beta}\mathcal{K}a_1$ is radiation-pressure noise. When we normalize $b_2$ such that coefficient in front of $F$ is unity, the shot noise will be  $\propto 1/\sqrt{\mathcal{K}}$, hence $\propto I_c^{1/2}$; the radiation-pressure noise will be $\propto \sqrt{\mathcal{K}}$, hence $\propto I_c^{1/2}$.  Combining the two types of noise,  we have a total noise spectral density of
 \begin{equation}
 S_F =\frac{1}{2}\left[ \frac{1}{\mathcal{K}} +\mathcal{K}\right]S^{F}_{\rm SQL} \ge S_{\rm SQL}^F\,.
  \end{equation} 
This is a direct consequence of the trade-off between shot noise and radiation-pressure noise: the former is inversely proportional to optical power, yet the latter is directly proportional.   However, the independence of $S_{\rm SQL}^F$ from the details of the experiments (e.g., the optical bandwidth $\gamma$) indicates a more universal origin~\cite{Braginsky:1992a,Danilishin:2012a,chen:2003}, as we shall discuss below in Sec.~\ref{subsec:SQL}.

The SQL can also be written for position, if we use the mechanical oscillator's response function as the converter, and we have
\begin{equation}
\label{sqlx}
S_{\rm SQL}^{x} =\frac{2\hbar}{M|\Omega^2-\omega^2|}\,.
 \end{equation}

When dealing with gravitational-wave detectors, whose test masses are suspended as nearly free masses, we often use the free-mass SQL,
\begin{equation}
\label{sqlxfm}
S_{\rm free\mbox{-}mass\,SQL}^{x} =\frac{2\hbar}{M\Omega^2}\,.
 \end{equation}
In the left panel of Fig.~\ref{fig:noise}, we show the noise spectral of several interferometers that are limited by the free-mass SQL.  {\it Henceforth in the paper, unless explicitly noted, we shall always use the SQL to indicate free-mass SQL.} See Sec.~\ref{subsub:sig:sql} below for the special significance of the free-mass SQL.

In gravitational-wave detection on earth,  in the long-wave regime ($\lambda_{\rm GW}$ much greater than the size of the detector), it is best to consider the effect of the gravitational wave as a classical force field acting on the test masses, one often quotes the strength of the wave by the strain $h$ it causes on an array of {\it free test masses} with separation much less than wavelength--- because that strain is also related to metric perturbation in the so-called Transverse Traceless (TT) gauge, in which wave propagation is most easily treated. Suppose the (polarization-dependent) conversion between $h$ and free-mass displacement is $x= Lh$, then the SQL for GW detection, for an oscillator with frequency $\omega$, is
\begin{equation}
S_{\rm SQL}^h =  \frac{S_{\rm SQL}^F}{M\Omega^4 L^2}=  \frac{2\hbar|\Omega^2-\omega^2|}{M\Omega^4 L^2}\,.
\end{equation}
In this way, SQL-limited GW sensitivity for an oscillator with frequency $\omega$ is much better than that of a free mass, in the narrowband of $|\Omega -\omega | \ll \omega$. 

\subsection{Linear Quantum Measurement Theory} 
\label{subsec:SQL}

After having introduced the SQL from a specific calculation, let us study some more fundamental issues of quantum measurement that have been used in the above calculations. 

\subsubsection{Significance  of the free-mass SQL.}
\label{subsub:sig:sql}

The free-mass SQL can often be used as a benchmark for the quantum-ness of a measurement process.   Suppose we have a measuring device with a classical position sensing noise better than the free-mass SQL, and suppose we focus on a frequency $\Omega$ --- then, after taking measurement for a time scale of $\tau \sim 1/\Omega$, our sensing position error will be 
\begin{equation}
\Delta x \approx \sqrt{\frac{S^x}{\tau}} < \sqrt{\frac{S_{\rm SQL}^x}{\tau}} \approx \sqrt{\frac{\hbar}{M\Omega}} \approx \Delta x_Q
\end{equation}
which is the  ground-state uncertainty of the position of a harmonic oscillator with resonant frequency $\Omega$.  Similarly, if we have a classical force noise below the free-mass force SQL, and if we allow the force to act on the mirror for a duration of $\tau \sim 1/\Omega$, the momentum uncertainty it causes will be
\begin{equation}
\Delta p \approx \sqrt{S_F \tau} < \sqrt{S_F^{\rm SQL} \tau} \approx \sqrt{{\hbar}{M\Omega}} \approx \Delta p_Q
\end{equation}
which is the ground-state uncertainty of momentum of the same oscillator.  This means, if within a broad frequency band, we have a device with {\it only classical} force noise and sensing noise, both below the level of the free-mass SQL, the device will be capable of localizing {\it both} position {\it and} momentum of the test mass below the ground-state level.  As a consequence, {\it  quantum noise will be the main enforcer of Heisenberg Uncertainty Principle for the test mass}. 

\subsubsection{Origin of the Standard Quantum Limit.}

Let us now discuss the origin of the free-mass SQL.  From a quantum measurement point of view, we suffer from the SQL because
\begin{equation}
\label{xxcomm}
\left[\hat x(t),\hat x(t')\right] =i\hbar(t'-t)/m \neq 0 
\end{equation}
This means we cannot determine $\hat x(t)$ continuously without worrying about the sequence of measurement and state reduction, because these operators we attempt to measure are not {\it simultaneously measurable}: subsequent measurements of $\hat x$ will {\it demolish} the quantum state prepared by previous measurements.  From Eq.~\eref{xxcomm}, one can derive a Heisenberg Uncertainty relation for position measurement at two different times, 
\begin{equation}
\Delta x(t) \Delta x(t')  \ge \frac{\hbar|t'-t|}{2m}
\end{equation} 
which eventually can lead to the SQL.

By contrast, if Heisenberg Operators $\hat A(t)$ of the observable $A$ we measure all commute with each other, namely, 
\begin{equation}
\left[\hat A(t),\hat A(t')\right]=0
\end{equation}
then we can find {\it simultaneous eigenstates} of all these operators $\{\hat A(t)\}$, and divide the system's Hilbert space into a direct sum of eigen-subspaces, each one corresponding to one particular measurement outcome.  We will only need to project the system's quantum state once,  into one particular simultaneous eigenstate of all these observables, and all measurement results can be  pre-determined, and pre-assigned with aprobability density.    Such a scenario has been called ``Quantum Non-Demolition'' (QND) measurements.  The signature of a QND measurement is the absence of back-action noise, and hence the absence of a Standard Quantum Limit. Obviously, if we sense a classical force by using a QND observable as the transducer, we will be immune from the SQL.  This is one approach toward devices that beat the SQL~\cite{Thorne:1978,Braginsky:1980,Caves:1980a,Bocko:1996,Braginsky:1996}.

\begin{figure*}
\centerline{\includegraphics[width=4.5in]{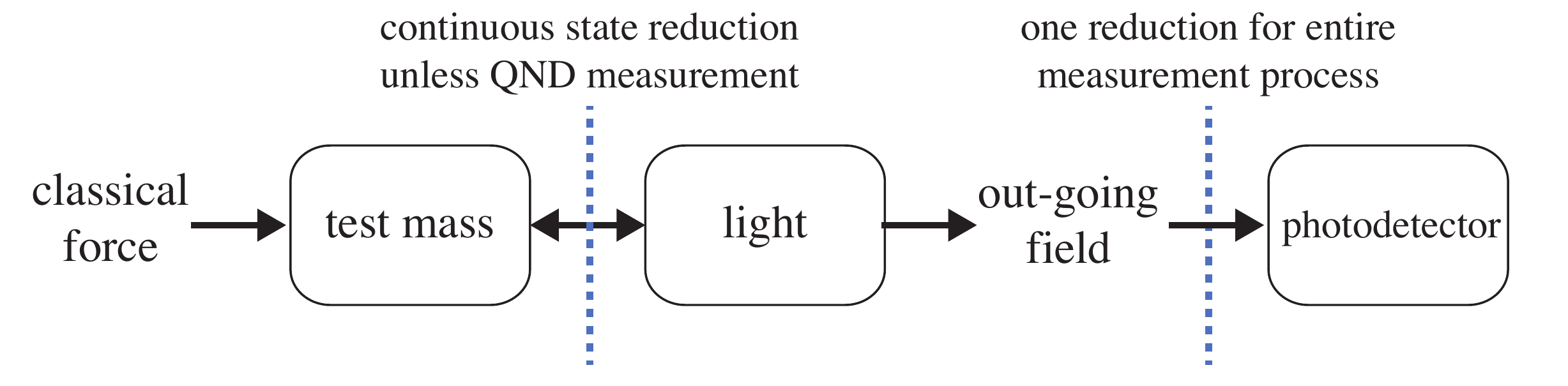}}
\caption{Two approaches toward the continuous quantum measurement of a test mass with light: The first approach places the quantum-classical cut between the test mass and light, and therefore the measurement process reduces the quantum state of the test mass continuously; reductions at later times often causing ``demolition'' of previously prepared quantum state.  The second approach places the cut between the light and the signal read out by the photodetector; demolition of quantum state is unnecessary because observables measured at different times have commuting Heisenberg operators.   }
\end{figure*}

\subsubsection{Continuous linear indirect measurements.}
\label{subsub:clim}

Coupling to a QND observable is not the only way to beat the SQL.  It was later realized that preparing the measuring device into appropriate quantum states can also achieve the same goal~\cite{Yuen:1983,Ozawa:1988}.  In order to better treat such possibilities, it was necessary to study ``indirect measurement'', which refers to the inclusion of (more of) the measurement device into the Hamiltonian of the system being measured.  In other words, instead of considering ``some observable of the test mass'' as being measured, one considers instead  the coherent quantum mechanical interaction between the test mass and the light --- forming an optomechanical system --- and {\it defer}, but not eliminate, the state reduction process to a further level --- inside of the photodetector --- where the out-going light is measured.  In this way, we can easily treat the more complex structure of quantum correlations that build up over time.

Taking the point of view of indirect measurement, instead of having to repeatedly carry out state reduction state-reductions and evolutions of the test mass, like 
\begin{eqnarray}
|\psi\rangle \rightarrow  P|\psi \rangle &\rightarrow& U P|\psi \rangle \nonumber\\
& \rightarrow& P U P|\psi \rangle  \rightarrow U P U P|\psi \rangle  \rightarrow\ldots\,.
\end{eqnarray}
we simply need to project out, from the initial state in the Heisenberg Picture, the simultaneous eigenstate of all the commuting  operators that are to be measured:
\begin{equation}
\label{projection}
|{\rm in}\rangle \rightarrow P_{\hat Z_H(t_1) =\xi_1, \ldots, \hat Z_H(t_n) =\xi_n } |{\rm in}\rangle
\end{equation}
This enlargement of the quantum system makes all continuous measurements QND. For example, the quadrature operators $\hat a_1(t)$ and $\hat a_2(t)$ for a freely propagating field would be QND observables --- because at different times they represent different degrees of freedom and therefore naturally commute.  In this way, the construction of QND observables is no longer the key issue.  

This point of view hides very well the state reduction~\eref{projection} that takes place in the indirect measurement process.   During our treatment in Sec.~\ref{subsec:spectrum}, the only use of the state-reduction postulate is in obtaining the spectrum of the quadrature fields $a_{1,2}$ --- probability distributions of different realizations of the successive stat-reduction processes are all summarized succinctly in these  random processes.  This simplicity and this de-emphasis of quantum-state reduction allows one to better focus on optimizing the device's sensitivity to the classical force it is supposed to measure.   In Sec.~\ref{sec:unified}, we will discuss how to recover the ``hidden'' measurement process from an indirect measurement process.

However, a QND measurement for the out-going fields does not directly convert into a sub-SQL sensitivity for the motion of the test mass.  In fact, having deferred treatment of state reduction, we must provide an alternative derivation and motivation for the SQL.  We will do this right now following the treatment of Braginsky and Khalili~\cite{Braginsky:1992a}.
For each continuous linear measurement, the observable that corresponds to measuring device's output must have Heisenberg Operators $\hat Z(t)$ that satisfy the simultaneous measurability condition:
\begin{equation}
\label{eq:commO}
\left[\hat Z(t),\hat Z(t')\right]=0\,.
\end{equation}
Due to linearity, $\hat Z$ must be the sum of $\hat x$, the observable we would like to measure (which could be the position of the test mass), and an operator $\hat N$ that arise from the device degrees of freedom:
\begin{equation}
\hat Z(t) = \hat N(t) + \hat x(t)\,.
\end{equation}
Because $\hat N$ is an observable from the device, it must commute with $\hat x$, and the only way Eq.~\eref{eq:commO} can hold is to have 
\begin{equation}
\left[\hat N(t),\hat N(t') \right]=-\left[\hat x(t),\hat x(t') \right]\,.
\end{equation}
Namely, device commutator must cancel test-mass commutator. 

Let us put the above argument into details.  First, let us assume  that the coupling between $\hat x$ and the measuring device is linear, through an interaction Hamiltonian of $\hat V = -\hbar \alpha\hat x \hat F$, with $\alpha$ the measurement strength, and $F$ an observable of the measuring device.  We shall make the simplifying assumption that 
\begin{equation}
\left[ \hat F^{(0)}(t),\hat F^{(0)}(t') \right] =0
\end{equation}
here the superscript ``(0)'' indicates before coupling is applied. As we will discuss in lengths in Sec.~\ref{rigidity}, this means the evolution of  $F$  will not be affected by the fact that it is now coupled with the system being measured.  On the other hand, the evolution of $x$ will in general be affected, and we can write
\begin{eqnarray}
\label{eq:xwithba}
\hat x (t) &=& \hat x^{(0)}(t) \nonumber\\
&+&\alpha  \int_{-\infty}^t \chi(t-t') \left[\hat F^{(0)}(t') +G(t')\right] dt'  
\end{eqnarray}
with $\chi$ the response function of $\hat x$ to an external force.  Moreover, we can write
\begin{equation}
\left[ \hat x^{(0)}(t), \hat x^{(0)}(t')\right] = i \hbar\chi(t'-t)\,.
\end{equation}

Having introduced a measurement strength, the linearity of the system and the device dictates that the operator $\hat Z$ which corresponds to the device's output must ``pick up''  $x$ with $\alpha$ as a constant of proportionality:  
\begin{equation}
\hat Z(t) =\hat Z^{(0)}(t) + \alpha \hat x(t)\,.  \end{equation}
Note that here $\hat x(t)$ has already been  acted back upon by $\hat F$, as in Eq.~\eref{eq:xwithba}.

The simultaneous measurability condition,
\begin{equation}
\left[\hat Z(t),\hat Z(t')\right] =0
\end{equation}
being valid at all orders of $\alpha$, requires that
\begin{equation}
\left[\hat Z^{(0)}(t),\hat Z^{(0)}(t')\right] =0
\end{equation}
we can also argue that only 
\begin{equation}
\label{commzf}
\left[\hat Z^{(0)}(t),\hat F^{(0)}(t')\right] =-i\hbar\delta(t-t')\,
\end{equation}
allows the commutator of $\hat x$ at difference times to be cancelled out by those of $Z$ and $F$ [See, e.g., \cite{Khalili:2012}]. This allows us to view $Z$ and $F$ as independent degrees of freedom brought to interact with the test mass at different times; the form of Eq.~\eref{commzf} makes us compare $Z$ and $F$ to the position and momentum of harmonic oscillators; mathematically, it leads to a frequency-domain Heisenberg Uncertainty Principle:
\begin{equation}
\label{HUP}
S_{ZZ} S_{FF} - |S_{ZF} |^2 \ge \hbar^2 +2\hbar |\mathrm{Im} S_{ZF}|\,.
\end{equation}

If we turn to the total position-referred measurement noise spectrum, we have
\begin{equation}
S_x =\frac{\hbar^2}{\alpha^2}S_{ZZ} + 2 \mathrm{Re} \left(\chi^*  S_{ZF}\right) + \alpha^2 | \chi|^2 S_{FF} +S_{x}^{(0)}
\end{equation}
where $S_{ZZ}$ is sensing noise, $S_{FF}$ is back-action noise, $S_{ZF}$ is their correlation, while $S_{xx}^{(0)}$ is noise caused by the zero-point fluctuation of the test mass. In gravitational-wave detection, we are often outside the frequency band in which $S_{xx}^{(0)}$ is important~\cite{Braginsky:2003} --- and we shall ignore this term here.  Further assuming {\it no correlation} between $Z$ and $F$, we obtain
\begin{equation}
S_{x} \ge 2 |\chi| \hbar \equiv S_x^{\rm SQL}
\end{equation}
which is the Standard Quantum Limit [Cf.\ Eqs.~\eref{sqlx} and \eref{sqlxfm}].  However, in general, if $S_{ZF}$ does not vanish but is instead chosen appropriately, the Uncertainty Principle~\eref{HUP} does not impose any bound on the total noise spectrum.  This means, beating the SQL requires correlations between the sensing noise $Z$ and back-action noise $F$.   We shall often refer to those schemes that beat the SQL as QND schemes, even though mathematically it is sometimes difficult to formally make them equivalent to a measurement of a QND observable of a test object.

\subsection{Beating the SQL in Linear Quantum Measurement Devices}
\label{subsec:sqlbeating}

\subsubsection{Gravitational-wave detectors and prototype experiments}

The need to beat the free-mass SQL in gravitational-wave detectors had long been recognized, at the conception of the LIGO project~\cite{Abramovici:1992}.  At this moment, the first generation of laser interferometer gravitational-wave detectors have completed their first round of runs, at sensitivities 10 times the free-mass SQL. Second-generations, such as Advanced LIGO~\cite{Harry:2010}, Advanced VIRGO~\cite{Accadia:2011} and KAGRA~\cite{Somiya:2012}, are under construction and will become operational in several years.  These detectors will operate very close to or moderately beat the free-mass SQL. The gravitational-wave community has already started designing third-generation detectors, e.g., the Einstein Telescope~\cite{Punturo:2010} and LIGO-3~\cite{Miao:2012c}; these detectors may have to surpass the SQL significantly in order to achieve a significant gain in sensitivity compared with the second generation.  

Before discussing techniques that allow the quantum noise to beat the SQL, we must realize that 
the SQL is used also because it is a easy benchmark that has a simple conceptual origin.  Major technical challenge in building sub-SQL interferometers exist in improving technology that first lower classical noise, and {\it bring the detector into the quantum regime}, these include: (i) isolating the test masses from ground motion~\cite{Abbott:2004}, and removing the influence of oscillating Newtonian gravity field~\cite{Driggers:2012}, (ii) building a suspension system that has low thermal fluctuations~\cite{Gonzalez:1994,Nawrodt:2011}, (iii) manufacturing mirrors that have less internal fluctuations~\cite{Levin:1998,Braginsky:1999b,Kondratiev:2011,Hong:2012}, (iv) minimize classical laser fluctuations~\cite{Kwee:2012}, and (v) building a control system that stabilizes the system from parametric instabilities that typically arise for high-power systems~\cite{Evans:2010}, and lock the interferometer at its working point.    We will not discuss these techniques here in detail, but refer to the review article by Adhikari~\cite{Adhikari:2012}.

Now returning to the SQL --- having recognized the correlation between sensing and back-action noise as the key to beating the SQL, when we examine Eq.~(\ref{c12compact}), we already see several major approaches towards beating the SQL:
\begin{enumerate}
\item  We can inject squeezed vacuum, while keeping measuring the output quadrature $b_{2}$~\cite{Caves:1981,Unruh:1982,Jaekel:1990}. This requires squeezing the input quadrature which is proportional to
\begin{equation}
b_2 \sim a_1 -\mathcal{K} a_2\,.
\end{equation}
The frequency dependence in $\mathcal{K}$ requires frequency-dependent squeezing angle, while in turn is realizable by filtering frequency-independent squeezing through detuned Fabry-Perot cavities, as realized by Kimble et al.~\cite{Kimble:2001}. See right panel of Fig.~\ref{fig:noise} for the noise spectrum of such a configuration (red dotted curve).  We have also shown the noise spectrum in presence of $1\%$ of optical loss (blue dotted curve). 
\item  
  We can also  detect a different quadrature than the phase quadrature, for example
  \begin{equation}
b_\zeta\propto  b_2 +\mathcal{K} b_1
 \end{equation}
  so that back-action noise would cancel out, leading the same fraction of both signal and shot noise. 
  This is often referred to as the {\it variational readout} scheme, and was originally proposed in the time domain by Vyatchanin et al.~\cite{Vyatchanin:1993,Vyatchanin:1995}. This is a {\it back-action evading} scheme --- instead of trying to ``squeeze'' the back-action noise, this scheme simply avoids looking at it.   Similar to (i), this {\it back-action evasion} approach requires detecting a frequency-dependent output quadrature --- also realizable by the filters invented by Kimble et al.~\cite{Kimble:2001}. See right panel of Fig.~\ref{fig:noise} for the noise spectrum of such a configuration (dashed curve).   We have also shown  the noise spectrum in presence of $1\%$ of optical loss (blue dashed curve).  
  A scheme with equivalent sensitivity, proposed by Tsang and Caves~\cite{Tsang:2010}, is to filter the out-going light with an additional ``squeezer'', which lets the signal go through but ``un-squeezes'' the quantum noise.  All these back-action-evading schemes are very susceptible to losses; Khalili has estimated that they are all limited by
  \begin{equation}
\sqrt{  S_h^{\rm BAE}/S_h^{\rm SQL} } \ge \left(e^{-2q}\epsilon\right)^{1/4}\,.
  \end{equation}
  Here $e^{-2q}$ is the squeezing factor, and $\epsilon$ is the total optical loss.  For this reason, 10\,dB squeezing and 1\% optical loss will be limited to factor $\sim 5.6$ below the SQL. 
  \item
   Finally, we may also modify the optical system so that $\mathcal{K}$ does not have much frequency dependence, so we do not have to rely (as much) on filters.  Some of these schemes~\cite{Purdue:2002a,Purdue:2002b,Chen:2003a,Danilishin:2004} were ``speed meters'', motivated by the fact that momentum of a free mass is a QND observable~\cite{Braginsky:1990}.  
   \end{enumerate}

As we can see from here, injection of squeezed vacuum is important for achieving sub-SQL sensitivity for a broad frequency band. Prior to application to gravitational-wave detection, squeezing~\cite{Xiao:1987} was mostly performed for sideband frequencies much higher than the most promising GW frequency band from astrophysical sources (up to $\sim 10\,$kHz~\cite{Cutler:2002}).   After McKenzie et al.\ demonstrated the possibility of squeezing within the GW band~\cite{McKenzie:2002}, low-frequency squeezing has been perfected~\cite{Vahlbruch:2006} and applied to prototype~\cite{Goda:2008}  and large-scale interferometers~\cite{Abadie:2011}.  Frequency-dependent squeezing and variational readout has been demonstrated in table-top experiments~\cite{Chelkowski:2005}.

As we have also seen from Fig.~\ref{fig:noise}, optical loss may pose a serious limitation to the application of squeezing.  The  second-generation detectors of Advanced LIGO (currently under construction) is projected to have a 20\% of optical loss, which will only be able to take advantage of a 6\,dB input squeezing, although efforts are being made to suppress optical losses in large-scale interferometers in anticipation of the application of squeezing~\cite{Dwyer:2012}. Interested readers are referred to these review articles on quantum noise of advanced gravitational-wave detectors~\cite{Schnabel:2008,Schnabel:2010,Sheon:2011,McClelland:2011,Danilishin:2012a}.    In the second part of this paper (Sec.~\ref{sec:exp_linear} and on), we shall see some of these SQL-beating techniques applied or adapted to the study of macroscopic quantum mechanics.

\subsubsection{Beating the SQL in other optomechanical systems}
\label{subsec:sql:other}

The above strategies of beating the SQL for GW detectors have focused on obtaining broadband sensitivity for a nearly free test mass --- although additional strategies exists for beating the SQL for high-$Q$ oscillators.   Most notable are the {\it variational}~\cite{Thorne:1978} and the {\it stroboscopic}~\cite{Braginsky:1978} approach. 

The variational approach attempts to measure a particular quadrature of the mechanical oscillator, 
\begin{equation}
X_\theta = x(t)\cos(\omega  t +\theta) - \frac{p(t)}{m\omega}\sin(\omega  t +\theta)
\end{equation} 
which is a QND observable, by varying measurement strength as a function of time.  This has been implemented experimentally  by Herzberg et al.~\cite{Hertzberg:2009} near the SQL.    The variational approach of beating the SQL has later been adapted to broad-band detectors~\cite{Vyatchanin:1995,Vyatchanin:1993,Kimble:2001}. 

The stroboscopic approach~\cite{Braginsky:1978} probes the position of an oscillator with pulses of measurements, separated by half period of oscillation.  This approach evades back action because any pulsed force exerted instantaneously onto the mechanical oscillator does not affect position of the oscillator at time delays that are exactly integer times the half oscillation period. 

\subsection{Adiabatic elimination of cavity mode}
\label{subsec:elimination}

The straw-man model described so far contains the interaction between the mirror and the cavity mode.  In fact, it is the cavity mode that couples directly to fields in the external continuum, which in turn gets detected.     In the limit when $\gamma$ and $\Delta$ are large compared with $\Omega$, the frequency of interest, the cavity mode will respond {\it almost instantaneously} to the motion of the mirror.  This kind of degrees of freedom can be {\it Adiabatically Eliminated} from system dynamics.   [In connection to this, certain {\it linear} degrees of freedom can be {\it non-adiabatically eliminated}, see Ref.~\cite{Yang:2012b}.]

In our problem, because of linearity, we can carry out adiabatic elimination of the cavity mode $(a,a^\dagger)$ in the Heisenberg Picture, converting Eqs.~\eref{eqx}--\eref{eqcoutd} into
\begin{eqnarray}
-i\Omega \hat x &=& \hat p/M\,, \\
-i\Omega \hat p &=& \left[-M\omega_m^2  +\frac{2G^2\Delta}{\gamma^2+\Delta^2}\right]\hat x\nonumber\\
& +&\frac{2\sqrt{\gamma} G(\gamma \hat a_1 +\Delta \hat a_2)}{\gamma^2+\Delta^2},\qquad\;
\label{adeqp}
\end{eqnarray}
and
\begin{eqnarray}
\left[\begin{array}{c}
\hat b_1\\ \hat b_2 
\end{array}\right]
&=&
\frac{1}{\Delta^2+\gamma^2}\left[\begin{array}{cc}
\Delta^2-\gamma^2 & -2\gamma\Delta\\
2\gamma\Delta & \Delta^2-\gamma^2
\end{array}\right]
\left[\begin{array}{c}
\hat a_1 \\
\hat a_2 
\end{array}\right]\nonumber\\
&-&
\frac{2\sqrt{\gamma} G }{\Delta^2+\gamma^2}\left(
\begin{array}{c}
\Delta \\
\gamma
\end{array}
\right) \hat x
\end{eqnarray}
We shall use this simpler model in most of our subsequent discussions. 

Note that the response of the mechanical oscillator's position $\hat x$ to the external force $F$ is now modified due to the existence of the second term in the bracket on the right-hand side of Eq.~\eref{adeqp} --- we shall explain the physical origin of this term in Sec.~\ref{rigidity} below. Here we only observe that this indicates a renormalized Hamiltonian with a shifted mechanical eigenfrequency:
\begin{equation}
\omega_m^2 \rightarrow \omega_{\rm opt}^2= \omega_m^2 +\frac{2G^2\Delta}{\gamma^2+\Delta^2}
\end{equation}
After re-defining the input and output quadratures by applying constant rotations (which corresponds to a microscopic propagation distance), we have a system with $\hat x$ measured, with an interaction Hamiltonian of
\begin{equation}
V_I = -\hbar \alpha \hat x \hat a_1\,,
\end{equation}
where
\begin{equation}
\alpha = 2G\sqrt{\frac{\gamma}{\gamma^2+\Delta^2}} =\sqrt{\frac{2 M \Theta^3\gamma}{\gamma^2+\Delta^2}}
\end{equation}
It is often convenient to write
\begin{equation}
\alpha^2 =  M\Omega_q^2
\end{equation}
with
\begin{equation}
\Omega_q^2 =\frac{2\Theta^2\gamma}{\gamma^2+\Delta^2}
\end{equation} 

If we apply a constant rotation to quadratures $a_{1,2}$ and $b_{1,2}$ it is possible to write an input-output relation in the same form as Eqs.~\eref{c12compact}: 
\begin{eqnarray}
\hat b_1 = \hat a_1 \,,\quad  \hat b_2 = \hat a_2 +\alpha \hat x
\end{eqnarray}
and
\begin{equation}
- M(\Omega^2-\omega_{\rm opt}^2)  \hat x =\alpha\hat a_1 +\hat F
\end{equation}
In other words, we have
\begin{equation}
\beta=0\,,\quad \mathcal{K} = \frac{\Omega_q^2}{(\Omega^2-\omega_{\rm opt}^2)}\,,
\end{equation}
and
\begin{equation}
\label{SFSQLspring}
 S_F^{\rm SQL}= 2M |\Omega^2 -\omega_{\rm opt}^2|
\end{equation}
For a free mass ($\omega_{\rm opt} \ll \Omega, \omega_q$), if we measure out-going field $b_2$, our sensitivity to $F$ will be limited by the SQL, touching it at frequency $\Omega = \omega_q$.  

In the following sections, we shall often use this simplified model as our strawman for studying quantum measurement. However, we should be aware that sometimes adiabatic elimination ignores features of the system that may become important.  One example will be discussed in Sec.~\ref{rigidity} below, where a full treatment reveals damping or anti-damping plus additional noise imposed onto the mechanical object.

\begin{figure}
\centerline{\includegraphics[width=2.5in]{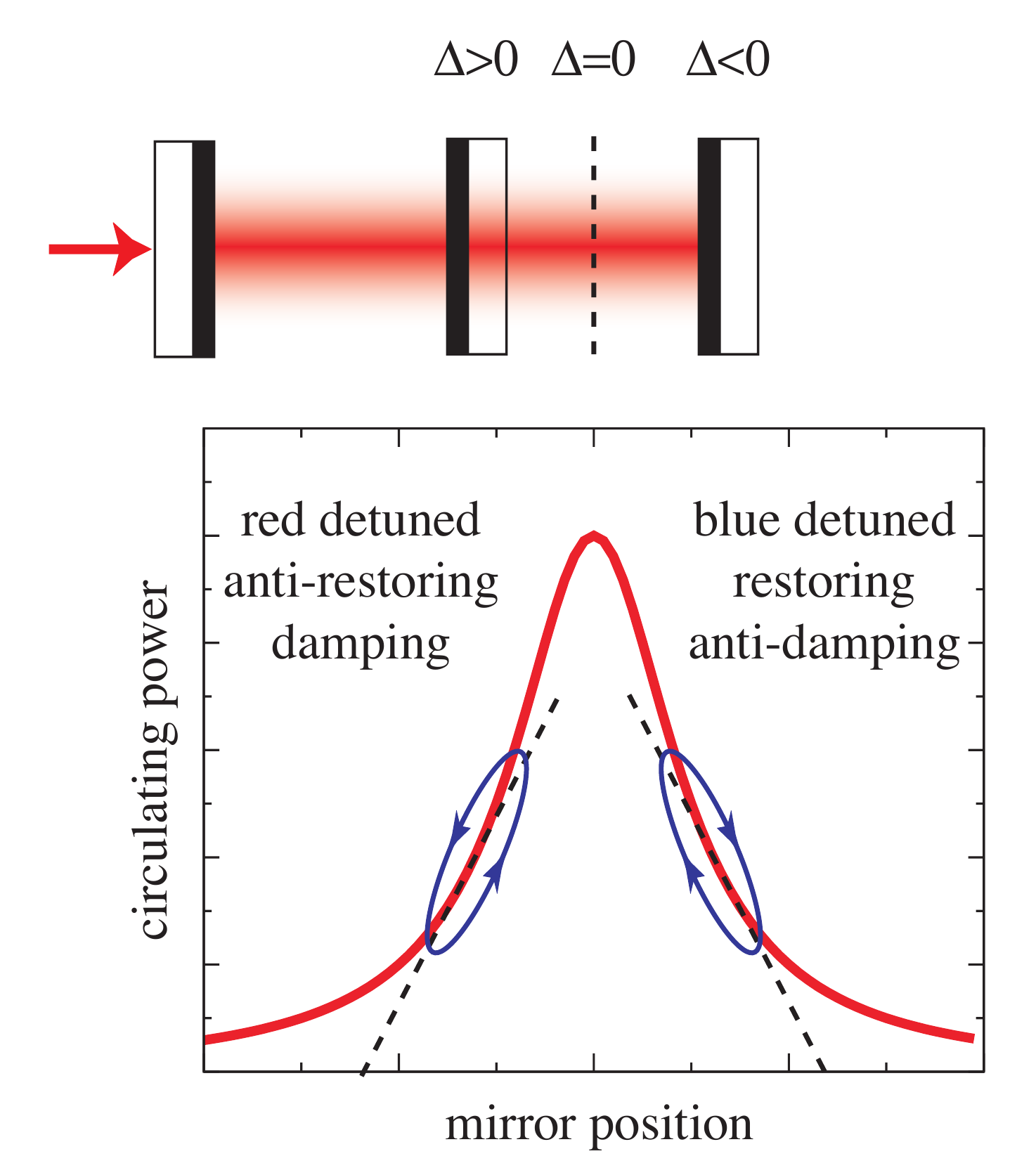}}
\caption{For a cavity locked at non-zero detuning, its movable end mirror not only feels a spring constant, which can be either positive or negative, but also gets damped or anti-damped.  For optical spring of a single carrier and at low frequencies, the sign of the damping can be worked out by assuming that the force lags the position by a small phase, in this way a restoring force always comes with a negative damping --- that is, the light always does positive work to the mirror within each cycle of motion. 
 \label{fig:rigidity}}
\end{figure}

\subsection{Optical rigidity ($\Delta \neq 0$)}
\label{rigidity}

As we have already noted in the above section, if the cavity is detuned ($\omega_c-\omega_0= \Delta \neq 0$), the dynamics of the mirror will be modified.  The simplest way to understand this is from a classical point of view.  Suppose we slowly vary the location of the mirror by a small amount (much less than the wavelength divided by the finesse of the cavity) --- because the zero point of the mirror is not on resonance with the injected carrier, the power inside the cavity, and hence the radiation-pressure force on the mirror, will depend linearly on $\delta x$.  (This is illustrated in Fig.~\ref{fig:rigidity}.)

The fact that $\hat F$ has a non-zero response is mathematically  equivalent to~\cite{Buonanno:2001,Buonanno:2001a,Buonanno:2002a,Buonanno:2003a},
\begin{equation}
\label{commFF}
\left[\hat F_H(t), \hat F_H(t')\right] \neq 0\,.
\end{equation}
In general, by solving for $\hat F_{\rm BA}$ from Eq.~\eref{eqA} and \eref{eqAd}, we can obtain a frequency-dependent spring constant
\begin{eqnarray}
\label{eqK}
K(\Omega) =  -\frac{M \Theta^3\Delta}{\Delta^2 +\gamma^2-2 i\gamma\Omega-\Omega^2} \,.
\end{eqnarray}

\subsubsection{Modification of dynamics}

Assuming that the mechanical oscillator starts off at a high quality factor, for low enough pumping power, this $K$ only shifts the oscillator's eigenfrequency weakly, and we can write
\begin{eqnarray}
\label{optoshift1}
\omega_m^2 &\rightarrow& \omega_m^2+\frac{\mathrm{Re}\left[K(\omega_m)\right]}{m} \,, \\
\gamma_m &\rightarrow& \gamma_m-\frac{\mathrm{Im} \left[K(\omega_m)\right]}{2 m\omega_m}\,.
\end{eqnarray}
On the other hand, for frequencies below optical frequency scales, or $\Omega \ll \sqrt{\Delta^2+\gamma^2}$, we return to the ``slowly varying'' case mentioned at the beginning of this section, and have 
\begin{eqnarray}
\label{eqKapprox}
K(\Omega) & \approx & - \frac{m\Theta^3 \Delta}{\Delta^2 +\gamma^2} - \frac{2 m\Theta^3\Delta\gamma }{(\Delta^2+\gamma^2)^2}(i\Omega)  \nonumber\\
&\equiv& K_0 + i\Omega K_1,\quad\;
\end{eqnarray}
then we basically find a shift in the oscillator's eigenfrequency and damping, given by 
\begin{equation}
\omega_m^2 \rightarrow \omega_m^2+\frac{K_0}{m}\,,\quad
\gamma_m \rightarrow \gamma_m-\frac{K_1}{2m}\,.
\label{optoshift2}
\end{equation}
which indicates that anti-damping ($K_1>0$) is always associated with positive rigidity ($K_0>0$, both takes place when $\Delta < 0$, or the light frequency $\omega_0$ is {\it higher} than cavity resonance $\omega_0+\Delta$, or {\it blue detuned} with respect to cavity resonance), while damping ($K_1<0$) is always associated with negative rigidity ($K_0<0$, both takes place  when the cavity is red detuned, or $\Delta > 0$, or the light frequency $\omega_0$ is {\it lower} than cavity resonance $\omega_0+\Delta$, or {\it red detuned} with respect to cavity resonance).    As illustrated by Fig.~\ref{fig:rigidity}, for such a simple optical spring (arising from one optical mode),  a restoring force with time lag will be associated with anti-damping.  As more complex optical modes are used to generate the optical spring, any combinations of restoring/anti-restoring vs damping/anti-damping could be possible~\cite{Tarabrin:2012}.

\begin{figure}
\centerline{\includegraphics[width=3in]{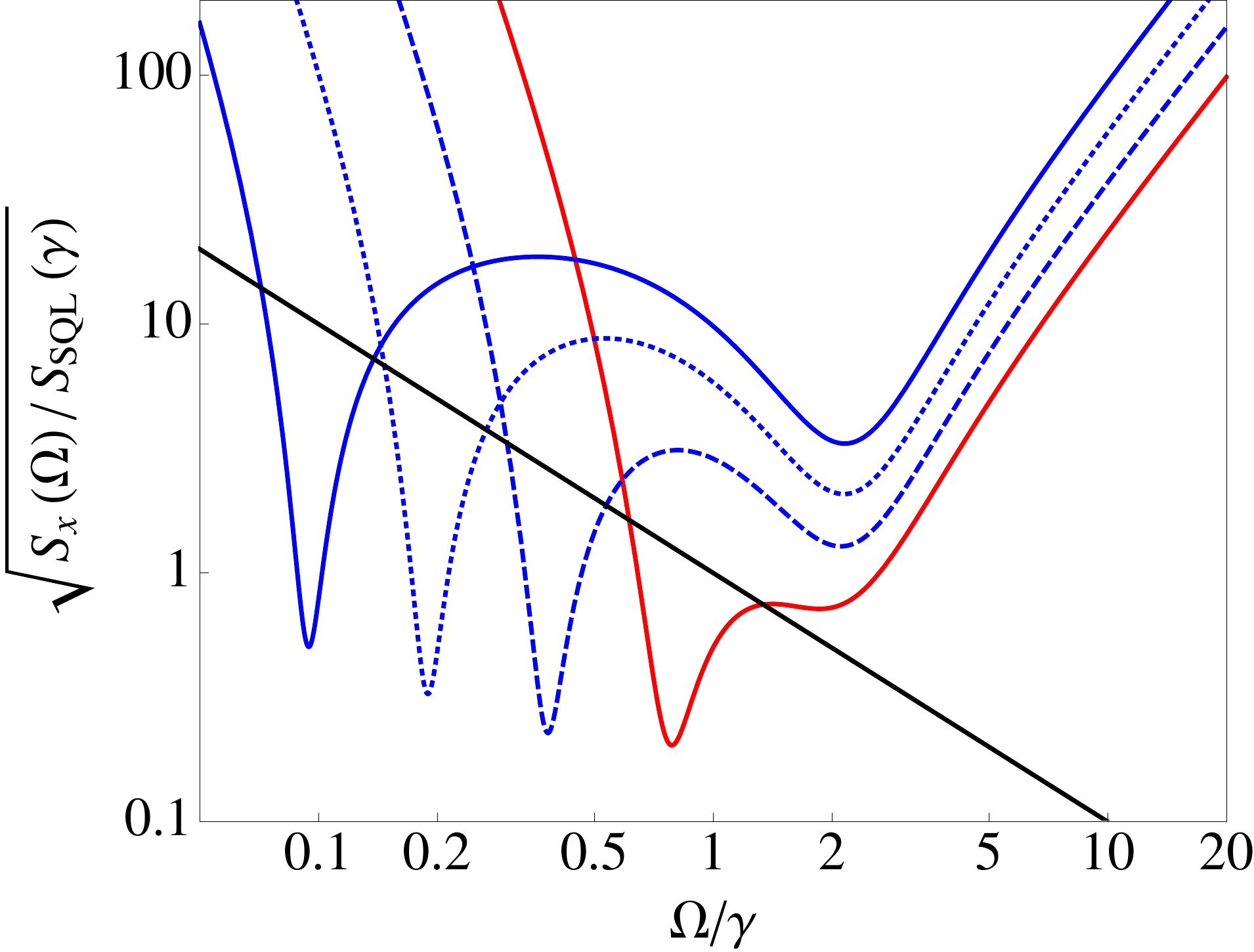}}
\caption{Noise spectrum of interferometers with optical springs [caused by non-zero input impedance, from the point of view of the test mass, see Eq.~\eref{commFF}].  We have again normalized frequency to $\gamma$ and noise spectrum to $S_{\rm SQL}(\gamma)$.  Interferometers here have $\Delta =-1.5\gamma$ (blue detuned) and $\Theta = \gamma$ (red curve) and $2^{-2/3}\gamma$ (blue dashed), $2^{-4/3}\gamma$ (blue dotted) and $2^{-6/3}\gamma$ (blue dashed).  These roughly corresponds to optical spring constant decreasing by factors of 4, and optical resonant frequency decreasing by factors of 2. \label{fig:noise_rigidity}}
\end{figure}



\subsubsection{Improving force sensitivity}

For gravitational-wave detection (a weak force measurement), where we start from a pendulum with very low suspension frequency (below the detection band), the {\it  blue detuned case} ($\Delta < 0$)  is useful: the positive rigidity upshifts the resonant frequency of test-mass translational motion, and allows us to improve sensitivity in the detection band, see Fig.~\ref{fig:noise_rigidity} for sample noise spectrum for configurations where $\Delta$ and $\gamma$ are the same order of magnitude, while $\Theta$ gradually increases from a low value to a value comparable to $\gamma$.  Here we have chosen to represent sensitivity not in terms of a noise spectrum referred to force $F_{\rm GW}$, but instead have referred to the displacement $x_{\rm GW}$ caused by the GW if the test masses were free. This suppression in noise is in fact due to the test mass's resonant response to the weak classical force, instead of a  suppression of quantum noise.
More specifically, for a strain $h$, $x_{\rm GW}$ for a free mass is $Lh$, while for an oscillator, we should note that 
\begin{equation}
\tilde F_{\rm GW}(\Omega) = -M\Omega^2 L \tilde h(\Omega)
\end{equation}
and therefore
\begin{equation}
S_h = \frac{S_F}{M^2\Omega^4 L^2}
\end{equation} 
and therefore
\begin{equation}
S_h^{\rm SQL} = \frac{S_F^{\rm SQL}}{M^2\Omega^4 L^2}= \frac{2 \hbar|\Omega^2-\omega_{\rm opt}^2|}{M \Omega^4 L^2}
\end{equation} 
This means, near the new, optomechanical resonance of the test mass, $\Omega\sim \omega_{\rm opt}$, the SQL itself is highly suppressed from the free-mass value~\eref{SFSQLspring} --- therefore simply following but not beating the new SQL would already indicate a strong improvement in sensitivity and a strong beating of the free-mass SQL.  

On the other hand, the {\it anti-damping} brought by the blue detuning may make the system dynamically unstable, yet the control system already in place in these detectors can be used to stabilize the system without (in principle) scarifying quantum-limited sensitivity~\cite{Buonanno:2002a}.

Although the optical spring's improvement of detector sensitivity can often be viewed as classical, it is not as trivial as simply replacing the free-mass detector by a resonant detector: the optical spring's shift of mechanical resonant frequency only comes with a low amount of radiation pressure noise (a large fraction of which can even be removed, as Korth et al.\ have later shown~\cite{Korth:2012}), many other mechanisms of doing so, e.g., through elasticity, will bring a significant amount of thermal noise.   When compared with back-action-evading techniques, optical spring's improvement of detector sensitivity is much more robust against optical losses.  Unfortunately, however, the improvements we are so far able to achieve have been narrowband --- around the resonant frequency of the optical spring.   Progress has been made toward constructing frequency-dependent optical springs that amplify weak force signal in broad frequency bands~\cite{Mueller-Ebhardt:2008a,Khalili:2011a,Tarabrin:2012}, yet a scheme suitable for broadband gravitational-wave detection has not been found. 

Finally, let us comment on the relation between the existence of the back-action noise (hence SQL) and the shift in test-mass dynamics. It is in fact  a generic feature of linear quantum measurement processes, that as we dial up the measurement strength from zero,  the change in dynamics of the system being measured is significant (due to input impedance of $\hat F$) when the back action noise (due to the fluctuations in $\hat F$) is comparable to sensing noise.  This originates from the fact that the noise spectrum arises from the symmetric Green's function of $\hat F$ (evaluated in vacuum state), while the rigidity arises from the anti-symmetric Green's function.  

\subsubsection{Experimental demonstrations of optical rigidity.}
\label{sub:ex:rigidity}

  Electrical and optical rigidity that arise due to parametric coupling was described theoretically by Braginsky and Manukin in the 1960s~\cite{Braginskii:1967}, and experimentally studied in the 1970s~\cite{Braginskii:1970}.  
It was later observed experimentally and studied theoretically for resonant-bar gravitational-wave detectors (although those experiments used electronic instead of optical readout)~
\cite{
Hirakawa:1979,
Linthorne:1990,Tobar:1993,Blair:1995,Cuthbertson:1996}. 
%
%
Optical spring in laser interferometers was later studied in the gravitational-wave community as an effort to prototype the Advanced LIGO interferometers~\cite{Sheard:2004,Somiya:2005,Miyakawa:2006}.  Optical rigidity and damping turns out to be very useful for studying macroscopic quantum mechanics.  We shall devote the entire Sec.~\ref{subsec:cooling} dealing with this issue.


\section{Stochastic approach}
\label{subsec:stochastic}

Having finished introducing the first approach toward quantum measurement, we now turn to the second approach, which focuses on the state of a quantum system that is being measured continuously.  We will illustrate this approach using the  ``adiabatically eliminated model'' of Sec.~\ref{subsec:elimination}, which is equivalent to having no cavity but a redefined mirror dynamics and a shifted coupling constant to incoming light.  We can write the linearized interaction Hamiltonian 
\begin{equation}
\hat V_I = -\hbar\alpha \hat x \hat a_1
\end{equation}
where $\hat x$ is the position of the mechanical oscillator, $\hat a_1$ the amplitude quadrature of the optical field, and $\alpha$ the coupling strength.   We will assume that the out-going phase quadrature $\hat b_2$ is measured.

We require that the incoming field degrees of freedom arriving at different times are not correlated.  This makes our measurement process {\it Markovian}.  During any small interval of time, e.g., $[t,t+\Delta t]$, a ``fresh'' (uncorrelated) section from the continuum of the incoming $\hat a$ field is brought to interact with the mechanical object, it then promptly returns to the outside world, and gets detected.  

To understand this elementary process better, suppose we have two quantum systems, $A$ (the incoming field) and $B$ (the mechanical object), initially at a pure product state:
\begin{equation}
|\Psi_0 \rangle = |\phi\rangle_A\otimes |\psi\rangle_B\,.
\end{equation}
The fact that $B$ is at a pure state is a mere Ansatz, while the fact that $A$ and $B$ are product states is due to Markovianity: the newly arriving field has no correlation with our current system because it has no correlations with the fields that had entered the system in the past. 
  Suppose $A$ and $B$ are brought to interact with each other, and the joint system evolves into a new state of
\begin{equation}
|\Psi_1\rangle =  \hat U |\phi\rangle_A\otimes |\psi\rangle_B
\end{equation}
which is in general not a product state. Here $\hat U$ is a unitary evolution operator. 
A projective measurement on system $A$ will bring system $B$ back into a {\it pure state}, although a {\it stochastic} one that depends on the measurement outcome: suppose we measure
$\hat O$ (of system $A$), which has eigenstates of $|\tilde \phi_k\rangle_A$ and eigenvalues $o_k$, then the probability of the $k$-eigenvalue to appear is
\begin{equation}
P_k = \left\|{}_A\langle \tilde \phi_k |\hat U |\phi_A\rangle \otimes |\psi \rangle_B\right\|^2\,.
\end{equation}
in which case the system $B$ will be left at a pure state of:
\begin{equation}
\frac{1}{\sqrt{P_k}}{}_A\langle \tilde \phi_k |\hat U |\phi_A\rangle \otimes |\psi \rangle_B\,.
\end{equation}
This will often be referred as the {\it conditional state} of $B$. The fact that the conditional state of $B$ remains pure prepares the Ansatz we need for the next time step, and assures the existence of a stochastic evolution of the system's conditional pure state as the measurement is to be repeated.

\subsection{Derivation of the Stochastic Schr\"odinger Equation}
\label{subsec:SSE}

In a continuous quantum measurement process, the above elementary process takes place repeatedly, therefore requires the mathematical machinery of stochastic calculus as we take the continuous-time limit.  Suppose a system starts out at a pure state independent from the measuring device, then its state will remain pure, but evolve over time depending on the measurement outcome.  Suppose  we have already reached  $t$, and would like to proceed towards $t+ \Delta t$.  During this time, let us define
\begin{equation}
\hat P \equiv -  \frac{1}{\sqrt{\Delta t}}\int_t^{t+\Delta t} \hat a_1(t') dt' \,,\;\;\hat Q \equiv \frac{1}{\sqrt{\Delta t}}\int_t^{t+\Delta t} \hat a_2 (t') dt'\,, 
\end{equation}
and
\begin{equation}
\left[\hat Q,\hat P\right] =  i,
\end{equation} a pair of canonical coordinate and momentum that satisfy the canonical commutation relation.  Suppose this degree of freedom is initially at ground state, and we measure its location after it interacts with our system, through the interaction Hamiltonian. 
The evolution operator here is 
\begin{equation}
\hat U =\exp\left( i\alpha \hat P \hat x \sqrt{\Delta t}\right)\,.
\end{equation}

The joint initial state of the system plus the measuring device is
\begin{equation}
|\Psi(t)\rangle = |\psi(t)\rangle \otimes |0\rangle
 = \int dQ 
 \frac{e^{-Q^2/2}}{\pi^{1/4}}|\psi(t) \rangle |Q\rangle\,, 
\end{equation}
with $|Q\rangle $ the $\hat Q$ eigenstate of the measuring device. After application of $\hat U$, we obtain [noting that when applied onto $Q$-wavefunctions, we have $\hat P =- i\partial/\partial Q$] 
\begin{equation}
|\Psi(t+\Delta t)\rangle 
=\int dQ \frac{\exp\left[-\frac{(Q- \alpha\hat x \sqrt{\Delta t})^2}{2}\right]}{\pi^{1/4} }|\psi(t)\rangle |Q\rangle \,.
\end{equation}
This new state needs to be projected onto a basis that corresponds to the observable we are going to measure.  Note that we measure the evolved quantity of 
\begin{equation}
\hat Q_{\rm new} =  \hat U^\dagger \hat Q \hat U =\hat Q + \alpha \hat x \sqrt{\Delta t}\,.
\end{equation}
This operator has an expectation value of $ \alpha \langle \hat x \rangle \sqrt{\Delta t}$, contributed only from $\hat x$ (because $\hat Q$ has zero mean) while its variation is at the level of $\sim O(\Delta t^0)$, therefore dominated by the uncertainty of $\hat Q$. Let us denote our measurement result as
\begin{equation}
\tilde Q = \alpha\langle \hat x\rangle \sqrt{\Delta t} +  \Delta W/\sqrt{2\Delta t}
\end{equation}
then $\Delta W$, in our limiting process of $\Delta t \rightarrow 0$, will become a Wiener increment because different $\Delta W$'s are independent, and each has variance $\Delta t$, see App.~\ref{app:wiener} for details on the Wiener process. [Also note that $\langle\hat Q^2\rangle = 1/2$.] We will then carry out the projection, and obtain the non-normalized {\it conditional state} of 
\begin{eqnarray}
&&|\psi(t+\Delta t)\rangle  \nonumber\\
&=& \langle \tilde Q |\Psi(t+\Delta t)\rangle \nonumber\\  
&=& \frac{1}{\pi^{1/4}}
\exp\left\{-\frac{1}{2}\left[\alpha( \hat x- \langle \hat x\rangle)\sqrt{\Delta t}+\frac{\Delta W}{\sqrt{2\Delta t}}\right]^2\right\}  |\psi(t)\rangle \nonumber\\
&=& \frac{e^{-\frac{\Delta W^2}{4\Delta t}}}{\pi^{1/4}} \Bigg[1+\frac{\alpha(\hat x-\langle\hat x\rangle)\Delta W}{\sqrt{2}} -\frac{\alpha^2(\hat x -\langle\hat x\rangle)^2}{4}\Delta t\Bigg] |\psi(t)\rangle\nonumber\\
%
\end{eqnarray}
Note that the factor in front represents a probability density for the value of $\Delta W$, while the evolution of the normalized conditional state at $t+\Delta t$ is given by the following Stochastic Schr\"odinger Equation (SSE):
\begin{eqnarray}
\label{eq:SSE}
d |\psi\rangle = - \frac{i}{\hbar} \hat H |\psi\rangle  dt &+&\frac{\alpha\left(\hat x- \langle \hat x \rangle\right)}{\sqrt{2}} |\psi\rangle {dW}   \nonumber\\&-&\frac{\alpha^2}{4}\left(\hat x -\langle \hat x\rangle\right)^2  |\psi\rangle dt\,.
\end{eqnarray}
The SSE describes the evolution of a conditional state, which always stays normalized.   The measurement outcome, in the continuous-time limit, also satisfies a stochastic differential equation:
\begin{equation}
\label{eq:SDE:outcome}
dy = \alpha \langle \hat x \rangle  + dW/\sqrt{2}\,.
\end{equation}
The SSE~\eref{eq:SSE} is meant to be used together with Eq.~\eref{eq:SDE:outcome} --- they are simultaneously determined by each realization of $W(t)$, which is a Wiener process.

In case there is already  uncertainty in the wave function, e.g., due to our initial ignorance and/or other channels of decoherence, using the same technique as above, we obtain the Stochastic Master Equation (SME) for the density matrix of the ensemble:
\begin{eqnarray}
\label{sme}
d\hat \rho &=& -\frac{i}{\hbar} \left[\hat H,\hat\rho\right] dt +\frac{\alpha}{\sqrt{2}}
\left\{\hat x-\langle\hat x\rangle ,\hat \rho\right\} dW \nonumber\\
&-&\frac{\alpha^2}{4}\left[\hat x,\left[\hat x,\hat\rho\right]\right]dt\,.
\end{eqnarray}
Here  we have used ``$\{...\}$''  to denote the anti-commutator:
$\{\hat A,\hat B\} =\hat A \hat B + \hat B \hat A$].  In Eq.~\eref{sme}, the second term on the right-hand side is due to the randomness of back action, while the third term describes dissipation --- it is often referred to as the {\it Lindblad Term}.   If we do not record measurement data, the $dW$ will be averaged out in Eq.~\eref{sme}, and we obtain a Master Equation, similar to one that describes a system coupled to a bath with zero temperature.   In other words, a measurement whose data were thrown out is no different from a source of dissipation.

Note that the SSE and SME are {\it nonlinear}, because the evolutions of $|\psi\rangle$ and $\hat \rho$  depend on the expectation value of $\hat x$ on $|\psi\rangle$.  For this reason, if we have two density matrices $\hat\rho_1$ and $\hat\rho_2$ (both normalized, with $\tr\hat\rho_{1,2} =1$) that are superimposed at $t=0$,  
\begin{equation}
\hat\rho(0)  =p_1 \hat\rho_1(0) + p_2 \hat\rho_2(0)
\end{equation}
with $p_1+p_2=1$. Even for the same realization of $\{W(t'):0<t'<t\}$, the initial state $\hat\rho(0)$ will not evolve to the same superposition of the final states that correspond to $\hat\rho_1(0)$ and $\hat\rho_2(0)$, or 
\begin{equation}
\rho(t)   \neq p_1 \hat\rho_1(t) + p_2 \hat\rho_2(t)\,.
\end{equation}
In fact, one can show~\footnote{Noting that the only nonlinear term in Eq.~\eref{sme} is the one containing $\langle \hat x\rangle$, which simply provides a time-dependent normalization factor for $\hat \rho$.} that there does exist $(p_1',p_2')$ with $p_1'+p_2'=1$ so that
\begin{equation}
\rho(t)   = p_1' \hat\rho_1(t) + p_2' \hat\rho_2(t)\,.
\end{equation}
but in general $(p_1',p_2')\neq (p_1,p_2)$.  One can make the interpretation that state reduction during $0<t'<t$ changes our understanding of the  probability for the system to have started out from the ensemble represented by $\hat \rho_1$ versus the ensemble represented by $\hat \rho_2$ --- this is the origin of  the nonlinearity. In other words, $(p_1,p_2)$ are the prior probabilities, while $(p_1',p_2')$ are the posterior probabilities --- and they naturally differ after observation.  {\it Within} each ensemble, however, the individual SME starting from $\hat\rho_1(0)$ and $\hat \rho_2(0)$ still predicts the correct evolution.

There are two specific uses for the SSE and SME. The {\it first} follows our derivation: the SSE/SME  {\it simulates} the continuous state-reduction process by producing a stochastic evolution of the true state $\hat \rho$ of the system/ensemble, driven by the randomness of state reduction --- described mathematically  by $dW$, which has an {\it a priori} probability distribution.  This treats wave function as known, {\it a priori}, and therefore  corresponds to a {\it frequentist} approach to statistics. 

The {\it second} treats $\hat\rho$ as describing the experimenter's knowledge (and uncertainty) of the quantum system --- this includes quantum uncertainty contained {\it within the wave function} as well as his/her {\it classical ignorance} regarding the wave function of the system itself.  $dW$ is the error between the measurement result and its conditional expectation from $\hat \rho$ (which is a ``best guess'') --- the experimenter then should use the SME to update $\hat\rho$ continuously.  This corresponds to the {\it Bayesian} approach to statistics. 


\subsection{Filter Equations for Linear Systems with Gaussian States}

For linear systems initially at Gaussian states, the states will remain Gaussian, which can in turn be characterized by the first and second moments of position and momentum.  In order to derive equations for those moments, we can  use the stochastic differential equation satisfied by the Wigner function, which can be obtained using the generating function approach (see \myapp\  \ref{app:wig}, although this equation also applies to  non-Gaussian states):
\begin{eqnarray}
\label{SMEW}
&&d\mathcal{W}(x,p)+\left[\frac{p}{M}\partial_x  -M\omega^2 x \partial_p \right]\mathcal{W}(x,p) dt \nonumber\\
&=&\sqrt{2} \alpha(x-\langle \hat x\rangle) \mathcal{W}(x,p) dW +\frac{\alpha^2}{4}
\partial_p^2 \mathcal{W}(x,p) dt\,.
\end{eqnarray}
Terms on the left-hand side arise from the free evolution of a simple harmonic oscillator (which can be replaced to the free evolution of an arbitrary system), while those on the right-hand side arise from the measurement.  Suppose we have a Gaussian state, Eq.~\eref{SMEW} can be converted into a self contained evolution of the conditional first moments ($\langle x \rangle_c$, $\langle p \rangle_c$) 
\begin{eqnarray}
d\langle  x\rangle_c &=&\frac{ \langle p\rangle_c}{M} dt +\sqrt{2} \alpha V_{xx} dW \,,\\
d\langle  p\rangle_c &=& -M\omega^2_m \langle  x\rangle_c dt +\sqrt{2}\alpha V_{xp} dW 
\end{eqnarray}
and second moments $(V_{xx}^c,V_{xp}^c,V_{pp}^c)$
\begin{eqnarray}
\dot{V}_{xx} &=& \frac{2V_{xp} }{M} - 2 \alpha^2 V_{xx}^2  
\label{vxx}
\\
\dot{V}_{xp} &=& \frac{V_{pp}}{M} -M\omega_m^2 V_{xx} - 2\alpha^2 V_{xx}V_{xp}\\
\label{vxp}
\dot{V}_{pp} &=& -2 M \omega_m^2 V_{xp} + \frac{\alpha^2}{2}  - 2\alpha^2 V_{xp}^2 
\label{vpp}
\end{eqnarray}
which completely characterize the evolving conditional state:  for any two {\it linear combinations} of $\hat x$ and $\hat p$, for example $\hat A$ and $\hat B$, we have 
\begin{equation}
\langle \hat A \hat B+\hat B \hat A\rangle/2 =\int\int A(x,p) B(x,p) \mathcal{W}(x,p) dx dp
\end{equation}
therefore defining covariance by symmetrizing the expectation value is the same as using Wigner function as quasi-probability distribution.

Note that measurement affects the evolutions of first moments, i.e., the expectation values $\langle x\rangle$ and $\langle p \rangle$, by inserting terms proportional to $dW$, which generate random driving; for the second moments, the diffusion in momentum due to back action leads to the $\alpha^2$ term that tends to increase $\dot V_{pp}$ in Eq.~\eref{vpp}. Measurement also leads to decrease of $V_{xx}$ and $V_{pp}$ through terms in Eqs.~\eref{vxx}-\eref{vpp} that contain $\alpha^2$ as well as the product of second moments --- these are due to the gathering of information about the oscillator.   The second kind of terms arise from the use of the Ito rule; for example, 
\begin{eqnarray}
dV_{xx} &=& d\langle x^2\rangle -d \langle x\rangle^2 \nonumber\\
&=& d\langle x^2\rangle-2\langle x\rangle \cdot d\langle x\rangle -  d\langle x\rangle \cdot  d\langle x\rangle
\end{eqnarray}
and the last term on the right-hand side will contain terms like $dW^2 =dt$.

In Appendix \ref{kalman}, we review the classical theory of {\it Kalman Filtering}, which describes a classical system driven by random force and under continuous measurement.   As it turns out, Kalman filtering allows us to obtain the same Eqs.~\eref{vxx}--\eref{vpp}. In fact, in the case of linear quantum systems with Gaussian states, there is always a classical system whose Kalman filter equations are exactly the same as the evolution equations for the first and second moments of the Gaussian state.   The reason why we can do so will be explained in Sec.~\ref{sec:unified}.

Equations~\eref{vxx}--\eref{vpp} can be viewed as a matrix Riccati equation, and this particular set can be solved analytically.  Here we only write down the steady-state solution, which can be obtained by setting their left-hand sides to zero~\cite{Hopkins:2003}:
\begin{eqnarray}
V_{xx} &=&\frac{\hbar}{\sqrt{2}M\omega_m}\cdot \frac{1}{\sqrt{1+\sqrt{1+\Lambda^4}}}\,,\\ V_{xp} &=& \frac{\hbar}{2}\cdot \frac{\Lambda^2}{1 + \sqrt{1+\Lambda^4}}\,,\\ 
V_{pp} &=& \frac{\hbar M\omega_m}{\sqrt{2}}\cdot \frac{ \sqrt{1+\Lambda^4}}{ \sqrt{1+\sqrt{1+\Lambda^4}}}\,,
\end{eqnarray}
where we have continued to use the definition of
\begin{equation}
\label{alphaomegaq}
\alpha^2 =  M\Omega_q^2 \,,\quad \Omega_q \equiv \Lambda \omega_m\,.
\end{equation}
Note that $V_{xp} >0$, and 
\begin{equation}
V_{xx}V_{pp}-V_{xp}^2 =\frac{\hbar^2}{4}\,,
\end{equation}
which means the steady-state conditional state is always minimum Gaussian, and hence a pure state. We note that as the measurement strength ($\alpha$, or $\Omega_q$) grows, the position uncertainty of the conditional state decreases, while the momentum uncertainty grows --- and there is always non-trivial correlation between position and momentum.  In fact, in the limit of strong measurement, or $\Omega_q \gg \omega_m$ (e.g., for free masses), we have
\begin{eqnarray}
V_{xx} &\sim&\hbar/(\sqrt{2} M\Omega_q),\; \\ 
V_{pp} &\sim& \hbar M\Omega_q/\sqrt{2},\; \\ 
V_{xp} &\sim& \hbar/\sqrt{2}.
\end{eqnarray} 
This is a {\it highly squeezed state} for an oscillator with eigenfrequency $\omega_m$, but a {\it mildly squeezed state} for an oscillator with eigenfrequency $\Omega_q$. 
As we recall from Sec.~\ref{subsec:elimination}, when $\Omega_q \gg \omega_m$, and if we use our device to measure a weak force  --- detecting the output field quadrature in which the entire signal is contained --- then $\Omega_q$ is also the frequency at which the quantum noise touches the SQL.  In other words, the ``frequency scale'' of the quantum state we prepare, is set by the measurement frequency scale. On the other hand, for very weak measurement, or $\Lambda \rightarrow 0$, the steady state of the oscillator is the ground state.  


\section{Non-Markovian linear systems}

\label{sec:unified}

In Sec.~\ref{sec:linear}, we presented the input-output Heisenberg formalism for linear optomechanical systems, and applied it to the calculation of noise spectra when measuring a weak classical force.  In Sec.~\ref{subsec:stochastic}, we presented the stochastic calculus approach, and applied it to describe the evolution of a general system under continuous measurement.   These two approaches looked very different --- although we know that for linear systems, they are really two very different ways of dealing with the same system, and therefore should provide the same answers, if we were to ask the the same questions. Since in this article, we are less interested in estimating a weak classical force acting on a quantum system, let us only show how the Heisenberg formalism in Sec.~\ref{sec:linear} can be used to calculate quantum-state evolution in a continuous measurement.  In some sense, we will be exposing features of quantum measurement that is ``hidden'' so well under the disguise of the Heisenberg formalism.   This is not only a sanity check:  for linear systems, this exercise will actually turn out answers that are often analytically simpler, especially for non-Markovian systems which are driven by classical noise and/or quantum fluctuations that are correlated in time (see Sec.~\ref{subsec:nongaussian} for the latter case). 

\subsection{State Reduction in an Indirect Quantum Measurement}
\label{subsec:path}

As in Sec.~\ref{sec:linear}, the formalism we shall present in this section best fits linear systems --- but unlike in Sec.~\ref{subsec:stochastic}, we do not require the system to be Markovian (driven by white noise).  Let us return to the formalism in \ref{subsec:linham}, and assume that an out-going quadrature field $b_\zeta$ is measured [Cf.~Eq.~\eref{eq:bzeta}]. Here $\zeta$ can be a function of time.  

Let us now  expose the state reduction process that has been hidden from our view in Sec.~\ref{sec:linear} --- by putting Eq.~\eref{projection} into use.   Now, assuming $b_\zeta$ to have been projectively measured during $0<t'<t$, and the measurement result has been $\{\xi(t'):0<t'<t\}$, then in the Heisenberg picture, Eq.~\eref{projection}  indicates that we should have projected out the conditional state of 
\begin{equation}
\label{psic_path}
|\psi_c\rangle = \int D[k(t')] e^{i \int_0^t k(t') \left[\hat b_\zeta(t') - \xi(t')\right]dt'}  |\mathrm{ini}\rangle\,,
\end{equation}
where $|\mathrm{ini}\rangle$ is the initial state of the optomechanical system at $t=0$, $\hat b_\zeta(t)$ is the Heisenberg operator of the quadrature of the out-going field which we measure, while $k(t')$ is an auxiliary function over which a path integral is performed in order to yield a projection operator at every instant of time. We shall often ignore writing $|\mathrm{ini}\rangle$ when using the Heisenberg picture, and simply use ``$\langle \ldots \rangle$'' when calculating expectation values at this initial state.  We are able to write Eq.~\eref{psic_path} this way also because 
\begin{equation}
\label{eq:bb}
\left[\hat b_\zeta(t') , \hat b_\zeta(t'')\right]=0\,.
\end{equation}
This means the Heisenberg operators which we  claim to measure projectively at different times are indeed {\it simultaneously} measurable, and therefore the entire measurement process simply projects out a simultaneous eigenstate of all these operators.  In fact, given $t>t'$, we can also write
\begin{equation}
\label{eq:ob}
\left[\hat O(t),\hat b_\zeta(t')\right]=0\,,\quad t>t'
\end{equation}
for any operator $\hat O$  (e.g., position and momentum of the mirror, excitation of the cavity mode, etc., and any of their linear combinations) that belongs to the optomechanical system~\cite{Buonanno:2002a}.  This is valid because $ b_\zeta$ is an out-going field, and anything acting on $b_\zeta$ at a certain earlier time $t'$ would not  propagate back to the system at a later time $t$; {\it vice versa}, anything that acts on $O$ at a certain later time   wouldn't retroactively be reflected by $b_\zeta$ at an earlier time. 

Because of Eqs.~\eref{eq:bb} and \eref{eq:ob}, at time $t$, we can write the generating function for the mechanical oscillator's Wigner function as
\begin{eqnarray}
&& J_c\left[u,v|\xi(t')\right] \nonumber\\ &\propto  &\langle \psi_c| e^{i\left[\mu \hat x(t) +\nu \hat p(t)\right] }|\psi_c\rangle  \nonumber\\
&=& 
\int D[k(t')] \Big\langle e^{
i\mu\hat x(t) +i\nu\hat p(t)+
i \int_0^t k(t') \left[\hat b_\zeta(t') - \xi(t')\right]} \Big\rangle,
\end{eqnarray}
and the conditional Wigner function $W$ can be written as an inverse Fourier transform:
\begin{equation}
W_c\left[x,p|\xi(t')\right] = \int  J_c\left[\mu,\nu|\xi(t')\right] e^{-i(\mu x +\nu p)}\,\frac{d\mu}{2\pi}\frac{d\nu}{2\pi}\,.
\end{equation}
This can be readily calculated for Gaussian states (see App~\ref{app:causal} for details);  as we shall see in Sec.~\ref{subsec:nongaussian}, the existence of such compact formulas also allow us to obtain analytical results when non-Gaussian quantum states including individual photons are injected.   

Our path-integral formulation here is an extension of the earlier work of Caves~\cite{Caves:1986}; the difference is that we consider a sharp projection of the out-going field, instead of a smeared out measurement of a system observable. For simple measurement processes, the two approaches are the same, but our extended approach will be able to accommodate more complex quantum states of the injected light.

\subsection{Relation to linear regression.}

We note that the commutation relations~\eref{eq:bb} and \eref{eq:ob}  tell us that for any $\theta$,  the oscillator's $\theta$ quadrature~\footnote{To emphasize the  distinction, henceforth in the paper we shall refer to $\hat x_\theta$ as ``oscillator quadrature'' or ``mass qudrature'', while $\hat a_\zeta$ and $\hat b_\zeta$ as ``optical quadratures''.}
\begin{equation}
\hat x_\theta (t) \equiv \hat x (t)\cos\theta + \frac{\hat p(t)}{M\omega_m} \sin\theta
\end{equation}
and the quadratures $\{b_\zeta(t'): t'<t\}$ of the out-going field which had chosen to measure, all commute with each other and hence can be treated as a classical random variable plus a classical random process. [This equivalence is proven rigorously in Appendix \ref{app:causal}.]   We can therefore simply use linear regression theory [see Appendix \ref{app:subsec:gauss} for details] to write down a predictor of any $\hat x_\theta(t)$:
\begin{equation}
\label{eq:cond:bzeta}
E\left[\hat x_\theta(t)|\hat b(t'):0<t'<t\right] = \int_0^t G_\theta(t,t') \hat b_\zeta(t') dt'
\end{equation}
with the predictor's kernel given by
\begin{equation}
G_\theta(t,t') =\int_0^t u_\theta(t,t'') V^{-1}_b(t'',t') dt''\,,
\end{equation}
where the covariance matrices have been defined as
\begin{equation}
\label{uandV}
u_\theta(t,t')\equiv \langle \hat x_\theta(t) \hat b_\zeta(t')\rangle\,,  \quad
V_b(t,t') \equiv \langle b_\zeta (t) b _\zeta(t')\rangle
\end{equation}
Here $V_b^{-1}$ is defined as
\begin{equation}
\int_0^t dt_2 V_b(t_1,t_2) V_b^{-1}(t_2,t_3) =\delta(t_1-t_3)\,.
\end{equation}
Note that given a set of measurement results,  $\hat b_\zeta$ in Eq.~\eref{eq:cond:bzeta} should be replaced by the results.

The conditional variance is independent from the measurement outcome, and can be written as
\begin{eqnarray}
&&\mathrm{Var}\left[\hat x_\theta(t)|b_\zeta(t'):0<t'<t\right] \nonumber\\
&=&
\mathrm{Var}\left[\hat x_\theta(t)\right] - \int_0^t u_\theta(t,t_1) u_\theta(t,t_2) V_b^{-1}(t_1,t_2) dt_1 dt_2.\quad
\end{eqnarray}
In this way, we have been able to get compact formulas for conditional expectations and conditional variances --- even when the process is non-Markovian, and during transients.  This approach is equivalent to the SME/SSE approach in the previous section, although directly applicable to non-Markovian systems. 

Here we note the only difference from classical linear regression: although we can calculate the symmetrized conditional covariance between $\hat x_{\theta_1}$ and $\hat x_{\theta_2}$ by 
\begin{eqnarray}
&&\mathrm{Cov}\left[\hat x_{\theta_1}(t),\hat x_{\theta_2}(t)|b_\zeta(t'):0<t'<t\right] \nonumber\\
&=&
\frac{1}{2}\langle\hat x_{\theta_1}(t) \hat x_{\theta_2}(t)+ x_{\theta_2}(t) \hat x_{\theta_1}(t)\rangle \nonumber\\ &-& \int_0^t u_{\theta_1}(t,t_1) u_{\theta_2}(t,t_2) V_b^{-1}(t_1,t_2) dt_1 dt_2\,,
\end{eqnarray}
this covariance does not have a direct physical meaning because $\hat x_{\theta_1}$ and $\hat x_{\theta_2}$ do not commute with each other, and cannot be measured simultaneously.  However, this covariance  can be written as being obtained from a covariance matrix:
\begin{eqnarray}
&&\mathrm{Cov}\left[\hat x_{\theta_1}(t),\hat x_{\theta_2}(t)|\hat b_\zeta(t'):0<t'<t\right]  \nonumber\\
&=& 
(\cos\theta \;\; \sin\theta)
\left[
\begin{array}{cc}
V_{xx}^{\rm cond} &  V_{xp}^{\rm cond} \\
V_{xp}^{\rm cond} &  V_{pp}^{\rm cond} 
\end{array}
\right]
\left(
\begin{array}{c}
\cos\theta \\
\sin\theta
\end{array}
\right)\,,
\end{eqnarray}
This possibility can be traced back to the linearity of the estimator [Cf.~Eq.~\eref{uandV}].   This covariance matrix enters the conditional Wigner function of the mechanical object; even though we cannot directly measure this matrix by measuring the covariance between $x$ and $p$, it can be fully determined if we measure the variances of all quadratures. 

\subsection{Steady-state solution: causal Wiener filtering.}
\label{subsec:wiener}

Suppose we measure a constant out-going quadrature, then $V_b(t,t') = V_b(t-t')$. If both $t$ and $t'$ run from $-\infty$ to $+\infty$, then $V_b$ can be inverted simply using Fourier transform.  The other case in which we do not always have to resort to numerical computation is  when we consider the half-infinity region of $-\infty < t' < t$.

In this case, we can first approximate the spectrum of $\hat b_\zeta$ as a rational function of $\Omega$, and write
\begin{equation}
S_b(\Omega) =\phi_+(\Omega) \phi_-(\Omega) \,,
\end{equation}
with  $\phi_+$ having no zeros or poles on the upper-half complex plane, $\phi_-$ having no zeros or poles on the lower-half complex plane, and
\begin{equation}
\phi_+(\Omega) =\phi_-^*(\Omega^*)\,.
\end{equation}  
This allows us to obtain a causal filter $\phi_+^{-1}$ (meaning
$\phi_+^{-1} (t)=0$ for $t<0$), which yields
\begin{equation}
\hat z(t) =\int_{-\infty}^t dt' \phi^{-1}_+(t-t') \hat b_\zeta(t')
\end{equation}
a white noise.  
%
It is then easy to obtain
\begin{eqnarray}
&& E[\hat x_\theta(t)|\hat b_\zeta(t'):t'<t]\nonumber\\ &=&\int_{-\infty}^t dt' \int_{-\infty}^{t'} dt'' g_\theta(t-t')\phi_+(t'-t'')\hat b_\zeta(t'')
\end{eqnarray}
with 
\begin{equation}
g_\theta(t,t') = g_\theta(t-t') = \langle \hat x_\theta(t) \hat z(t')\rangle
\end{equation}
and
\begin{eqnarray}
&&\mathrm{Var}[\hat x_\theta(t)| \hat b_\zeta(t'):0<t'<t] \nonumber\\
&=&\langle \hat x^2_\theta(t)\rangle - \int_{-\infty}^t dt' g^2_\theta(t-t')\,.
\end{eqnarray}
In the Fourier domain, we can write 
\begin{equation}
\tilde g_\theta(\Omega) =\left[\frac{S_{xb}(\Omega)}{\phi_-(\Omega)}\right]_+
\end{equation}
where $\left[...\right]_+$ indicates taking the causal part of a rational function of $\Omega$, which either: (i) inverse transform into time domain, multiply by the Heaviside step function $\Theta(t)$, then transform back to the frequency domain, or (ii) express formula as a sum of fractions $\alpha/(\Omega-\beta)$, and eliminate ones that have poles in the upper-half complex plane. [See App.~\ref{app:causal} for details.]  This allows us to represent the conditional covariance matrix in terms of cross spectra: 
\begin{eqnarray}
&&V_{x_{\theta_1}x_{\theta_2}}^{\rm cond}\nonumber\\
&=&
\mathrm{Re} \int_0^{+\infty} \frac{d\Omega}{2\pi}\left[S_{x_{\theta_1}x_{\theta_2}}
-
\left[\frac{S_{x_{\theta_1}b}}{\phi_b^-} \right]_+ \left[\frac{S_{x_{\theta_2}b}}{\phi_b^-}\right]_+^* 
\right]\,.
\end{eqnarray}

Finally, let us show that, $V^{\rm cond}_{xp}$ must always be positive.  For this, we note that because we are at steady state, 
\begin{equation}
\langle \hat x(t) \hat p(t) + \hat p(t) \hat x(t)\rangle=0\,,
\end{equation}
 and therefore  
\begin{eqnarray}
&& V_{xp}^{\rm cond} \nonumber\\
&=& 
\left\langle\left[\hat x (t) -\int_{-\infty}^t g_x(t-t') \hat z(t') dt'\right]
\right. \nonumber\\
&&\qquad\left.
\cdot\left[p (t) -\int_{-\infty}^t g_p(t-t') z(t') dt'   \right]\right\rangle_{\rm sym}\nonumber\\
&=& -\int_{-\infty}^t g_x(t-t') g_p(t-t') dt'
\end{eqnarray}
However, for any $t'<t$, we can show that
\begin{equation}
\label{gpgx}
g_p(t-t')  = \frac{d}{dt}g_x(t-t') = -\frac{d}{dt'}g_x(t-t')
\end{equation} 
which gives
\begin{equation}
\label{eq:vxp}
V_{xp}^{\rm cond} = \frac{1}{2}g_x^2(0) \ge 0\,.
\end{equation}
This means the conditional correlation between position and momentum is always non-negative; its magnitude is related to the rate at which the last bit of data carries information about  $x$.  This fact will become useful when we discuss optimal feedback control. 

\begin{figure*}
\centerline{\includegraphics[width=4.25in]{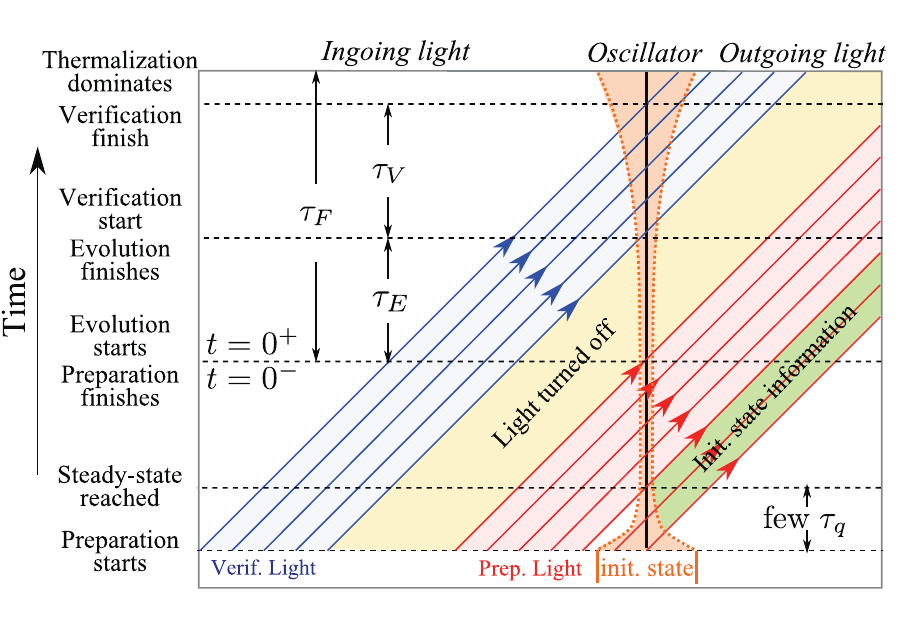}}
\caption{A space-time diagram that illustrates a three-stage, preparation-evolution-verification experiment. Light is incident from the left, the mechanical object (mirror)  stays along a nearly vertical world line (with position uncertainty illustrated by an orange-colored region), while out-going light is shown to propagate to the right (although in reality it may be reflected).   Initial state of the mirror quickly goes away, with information contained in the green region; the pink region represents {\it state preparation}, in which measurement on the out-going light induces state collapse of the mirror (see Sec.~\ref{subsec:conditional}); the yellow region represents {\it evolution}, during which no measurement is made (``turning light off'' is an idealization; we may measure the back action force imposed onto the mirror during this stage and remove this from post-processing of data, see Sec.~\ref{sec:tomography} for details); finally,   the blue region represents {\it verification} during which we carry out a tomography of the mirror's quantum state (see Sec.~\ref{sec:tomography} as well).  As shown in the figure, duration of the evolution and verification stage is limited by thermal decoherence timescale. 
\label{fig:threestage}}
\end{figure*}

\section{Quantum State Preparation and Verification in linear systems}
\label{sec:exp_linear}

After having reviewed various theoretical techniques that are developed to treat continuous quantum measurement processes, in this section, we will discuss basic building blocks of macroscopic quantum mechanics experiments.   We will first discuss how an optomechanical system can be used to prepare the moving mirror into a nearly pure quantum state --- this is arguably the first step if one would like to study macroscopic quantum mechanics.  We will then discuss how the quantum state of the mirror can be reconstructed, in such a way that crucial features of the quantum theory might be illustrated and perhaps tested.

 We shall start from {\it optomechanical cooling}, which describes the process of achieving an {\it unconditional quantum state} that is nearly pure --- by coupling the mechanical object with an optical field which we do not read out.  
We will then go on to discuss {\it conditional state preparation}, which uses measurement-induced state reduction to transfer the quantum-ness of light into the quantum-ness of the test mass.  In such a process, the quantum state of the test mass, conditioned on the measurement result, undergoes a random walk --- such a time-dependent state-preparation strategy could be experimentally more challenging than a steady state one. After discussing conditional state preparation, we will return to a type of cooling that relies on measurement and feedback~\footnote{This structure allows us to provide the best possible treatment for feedback cooling, because, as we shall show, an optimal feedback strategy is crucial for feedback cooling, and more easily obtained after we build an understanding about the conditional state of the mechanical object.}. 

We will finally discuss verification, a set of processes, each will be repeated many times, so that the data collected in these executions will be combined together to yield a {\it tomography} of the quantum state of the mechanical object.  From this we will obtain the final state of the macroscopic mechanical object.  We will also discuss the possibility of leaving a time gap between preparation and verification, during which the mechanical object is freely evolving.  We shall refer to this as {\it evolution}.  We illustrate a three-stage {\it preparation-evolution-verification} experiment in Fig.~\ref{fig:threestage}, using a space-time diagram.

In this paper, we will focus on optomechanical systems pumped by continous beams of light.  A very different set of strategies has been formulated by Vanner et al.~\cite{Vanner:2011}, in which pulsed beams are used to prepare and verify quantum states of mechanical oscillators~\cite{Machnes:2012,Hofer:2011}.  These schemes were  inspired by stroboscopic quantum measurements~\cite{Braginsky:1978} [Cf.~Sec.~\ref{subsec:sql:other}].

\subsection{Cooling Without Measurement}

\label{subsec:cooling}

It is not only conceptually easier, but also practically advantageous, if one can achieve {\it unconditional state preparation}, in which the mechanical object is prepared into an acceptable quantum state regardless of measurement outcome.   However, we should already keep in mind here that techniques that allow cooling may not always allow us to extract the quantum state of the mirror.  In fact, as we shall see, an auxiliary optical system's ability to cool the system to a pure state relies on the system's inefficiency to extract information from the mirror.

\subsubsection{Radiation-Damping Cooling.} 
\label{subsec:radcooling}

As we have discussed in Sec.~\ref{rigidity}, optical (or lower-frequency electromagnetic) field coupled to mechanical resonators can cause optical rigidity, which can provide additional restoring force and damping force.  This has been observed experimentally in gravitational-wave experiments (see Sec.~\ref{sub:ex:rigidity}).  If the optical field introduces damping to the mechanical oscillator, the fluctuation it brings is likely to have an effective temperature much lower than the environment, which will lead to an effective cooling of the oscillator.  This idea has been applied to laser cooling of atoms~\cite{Phillips:1998}; in 1977, Zeldovich proposed using such radiation damping to cool mechanical oscillators, and this idea was analyzed by Vyatchanin~\cite{Vyatchanin:1977}.   

Historically, cold load (e.g., a resistor submerged in liquid helium) has been inserted into linear electromechanical circuits to lower the effective temperature of the entire circuit, as discussed, e.g., by Hirakawa et al.~\cite{Hirakawa:1977}. It was also shown that by feedback using amplifiers with low effective noise temperature, such cold load can be created electronically~\cite{Hirakawa:1978,Oide:1978}. This cooling mechanism was referred to as ``cold damping'' by the gravitational-wave community in the late 1970s.  Although cold damping did not improve the signal-to-noise ratio of resonant bar detectors, they were able to increase their response time~\cite{Oide:1979}.

For modern optomechanical systems, radiation-damping cooling was proposed more specifically by Marquardt et al.\ and Wilson-Rae et al.~\cite{Marquardt:2007,Wilson-Rae:2007}. This has become a widely used way to prepare mechanical oscillators into nearly pure quantum states; see Refs.~\cite{Schliesser:2006,Arcizet:2006b,Gigan:2006,Teufel:2008,Groblacher:2009,Groblacher:2009b} for initial experimental implementations.  Up till now, several experimental groups has successfully cooled mechanical oscillators close to the ground state using this approach~\cite{Rocheleau:2009,Schliesser:2008,Schliesser:2009,Teufel:2011,Chan:2011,Riviere:2011}; here ``close to the ground state'' means effective thermal occupation number $\bar n \stackrel{<}{_\sim} 1$, and this marks a convenient starting point of macroscopic quantum mechanics.

One way to view this process is that the mechanical oscillator is now not only coupled to its original heat bath with original width $\gamma_m$ and temperature $T_0$, but also coupled to an optical bath which creates an additional damping rate $\gamma_{\rm opt}$ which is usually much greater than $\gamma_m$, but with a temperature $T_{\rm opt}$ much lower than $T_0$.   

In general, for an oscillator that is damped by multiple baths, we can write down  its equation of motion as
\begin{equation}
M\left(\ddot {x} + 2\sum_j \gamma_j \dot x -\omega_m^2 x\right) =\sum_j F_j\,,
\end{equation}
where $\gamma_j$ is the damping rate toward the $j$-th bath, and $F_j$ is the fluctuating force from the $j$-th bath. If the $j$-th bath has a temperature $T_j$, then the Fluctuation-Dissipation Theorem~\cite{Callen:1951} states that
\begin{eqnarray}
\label{eq:fdt}
S_{F_j} &=& 4 M\gamma_j \hbar\omega_m \coth\left(\frac{\hbar\omega_m}{2k_B T_j}  \right) \nonumber\\
&=& 8 M \gamma_j \hbar \omega_m \left(\bar n_j+\frac{1}{2}\right)
\end{eqnarray}
where $\bar n_j$ is the mean occupation number of an oscillator with frequency $\omega_m$ if the temperature is $T_j$, or
\begin{equation}
\bar n_j= \frac{1}{e^{\frac{\hbar\omega_m}{k_B T_j}}-1}
\end{equation}
Assuming  the oscillator to remain weakly damped (i.e., high-$Q$), and that the oscillator's frequency not to be significantly shifted by optical rigidity, the combined effect of all baths on the oscillator is an occupation number of 
\begin{equation}
\bar n =\frac{\sum_j \gamma_j \bar n_j}{\sum_j \gamma_j }
\end{equation}

 
For optomechanical cooling, as suggested by Marquardt et al.~\cite{Marquardt:2007} and Wilson-Rae et al.~\cite{Wilson-Rae:2007}, one often choses the cavity detuning $\Delta$ to be equal to the  mechanical resonant frequency $\omega_m$,  and $\gamma \ll  \Delta$. The motivation underlying these choices is that: incoming photons, after scattering off the mirror, create photons with frequencies $\omega_0 \pm \omega_m$, and one would like the cavity to be resonant with $\omega_0+\omega_m$ (i.e., $\Delta=\omega_m$), so that these higher-frequency photons are more preferably emitted, and therefore extracts energy from the motion of the mirror. In this regime, if we only consider up to the leading order effect of the optical power then, around $\Omega\sim \Delta=\omega_m$ [Cf.~\eref{eqK}, \eref{eqp}],  
\begin{eqnarray}
K(\Omega \approx \omega_m) &\approx& -i \frac{G^2}{\gamma}\left(1-\frac{\gamma^2}{4\Delta^2}\right) -\frac{G^2}{2\Delta}\,,
\end{eqnarray}
and [Cf.~\eref{eqA}, \eref{eqAd}, and Eq.~\eref{eq:Fba}]
\begin{eqnarray} 
\label{eq:sfopt}
S_{F_{\rm opt}}(\Omega \approx \omega_m) &=& \frac{2G^2}{\gamma }\left(1+\frac{\gamma^2}{4\Delta^2}\right)
\end{eqnarray}
Because the $\Delta >\gamma$ and the oscillator starts off (before applying radiation-pressure damping) as a high-$Q$ oscillator, as we increase $G$ from 0 towards higher values, the increase in the damping of the oscillator is much more significant than the shift in the real part of its oscillation frequency, and we have
\begin{equation}
\gamma_{\rm opt} = \frac{G^2}{2 m \gamma \omega_m}\left(1-\frac{\gamma^2}{4\Delta^2}\right)
\end{equation}
The oscillator's mean occupation number associated with the optical bath can be converted from the spectrum of $F_{\rm rad}$ [Eq.~\eref{eq:sfopt}] and the fluctuation-dissipation theorem~\eref{eq:fdt}, and reads: 
\begin{equation}
\bar n_{\rm opt}  = \frac{\gamma^2}{4\Delta^2}\,.
\end{equation}
In this way, $\bar n_{\rm opt}$ is the limit of radiation-pressure damping. This only approaches zero in the {\it resolved side-band limit}, i.e., when the cavity linewidth $\gamma$ is much narrower than detuning $\Delta$. Another way to understand such a requirement  is that when $\gamma$ is comparable to $\omega_m$, information about the mirror leaks out into the outside continuum.  Even when we do not have any thermal fluctuations, the out-going light is entangled with the mechanical object, which means the mechanical object, when viewed alone, must be in a mixed state.  As a consequence, one way to recover a  more pure state is to measure the out-going light and feedback, as described by Miao et al.~\cite{Miao:2010d}, which we shall discuss in more details in Secs.~\ref{subsec:conditional} and \ref{subsec:colddamping}.   

A purely coherent way to circumvent the resolved sideband limit is to use the so-called {\it dissipative coupling} scheme, in which the damping rate, instead of the resonant frequency, of the optical mode depends parametrically on the position of the mirror~\cite{Elste:2009,Xuereb:2011}. 

An  extension of resolved side-band cooling into the more quantum regime is to prepare the oscillator to a squeezed state by injecting squeezed vacuum centered at the cavity resonant frequency, as proposed by J\"ahne et al.~\cite{Jahne:2009}.  In this scheme, the quantum radiation-pressure force felt by the mirror is non-stationary, which leads to a squeezed state that is also non-stationary, which features ``breathing'', with position and momentum uncertainty both oscillating at twice the oscillator's resonant frequency.

If we ignore quantum fluctuations of light, and assuming that finally radiation-pressure damping dominates, then we can write:
\begin{equation}
\bar n_{\rm tot} \approx \frac{k_B T_m}{\hbar \omega_m}\frac{\gamma_m}{\gamma_{\rm opt}}   \stackrel{>}{_\sim} \frac{k_B T_m}{\hbar \omega_m}\frac{\gamma_m}{\omega_m}  
=  \frac{k_B T_m}{\hbar \omega_m Q_m} \,.
\end{equation}
Note that the inequality sign comes from $\gamma_{\rm opt} \stackrel{<}{_\sim} \omega_m$, namely the damping rate cannot be greater than oscillation frequency.  We can then esimate that an oscillator can be cooled to near the ground state if it satisfies:
\begin{equation}
\frac{\hbar\omega_m Q_m}{k_B T} \stackrel{>}{_\sim} 1\,.
\end{equation}
At room temperature, we recover the bound of 
\begin{equation}
\label{Qfprod}
Q_m \cdot f_m \sim 6\times 10^{12} \left(\frac{T_m}{300\,{\rm K}}\right)
\end{equation}
which is widely quoted as a benchmark for ground-state cooling.

It is worth mentioning that the mechanism of radiation-damping cooling also applies when mechanical motion is coupled to other baths that have low effective temperature. For example, phonon-exciton coupling in  semiconductors has been shown to be an effective cooling mechanism~\cite{Xuereb:2012,Usami:2012}; ground-state cooling using cooper-pair box has also been proposed~\cite{Jahne:2008}.

\begin{figure}
\centerline{\includegraphics[width=2.25in]{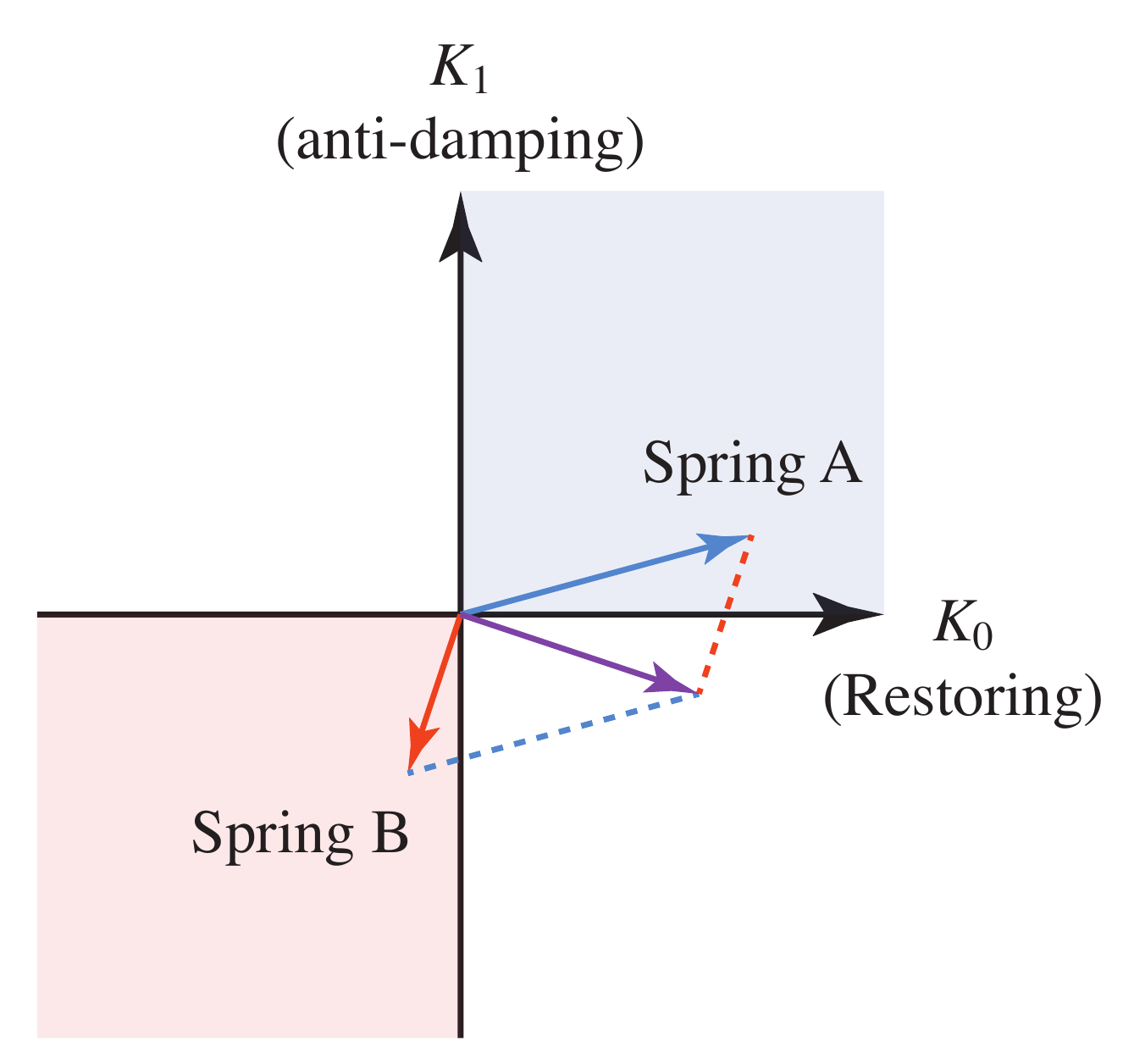}}
\caption{A stable optical spring can be formed by two individual springs, each associated with a detuned beam.  In this figure, we consider optical springs formed by beams with $\Omega \ll \sqrt{\Delta^2+\gamma^2}$, and the spring constant is well-approximated by Eq.~\eref{eqKapprox}.  In this case each individual spring constant can be expressed as $K(\Omega) = K_0 +  i\Omega K_1$, with $K_0$ describing rigidity and $K_1$ describing anti-damping.  As shown in the figure, spring A is responsible for restoring, and Spring B is responsible for damping; the composite spring is stable while each individual one is not.  \label{fig:DOS} }
\end{figure}

\subsubsection{Trapping and Cooling}
\label{subsec:trapping}

In addition to providing a bath with very low effective temperature, another way the optical field can help preparing a nearly pure quantum state is that optical rigidity can be used to increase the mechanical resonant frequency, so that the same thermal energy would correspond to lower occupation number. This was proposed by Braginsky and Khalili in the late 1990s~\cite{Braginsky:1999}, and later independently elaborated by other authors~\cite{Buonanno:2003a,Bhattacharya:2007,Corbitt:2007,Bhattacharya:2007b,Bhattacharya:2008}.  This is somewhat analogous to the effect of ``dilution'', which refers to the fact that a suspended pendulum has a very high $Q$ value compared with the $Q$ of the material of the suspension wire, because most of the restoring force of that pendulum comes from gravity (which is lossless), instead of elastic deformations of the wire~\cite{Gonzalez:1994}. 

 If we consider the thermal noise of the mechanical oscillator, we find that
\begin{eqnarray}
&&M \ddot{x} + 2M\left(\gamma_m +\gamma_{\rm opt}\right) \dot{x} + M \big(\omega_m^2 +\omega_{\rm opt}^2\big) x  \nonumber\\
&=& F_{\rm th} +F_{\rm opt}\,.
\end{eqnarray}
In the limit of $\gamma_{\rm opt} \gg \gamma_m$ and $\omega_{\rm opt} \gg \omega_m$, and assuming that the spectrum of $F_{\rm th}$ is constant with
\begin{equation}
S_F = 8 M\gamma_{\rm m} k_B T_m\,,\quad k_B T_m\gg \hbar\sqrt{\omega_m^2+\omega_{\rm opt}^2}\,.
\end{equation}
Ignoring radiation-pressure noise from the light, the  thermal occupation number now becomes
\begin{equation}
\label{neffdilution}
\bar n = \frac{k_B T_m}{\hbar\omega_m}\frac{\gamma_m}{\gamma_{\rm opt}}\frac{\omega_m}{\omega_{\rm opt}} \stackrel{>}{_\sim} \frac{k_B T_M}{\hbar\omega_m Q_m}\left(\frac{\omega_m}{\omega_{\rm opt}}\right)^2\,,
\end{equation}
this is reduced from cooling-only schemes by the dilution factor of $\left(\omega_{\rm opt} /\omega_m\right)^2$, which can be quite very substantial --- it is also {\it in principle} unlimited if higher optomechanical coupling (e.g., higher optical power) is always available\footnote{Here we have assumed that the additional trapping force is directly related to $x$, yet in many cases, the force senses an additional displacement, e.g., due to thermal fluctuations of the coating applied onto the mechanical oscillator. This will limit the dilution factor, as we shall discuss in Sec.~\ref{subsubsec:relationQF}.}.  [The second inequality in Eq.~\eref{neffdilution} becomes equality if $\gamma_{\rm opt} \sim \omega_{\rm opt}$.]
  This provides an additional way to get closer to a pure state, and the requirement~\eref{Qfprod} becomes
\begin{equation}
\label{Qfproddilute}
Q_m \cdot  f_m \sim 6\times 10^{12} \left(\frac{T_m}{300\,{\rm K}}\right)\left(\frac{\omega_m}{\omega_{\rm opt}}\right)^2
\end{equation}

Such an ``optical dilution'' has been demonstrated by various experiments in the gravitational-wave community; for example, Corbitt et al.~\cite{Corbitt:2007b} demonstrated a dilution from $\omega_m = 2\pi\times 12.7$\,Hz to $\omega_{\rm opt} = $1\,kHz for a 1\,g suspended mirror, a dilution factor of nearly 80, which relaxes requirement on the $Q\cdot f$ product by $6\times 10^3$. 

The ultimate trapping strategy is to levitate dielectric objects~\cite{Romero-Isart:2010,Pflanzer:2012,Chang:2010,Singh:2010,Pender:2012,Monteiro:2012,Chang:2012,Lechner:2012}.  Such strategies will completely get rid of {\it mechanical dissipations} due to the existence of a suspension system, and replace that with dissipations due absorption, scattering etc., which would likely bring much lower levels of noise. 


As we consider the trapping situation in more details, we will have to be careful about: (i) possible dynamical stability of the system, and (ii) the quantum noise brought by the optical fields that generates rigidity.   As for (i), because a restoring force is usually associated with an anti-damping, a single optical field brings an optical spring that, although is capable to increasing the mechanical resonant frequency, brings instability.  As can be shown~\cite{Corbitt:2007,Rehbein:2008}, this instability can be eliminated by using two optical springs at the same time, one mainly provides restoring force, while the other mainly provides the damping --- as explained in Fig.~\ref{fig:DOS}. As for point (ii), most straightforward applications of a {\it double optical spring} on a free mass does not provide a very pure unconditional state; radiation-pressure noise from the springs, especially the one providing the damping, contributes to an effective occupation number above 1 but below 10 --- this is due to the information about the test-mass being carried away by the out-going light.  One can detect the out-going light and perform a conditional state preparation (see Sec.~\ref{subsec:conditional} below), or perhaps use a third light beam~\cite{Korth:2012}.


\subsection{Conditional State Preparation}
\label{subsec:conditional}
In this and the next section, we shall use formalisms introduced earlier in this paper to treat the issue of conditional state preparation and optimal feedback control --- which allows incorporation of colored noise and complex optical systems.  However, for pedagogical reasons, we shall mostly focus on white noise, and our results will be largely equivalent to  those of Clerk et al.~\cite{Clerk:2008}.

Unconditional state preparation must not allow information of the mechanical object to leak out in order to optimize the efficiency of cooling, and this seriously limits the efficiency.   In this section, we consider the measurement process as a state-preparation process, via measurement-induced quantum-state collapse.  This often has a higher efficiency ``on paper'', although it would be  technically more difficult to implement. In addition, the state prepared must be verified using a non-steady-state measurement, because the conditional state often evolves in time in a random manner (if it doesn't then the unconditional state is as good as the conditional state).

\begin{figure}
\centerline{\includegraphics[width=1.75in]{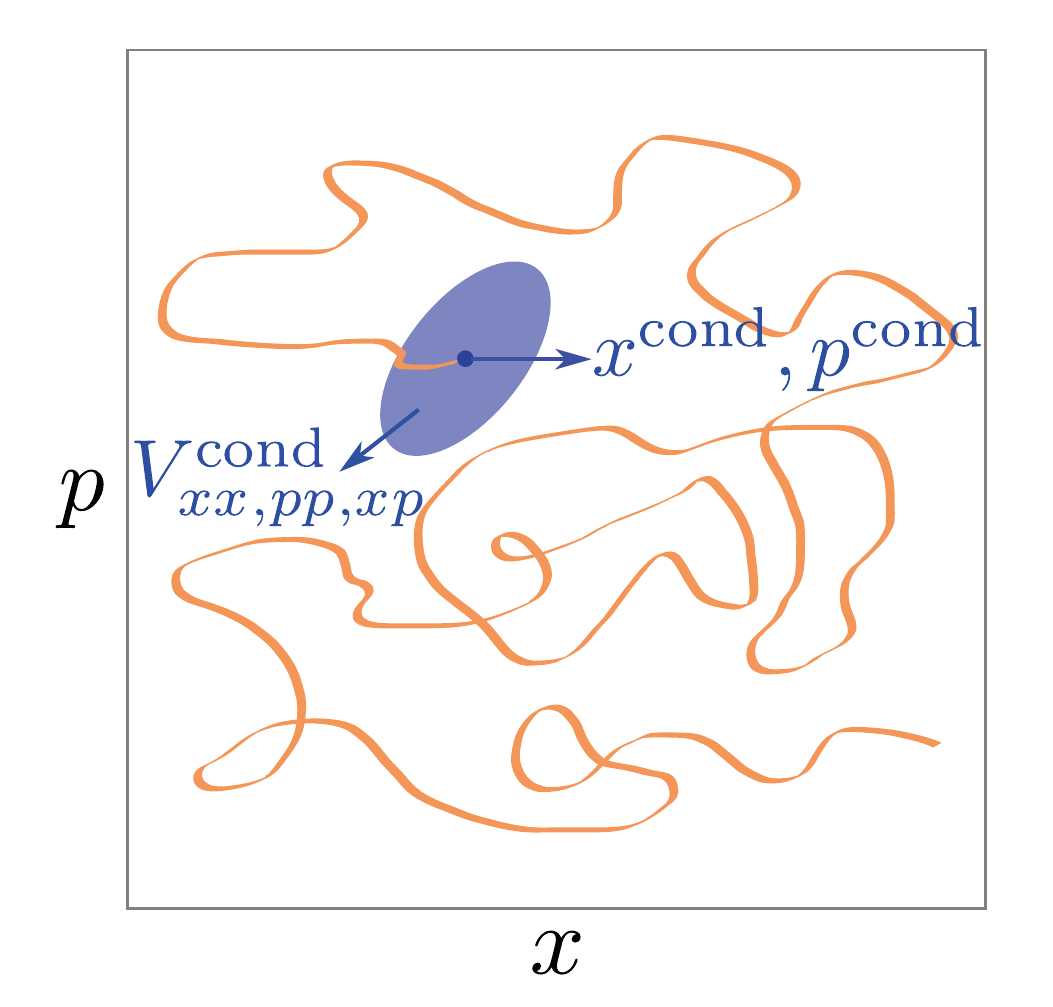}}
\caption{Illustration of conditional state preparation in the phase space.  The conditional expectations of position and momentum undergo random walk in the two-dimensional phase space, driven by quantum state reduction; experimentally, they can be obtained by applying Wiener/Kalman filter to the measurement outcome.  The conditional covariance matrix, after an initial transient, approaches a steady state --- it is represented as a noise ellipse in the figure.  In absence of classical noise, this covariance matrix is Heisenberg Limited (has determinant of $\hbar^2/4$), and the conditional state is a pure Gaussian state; in presence of classical noise, the covariance matrix has a determinant greater than  $\hbar^2/4$ --- but very close to this value if classical noise of the experiment, around the frequency of measurement, is below the SQL.   See Sec.~\ref{subsec:conditional} for details, and see Fig.~\ref{fig:conditional} for further illustration.  \label{fig:randomwalk}}
\end{figure}

\subsubsection{Conditional state preparation in presence of noise}

\begin{figure*}
\centerline{\includegraphics[width=6in]{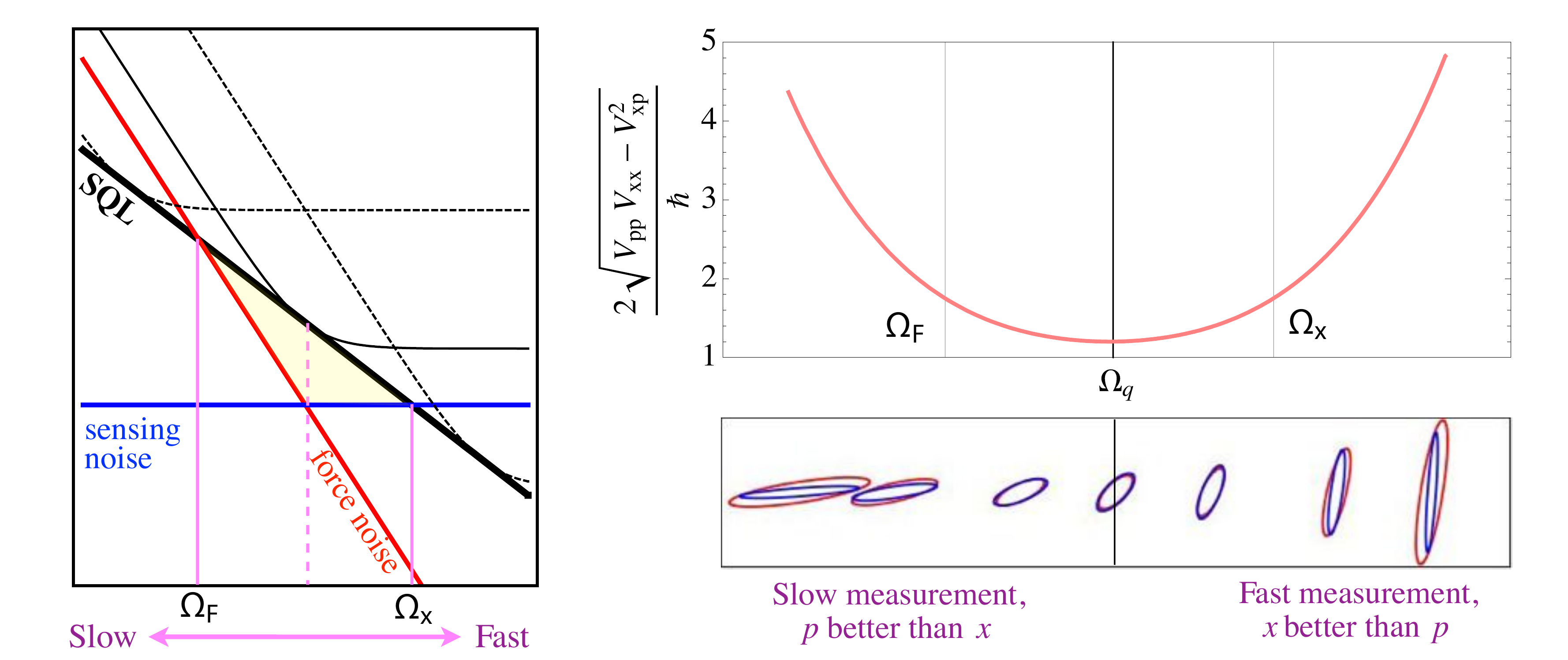}}
\caption{Illustration of conditional state preparation.  Left panel: noise budget of a device that measures the position of a free mass.  The free-mass SQL is shown in solid black curve, sensing noise in blue curve (and crosses the SQL at $\Omega_x$), position-referred force noise in red curve (and crosses the SQL at $\Omega_F$), and possible quantum noise budget of different measurement schemes are shown in black dashed curves (they each touch the SQL at $\Omega_q$).  Upper right panel: $U$ as a function of $\Omega_q$; the ``sweet spot'' is at $\Omega_q = \sqrt{\Omega_F \Omega_x}$.  Lower right panel: noise ellipse for different $\Omega_q$, blue for no classical noise, and red for classical noise as shown in the left.  Because we start with an idealized free mass, the scale of the quantum state is determined by $\Omega_q$: larger $\Omega_q$ tends to produce more position squeezed state, while smaller $\Omega_q$ tends to produce more momentum squeezed state;  states are pure when $\Omega_q$ is around the window opened by $\Omega_F$ and $\Omega_x$. \label{fig:conditional}}
\end{figure*}

In this section, in addition to the idealized system considered in Sec.~\ref{subsec:elimination}, we shall add simplified models of sensing and force noise, but we shall consider the example of a free mass, or the case when the measurement frequency $\Omega_q$ is much higher than the eigenfrequency $\omega_m$ of the mechanical oscillator (this assumption will be relaxed in Sec.~\ref{subsubsec:connection}, when we broadly discuss the efficiency of conditional-state preparation):
\begin{eqnarray}
-M\Omega^2 \hat x  = \alpha \hat a_1 + \hat n_F \,,\quad   \hat  b_2 = \hat  a_2 + \alpha (\hat x +\hat n_x)\,.
\end{eqnarray}
Here $\hat n_F$ is a white force noise, and $\hat n_x$ a white sensing noise; we shall write
\begin{equation}
S_{n_F} = 2\hbar M \Omega_F^2\,,\quad S_{n_x} =\frac{2\hbar M}{\Omega_x^2}\,.
\end{equation}
Note that $\alpha^2 =M\Omega_q^2$, as in Eq.~\eref{alphaomegaq}. We have chosen this way to represent the noise spectra so that when we refer to displacement $x$, the effect of $S_{n_F}$ is a $\Omega^{-4}$ curve that crosses the displacement SQL at frequency $\Omega_F$, while the effect of $S_{n_x}$ is a white displacement noise that crosses the  SQL at frequency $\Omega_x$, and the total classical noise is
\begin{equation}
S_x^{\rm cl} = 2\hbar M\left[\frac{1}{\Omega_x^2}+\frac{\Omega_F^2}{\Omega^4}\right] 
\end{equation}
and we have
\begin{equation}
\label{eq:sqlbeating}
\min_\Omega\left[\frac{S_{\rm cl}(\Omega)}{S_{\rm SQL}(\Omega)}\right] =\frac{2\Omega_F}{\Omega_x}
\end{equation}
this means if $ \Omega_x > 2 \Omega_F $, then the total classical noise leaves a window below the SQL (see the left panel of Fig.~\ref{fig:conditional} for a sketch).  In gravitational-wave detectors, the source of force noise in the detection band is dominated by suspension thermal noise (center of mass motion of the test mass driven by thermal fluctuations in the suspension system), while sensing noise arises from optical losses and internal thermal noise (fluctuations in the distance between the mirror's center of mass and the effective surface the light reflects from)~\cite{Abbott:2009}.

The basics of conditional state preparation is discussed in Sec.~\ref{subsec:SSE}, where we have obtained the SSE and SME for a continuous position measurement (see Fig.~\ref{fig:randomwalk}).  Let us only consider the simplest system mentioned above, and after the initial transients have died down, and we have reached a steady state --- at which the conditional expectation values of $x$ and $p$ still undergo a random walk, but the statistical characteristic of the random walk, as well as the conditional covariance matrix of $x$ and $p$, has reached constant values.   In this simplest case, using our formalism of Wiener filtering (Sec.~\ref{subsec:wiener}), we obtain 
\begin{eqnarray}
\label{condvxx}
V_{xx}^{\rm cond} &=& \frac{\hbar}{\sqrt{2}M\Omega_q} (1+2\xi_x^2)^{3/4} (1+2\xi_F^2)^{1/4} \\
\label{condvxp}
V_{xp}^{\rm cond} &=&  \frac{\hbar}{2}  (1+2\xi_x^2)^{1/2} (1+2\xi_F^2)^{1/2} \\
\label{condvpp}
V_{pp}^{\rm cond}&=& \frac{\hbar M\Omega_q}{\sqrt{2}}(1+2\xi_x^2)^{1/4} (1+2\xi_F^2)^{3/4} 
\end{eqnarray}
where we have defined
\begin{equation}
\label{eq:xiFxix}
\xi_F = \frac{\Omega_F}{\Omega_q}\,,\quad \xi_x \equiv \frac{\Omega_q}{\Omega_x}\,,
\end{equation}
and we prefer both $\xi_F$ and $\xi_x$ to be small, because that means: (i) during the measurement time scale, the classical force noise does not have time to create disturbance of the mirror comparable to its quantum uncertainty, and (ii) the measurement time is long enough so that classical sensing noise is below  quantum uncertainty. Together this means we should choose
\begin{equation}
\Omega_F < \Omega_q <\Omega_x
\end{equation}
in order to prepare a state that is nearly pure.    In the case classical noise is negligible, the Wiener (Kalman) filters for $x$ and $p$ read:
\begin{eqnarray}
G_x &=& \sqrt{2} \Omega_q e^{-\Omega_q t/\sqrt{2}}\cos\frac{\Omega_q t}{\sqrt{2}}\,,\\ G_p &=&\sqrt{2}m\Omega_q^2 e^{-\Omega_q t/\sqrt{2}} \cos\left(\frac{\Omega_q t}{\sqrt{2}}+\frac{\pi}{4}\right)
\end{eqnarray}
This means the conditional expectation for position and momentum at any time depends mostly on data taken in the past within a time scale of $1/\Omega_q$.  This literally confirms $\Omega_q$ as the measurement frequency scale.

\subsubsection{Figure of merit for conditional state preparation}

\label{subsec:fomconditional}

One figure of merit for the quality of the conditional state is 
\begin{equation}
U =\frac{2}{\hbar}\sqrt{V_{xx}V_{pp}-V_{xp}^2}
\end{equation}
to measure how pure the state is: in general $U \ge 1$, and the state is more pure when $U$ is more close to $1$. It is related to the ``linear entropy'' via
\begin{equation} 
\mathcal{S}_{\rm lin} =1-\mathrm{tr} \hat\rho^2 = 1-\frac{1}{U}
\end{equation}
A related figure of merit would be the  {\it effective occupation number}, $\mathcal{N}_{\rm eff}$:
\begin{equation}
\mathcal{N}_{\rm eff} \equiv U/2-1/2
\end{equation}
and the von Neumann entropy is given by
\begin{equation}
\mathcal{S}_{\rm vN} = -\mathrm{tr}\left[ \hat\rho \log\hat\rho\right] =
\left(\mathcal{N}_{\rm eff}+1\right)\log\left( \mathcal{N}_{\rm eff} +1\right) -
\mathcal{N}_{\rm eff}\log \mathcal{N}_{\rm eff} \,.
\end{equation}
This means a conditional state with $\mathcal{N}_{\rm eff}$ has the same von Neumann entropy as a thermal state with thermal occupation number $\bar n=\mathcal{N}_{\rm eff}$.   Note that by using $U$ and $\mathcal{N}_{\rm eff}$ we are not giving additional weight to whether the state is ``squeezed''.  

\subsubsection{Connection between purity of conditional state and device sensitivity.}  \label{subsubsec:connection}

Returning to our case, Eqs.~\eref{condvxx}--\eref{condvpp} give, for $\omega_m \ll \Omega_q$, 
\begin{equation}
\label{Neff_cond}
\mathcal{N}_{\rm eff} =\frac{\sqrt{(1+2\xi_x^2)(1+2\xi_F^2)} - 1}{2}\,.
\end{equation}
The optimal choice to minize $\mathcal{N}_{\rm eff}$ is to choose $\Omega_q =\sqrt{\Omega_x\Omega_F}$, so that $\xi_x =\xi_F = \sqrt{\Omega_F/\Omega_x}$, and 
\begin{equation}
\mathcal{N}_{\rm eff}^{\rm min} = \frac{\Omega_F}{\Omega_x} =\frac{1}{2}\min_\Omega\left[\frac{S_{\rm cl}(\Omega)}{S_{\rm SQL}(\Omega)} \right]
\end{equation}
This means, for a device that has classical noise right at the SQL, we should expect to be able to prepare a conditional state that has a von Neumann entropy equivalent to a thermal state with mean occupation number of $1/2$. In this way, efforts in the gravitational-wave community to built detectors and prototype experiments that beat the free-mass Standard Quantum Limit~\cite{Danilishin:2012a,Chen:2009,Punturo:2010,Miao:2012c} will also be suitable for preparing nearly pure test-mass quantum states.

As discussed in Ref.~\cite{Mueller-Ebhardt:2009}, for mechanical oscillators, even for {\it a general} $\omega_m$, as long as the total sensing noise $Z$ and the force noise $F$ have white spectra,  and have real-valued correlation $S_{ZF}$, we will always have a simple but powerful relation:
\begin{equation}
\label{eq:purity:sql}
U= \frac{2}{\hbar}\sqrt{V_{xx}V_{pp}-V_{xp}^2}= \frac{\sqrt{S_{ZZ}S_{FF}-S_{ZF}^2}}{\hbar}\,.
\end{equation}
This means the ``purity of the measurement noise spectra'' (right-hand side) is equal to the purity of the conditional sate (left-hand side), and in particular Minimum Heisenberg Uncertainty measurement noise spectra (indicating absence of classical noise) leads to minimum Heisenberg conditional states (i.e., pure states).

Equation~\eref{eq:purity:sql} implies that, if we use an optomechanical device with broad measurement bandwidth and detect the out-going optical quadrature that contains displacement signal, {\it assuming white sensing and force noise}, then the optimal purity is achieved when $S_{ZF}=0$, and in this case, {\it Eq.~\eref{Neff_cond} holds regardless of the value of $\omega_m$, namely:}
\begin{equation}
\mathcal{N}_{\rm eff}^{\rm min}=  \frac{\Omega_F}{\Omega_x}\,,\quad \Omega_q =\sqrt{\Omega_F \Omega_x}\,.
\end{equation}
In other words, assuming that measurement strength is not an issue, the optimal purity of the conditional state is {\it only} related to the {\it size of the frequency window} $(\Omega_F, \Omega_x)$ in which both the classical sensing noise and classical force noise are below the {\it free-mass SQL}.

The Wiener filtering formalism is capable to treating general colored noise spectra without much additional effort --- one simply has to approximate all spectra in rational functions of $\Omega$, although the answer will not be so simple, see Ref.~\cite{Mueller-Ebhardt:2008b,Mueller-Ebhardt:2009} for more detailed discussions of results. A more fundamental issue would be that finiteness of cavity bandwidth means correlations between the mirror and the cavity mode, and therefore the mirror alone will not be in a pure state even in absence of classical noise.  This is discussed in Sec.~V of Ref.~\cite{Mueller-Ebhardt:2009}, where they obtained algebraically rather complicated analytical results.  Here we merely quote the result when the cavity is relatively broadband $\Omega_q \stackrel{<}{_\sim} \gamma$, and we can still keep using 
\begin{equation}
\Omega_q^{\rm cav} = \sqrt{\hbar\alpha^{\rm cav}/M}
\end{equation}
and
\begin{equation}
\alpha^{\rm cav} \approx 2G\sqrt{\frac{\gamma}{\gamma^2+\Delta^2}} 
\end{equation}
with $G$ related to cavity length and power via Eqs.~\eref{eq:G}, \eref{eq:Ecavity} and \eref{eq:g}.  Expressed in these quantities, in absence of classical noise, we have a leading-order approximate effective occupation number of
\begin{equation}
\mathcal{N}_{\rm eff} \approx \frac{\Omega_q^{\rm cav}}{4\sqrt{2}\gamma}\,, \quad \Omega_q \stackrel{<}{_\sim} \gamma\,.
\end{equation}
due to finite cavity bandwidth.

\subsubsection{Relation between $Q\cdot f$ criterion and the $\Omega_x/\Omega_F$ crieteron.}
\label{subsubsec:relationQF}

In Sec.~\ref{subsec:cooling}, as we discussed cooling, we have used $Q\cdot f$ as the figure of merit to estimate a system's possibility of being cooled to an effective thermal occupation number below unity, see Eqs.~\eref{Qfprod} and \eref{Qfproddilute}.  Taking cooling without dilution as a special case of dilution factor equal to unity, the approximate condition for reaching a nearly pure state (occupation number approximately unity) there is
\begin{equation}
\label{qf_dilute}
\frac{Q_m \hbar \omega_m }{k_B T_m}  \stackrel{>}{_\sim} \left(\frac{\omega_m}{\omega_{\rm opt}}\right)^2 \,.
\end{equation}
or
\begin{equation}
\label{neff_dilute}
\bar n_{\rm eff} \approx \frac{k_B T_m}{Q_m\hbar\omega_m } \left(\frac{\omega_m}{\omega_{\rm opt}}\right)^2\,.
\end{equation}

In conditional state preparation, because we have measurement as an additional element, we have two numbers, $\Omega_F$ and $\Omega_x$ to represent a white force noise and a white sensing noise, and we require
\begin{equation}
\label{cond_cond}
\Omega_F \stackrel{<}{_\sim}  \Omega_q \stackrel{<}{_\sim}  \Omega_x
\end{equation}
to reach a nearly pure quantum state.   Assuming that the force noise originates from dissipative heat bath with temperature $T_m$ and damping rate of $\gamma_m$,  
we have 
\begin{equation}
\label{neff_cond_qf}
\mathcal{N}_{\rm eff} \approx \frac{k_B T}{Q_m \hbar \omega_m}\left(\frac{\omega_m}{\Omega_q}\right)^2
\end{equation}
Equation  \eref{neff_cond_qf} is compatible with Eq.~\eref{neff_dilute} if  $\Omega_q \sim \omega_{\rm opt}$. As we have discussed in Sec.~\ref{rigidity}, for our straw-man optomechanical system, if the mirror starts off as free mass, if optical parameters $\gamma$, $\Delta$ and the characteristic frequency $\Theta$ are all of the same order of magnitude, then $\omega_{\rm opt}$ will indeed be the same order as $\Omega_q$.   This means, in terms of dissipative thermal noise and optomechanical coupling strength, {\it requirements for cooling-trapping and conditioning to bring the mirror into a nearly quantum state are comparable}.  

If we summarize our conclusion so far using the frequency scales $\omega_m$, $\Omega_F$, $\Omega_q \sim \omega_{\rm opt}$ and $\Omega_x$: if we only consider thermal force noise arising from velocity damping, then both cooling-trapping and conditioning have a similar figure of merit of:
\begin{equation}
\bar n_{\rm eff}\approx \mathcal{N}_{\rm eff} \approx 
\left(\frac{\Omega_F}{\omega_{\rm opt}}\right)^2 \approx
\left(\frac{\Omega_F}{\Omega_q}\right)^2\,.
\end{equation} 
The scale of $\omega_m$ does not affect the effective occupation number we can reach, except that when $\omega_m \stackrel{>}{_\sim} \Omega_F$ then no optical  dilution is needed to reach an occupation number of unity.  [Note that this corresponds to the first $Q\cdot f$ requirement~\eref{Qfprod}, when we discussed radiation-damping cooling without dilution.] 


Accounting for classical sensing noise in conditional state preparation --- the second part of the inequality~\eref{neff_cond_qf} --- this imposes a limit beyond which further increase of $\Omega_q$ will no longer improve purity of the conditional state.  Assuming both classical force noise and classical sensing noise to be white, this limit is given by 
\begin{equation}
\label{omega_limit}
\Omega_q \le  \sqrt{\Omega_x \Omega_F}
\end{equation}
and therefore 
\begin{equation}
\mathcal{N}_{\rm eff} \approx \frac{\Omega_F}{\Omega_x}
\end{equation}

Because no measurement is required for the trapping-cooling method for state preparation, the second inequality in Eq.~\eref{cond_cond} involving sensing noise does not directly appear, and has been ignored in our discussions in the previous sections.  Nevertheless, a similar noise comes into play when one attempts to optically trapping mechanical objects with significant internal thermal noise, namely thermal fluctuations in the distance between the object's center of mass and its reflective surface, e.g., due to dielectric coatings applied onto the mechanical object.  In this case, although the internal thermal noise directly imposes phase fluctuations onto reflected light, but those phase fluctuations get converted into amplitude flucuations, and then get imposed onto the center-of-mass motion.  At the end, one has to impose the same limit on optical dilution (for $\omega_{\rm opt}$) as Eq.~\eref{omega_limit}.  As a consequence, we have
\begin{equation}
\bar n_{\rm eff} \approx \mathcal{N}_{\rm eff}  \approx \frac{\Omega_F}{\Omega_x} \,.
\end{equation} 
This shows that, with the same level of classical noise, conditional state preparation and trapping-cooling could achieve the same level of effective occupation number --- although for finally achieving a pure state, in the cooling-trapping case, one has to prevent quantum information of the mirror from leaking out.


\begin{figure*}
\centerline{\includegraphics[width=5.5in]{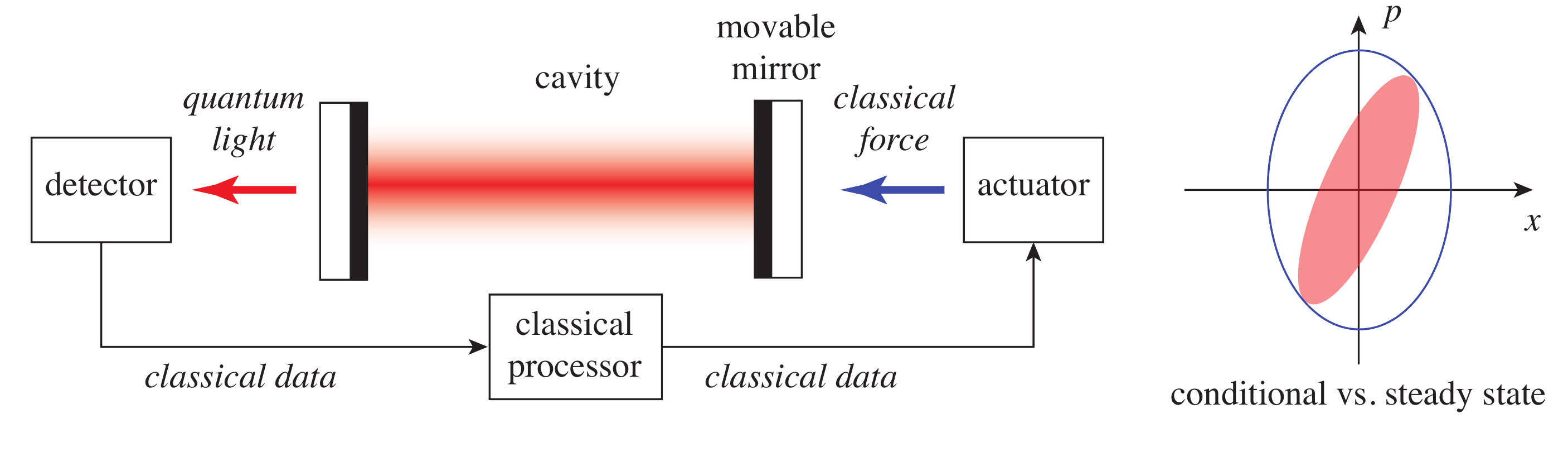}}
\caption{Left panel: the setup of a feedback cooling experiment, which can also be viewed as a quantum system under feedback control.  After detecting (and therefore measurement) the out-going light, the measurement result (classical data) is processed, and then fed back onto the mirror as a classical force.  Right panel: red ellipse represents the conditional variance of the measurement, and the blue ellipse represents one possible noise ellipse realizable by a control system, which: (i) has principal axes aligned with $x$ and $p$ directions, and (ii) is always outside the conditional noise ellipse.   \label{fig:damping}.}
\end{figure*}

\subsection{Quantum Feedback Control}
\label{subsec:colddamping}

Historically, classical feed back loops have been used to create cold loads, used in cold-damping experiments~\cite{Hirakawa:1977,Hirakawa:1978,Oide:1978} (see discussions at the beginning of Sec.~\ref{subsec:radcooling}).   However, those works were in the classical regime, and there was no role for quantum measurement. 

In this section, we enter the quantum regime, and consider explicitly feedback processes that are enabled by a quantum measurement, as shown in left panel of Fig.~\ref{fig:damping}.  In particular, we will focus on feedback schemes that will allow cooling of the mechanical oscillator.  For historical reasons, feedback cooling is sometimes also referred to as ``cold damping'' --- but that offers no distinction from radiation damping.  Therefore we shall only use the term ``feedback cooling'' in this paper, having in mind that ``feedback'' requires an input which would have to be obtained through quantum measurement.   

Feedback cooling was described and analyzed for optomechanical systems by many authors~\cite{Vitali:1998,Courty:2001,Genes:2008}.  Experimentally, it has been implemented by Corbitt et al.~\cite{Corbitt:2007b} and Mow-Lowry et al.~\cite{Mow-Lowry:2008} in experiments involving gram-scale test masses, and by Abbott et al.~\cite{Abbott:2009} kg-scale test masses; by Arcizet et al.~\cite{Arcizet:2006} in smaller scales.

Feedback cooling is interesting not only because it is an important way to prepare quantum states, but also because it is an important example of feedback control of a quantum system, when the measurement process is quantum-limited~\cite{Doherty:2000,Wiseman:2010}.   
Although a program involving conditional-state preparation and verification is more difficult to realize {\it experimentally}, the {\it theory} of conditional-state preparation is still indispensible if one would like to design an {\it optimal} unconditional state-preparation scheme using measurement and feedback. The optimality of the control kernel is crucial for the purity of the close-loop quantum state --- and optimization of this control kernel is best achieved when we know the conditional state of the test mass during the measurement.   This is why we have postponed discussion of feedback cooling until now.  

\subsubsection{Optimal Control Theory For Linear Mechanical Oscillators}
\label{subsubsec:optctrl}

It is clear that: (i) the steady state achievable by feedback cannot in any way be better than the conditional state, and (ii) there should be no steady-state correlation between position and momentum:  
\begin{equation}
V_{xp}^{\rm ctrl} =0\,.
\end{equation}
One such state is depicted on the right panel of Fig.~\ref{fig:damping}, with the feature that the controlled state has a noise ellipse that has $x$ and $p$ as its principal axes, and it is tangential to the conditional-state noise ellipse at two points.  The question would be, for any steady Gaussian state satisfying (i) and (ii), does there exist a controller that allow us to achieve it?  The answer is yes, and this is rather straightforward to prove, based on Ref.~\cite{Danilishin:2008} (see also Ref.~\cite{Doherty:2012}), in any stationary (i.e., optical power and detected optical quadrature does not depend on time) linear measurement processes, even for non-Markovian ones.

Suppose we have a causally whitened measurement outcome $z(t)$ (See Sec.~\ref{subsec:wiener}).  If we start from $t=0$, and keep track of what we had already fed back before --- which propagates through the system deterministically, we should be able to obtain the unperturbed $z(t)$ for any $t>0$ --- by systematically subtracting the part of the output signal that is due to our previous feedback force --- even though we had acted during the period of $0<t'<t$.  For this reason, we should be able to create a scheme for which, after reaching steady state, has 
\begin{eqnarray}
\label{eqxcontrol}
x^{(1)}(t) &=& x^{(0)}(t) - \int_{-\infty}^t dt' f(t-t') z^{(0)}(t')\\
p^{(1)}(t) &=& \dot{x}^{(1)}(t)
\end{eqnarray}
where we have used superscript  ``(0)'' for uncontrolled state and ``(1)'' for controlled state, and have denoted
\begin{equation}
f(t) =\int_{0}^t dt' \chi(t-t') G(t')\,.
\end{equation} 
Here $\chi$ is the response function of $x$; $\chi(0)=0$ and $\chi'(0)\neq 0$, because the mass does not respond instantaneously to a force, but must go through intergration.  The quantity $G$ is the transfer function from data to feedback force, which can have a $\delta$-function part (which represents instantaneous response) and a Heaviside step function part:
\begin{equation}
G(t) = g_0\delta(t) + g(t) \Theta(t)\,.
\end{equation}
Here $g_0$ is the coefficient for direct response, and $\Theta$ is the Heaviside step function. This means, although $\chi(0)=0$ dictates that $f(0)=0$,  the derivative $f'(0)$ does not need to vanish, and   $f$ can be any function that vanishes at $t=0$. 

\begin{figure}
\centerline{\includegraphics[width=3.25in]{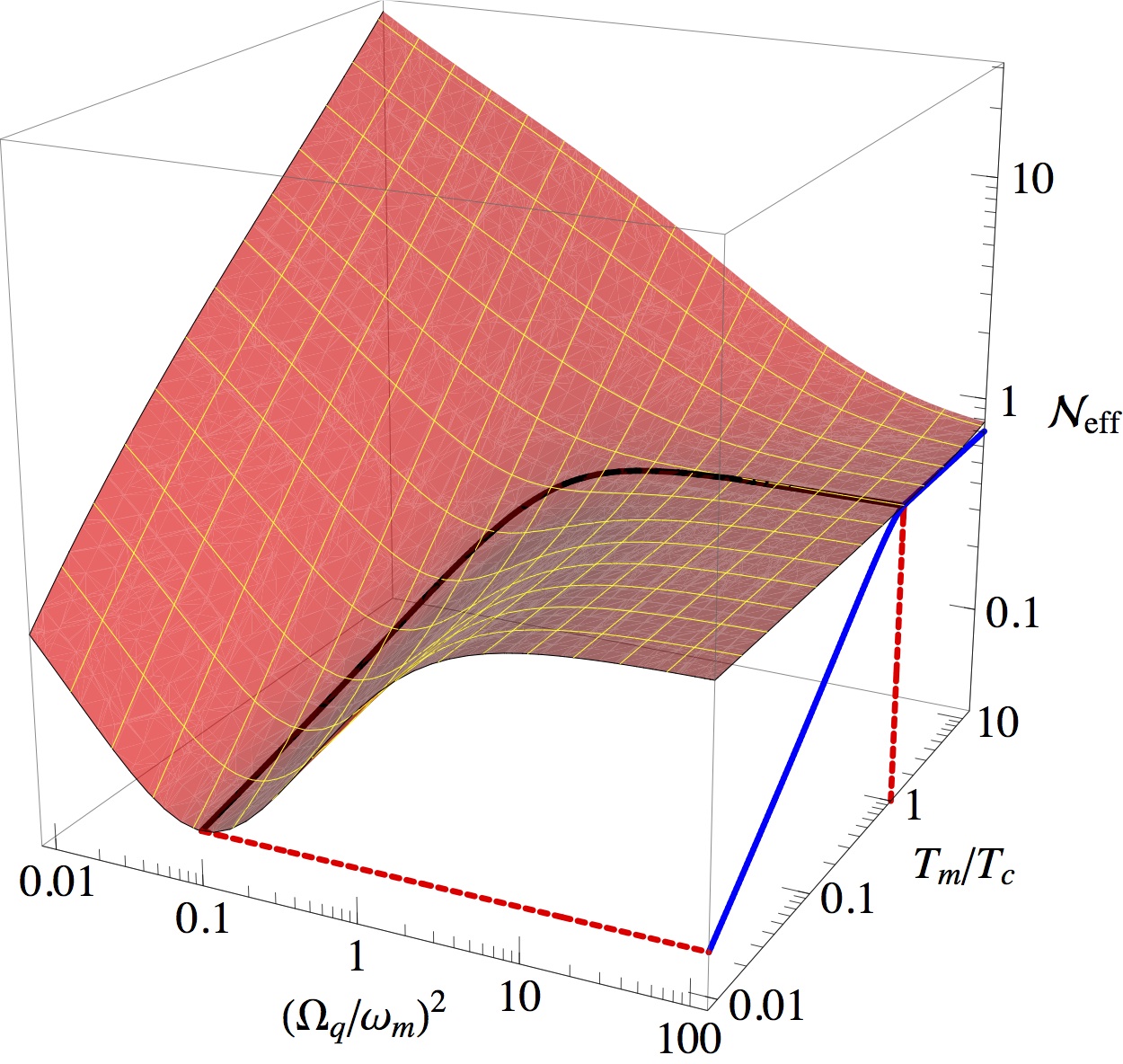}}
\caption{Critical behavior of {\it optimally controlled steady state} of a mechanical oscillator under a ``conventional'' continuous measurement (i.e., with phase-quadrature of out-going field measured and fed back). Assuming only thermal noise from dissipation to a heat bath with temperature $T_m$ (and no classical sensing noise or optical loss),  we plot the effective occupation number $\mathcal{N}_{\rm eff}$ as a function of measurement strength and the ratio $T_m/T_c$.  For $T_m/T_c<1$, there exists an optimal measurement strength at which the cooling is most efficient (black solid trajectory on the surface), while for $T_m /T_c>1$, the optimal cooling is always $\Omega_q\rightarrow +\infty$, reaching an occupation number of $1/\sqrt{2}$. The optimal $N_{\rm eff}$ as a function of $T_m/T_c$ has been projected onto the $T_m/T_c$-$\mathcal{N}_{\rm eff}$ surface as a blue trajectory. See Sec.~\ref{subsec:cd:fm} for details.
\label{fig:cd:critical}}
\end{figure}

In the steady-state limit of $t\rightarrow +\infty$, Eq.~\eref{eqxcontrol} readily becomes 
\begin{equation}
V_{xx}^{\rm ctrl} = V_{xx}^{\rm cond}  +  \int_0^{+\infty} h^2(t) dt \,,\quad h(t) = f(t)-g_x(t)\,,
\end{equation}
which shows that the steady state $V_{xx}^{\rm ctrl}$ cannot be less than the conditional variance. Because $f(0)=0$, $h$ is a function that satisfies $h(0)=-g_x(0)$. For momentum, we have [noting Eq.~\eref{gpgx}]
\begin{equation}
V_{pp}^{\rm ctrl} = V_{pp}^{\rm cond}  +  \int_0^{+\infty} [h'(t)]^2 dt\,.
\end{equation}
Recalling that $V_{xp}^{\rm ctrl}=0$, we have already obtained all the necessary steady-state covariances.  For any $x_\theta$, we can show that the steady state has
\begin{eqnarray}
\label{eq:vttctrl}
&&V_{\theta\theta}^{\rm ctrl}-V_{\theta\theta}^{\rm cond} \nonumber\\
&=& \int_0^{+\infty} \left[h^2\cos^2\theta + (h')^2 \sin^2\theta \right] dt - 2V_{xp}^{\rm cond} \sin\theta\cos\theta  \nonumber\\
&\ge& \left[ \int_0^\infty (-2 h h')dt -2V_{xp}^{\rm cond} \right]\sin\theta\cos\theta =0\,.
\end{eqnarray} 
Here we have recalled from Eq.~\eref{eq:vxp} that 
\begin{equation}
V_{xp}^{\rm cond} = \frac{1}{2}g_x^2(0) \ge 0\,.
\end{equation}
Equation~\eref{eq:vttctrl} means for any $x_\theta$, the steady state always has a larger covariance than the conditional state, as illustrated in the right panel of Fig.~\ref{fig:damping}.  In addition, if we choose
\begin{equation}
h(t)  = - g_x(0)e^{-\lambda t}\,,\;\lambda >0\,,
\end{equation}
then we have
\begin{equation}
\label{vcontrol}
V_{\theta\theta}^{\rm ctrl} = V_{\theta\theta}^{\rm cond} +\frac{(\cos\theta-\lambda\sin\theta)^2}{\lambda}V_{xp}^{\rm cond}
\end{equation}
then the noise ellipses of the conditional state and the steady state will be tangential to each other at precisely two antipodal points, with $\tan\theta=1/\lambda$~\footnote{Here $\theta$ is the inclination angle of the common tangent of the two ellipses.}.  

In this way, we have proved the statement at the beginning of this section, namely any steady state that has a noise ellipse that looks like the blue ellipse in the right panel of Fig.~\ref{fig:damping} can be achieved by an optimal controller.  The detailed ``kinematical'' reason why the controlled steady state's quality is limited by $V_{\rm xp}^{\rm cond}$ or $g_x^2(0)/2$ is the following:  because our feedback system cannot change $x$ instantaneously, but must go through $p$, we will not be able to make corrections to $x$ according to the last-minute information about $x$ --- therefore our steady state will never be as good as the conditional state.

Equation~\eref{vcontrol} is also equivalent to 
\begin{eqnarray}
V_{xx}^{\rm ctrl}& =& V_{xx}^{\rm ctrl} +\frac{V_{xp}^{\rm cond}}{\lambda} \\
V_{pp}^{\rm ctrl}& =& V_{pp}^{\rm ctrl} +\lambda{V_{xp}^{\rm cond}}
\end{eqnarray}
The effective occupation number $\mathcal{N}_{\rm eff}$ [Sec.~\ref{subsec:fomconditional}] is minimized with $\lambda = -\sqrt{V_{pp}^{\rm cond}/V_{xx}^{\rm cond}}$, with a minimum value of 
\begin{equation}
\mathcal{N}_{\rm eff}^{\rm ctrl} =\frac{2}{\hbar} \left[\sqrt{V_{xx}^{\rm cond} V_{pp}^{\rm cond}} +V_{xp}^{\rm cond}\right]
\end{equation}

\begin{figure}
\centerline{\includegraphics[width=3in]{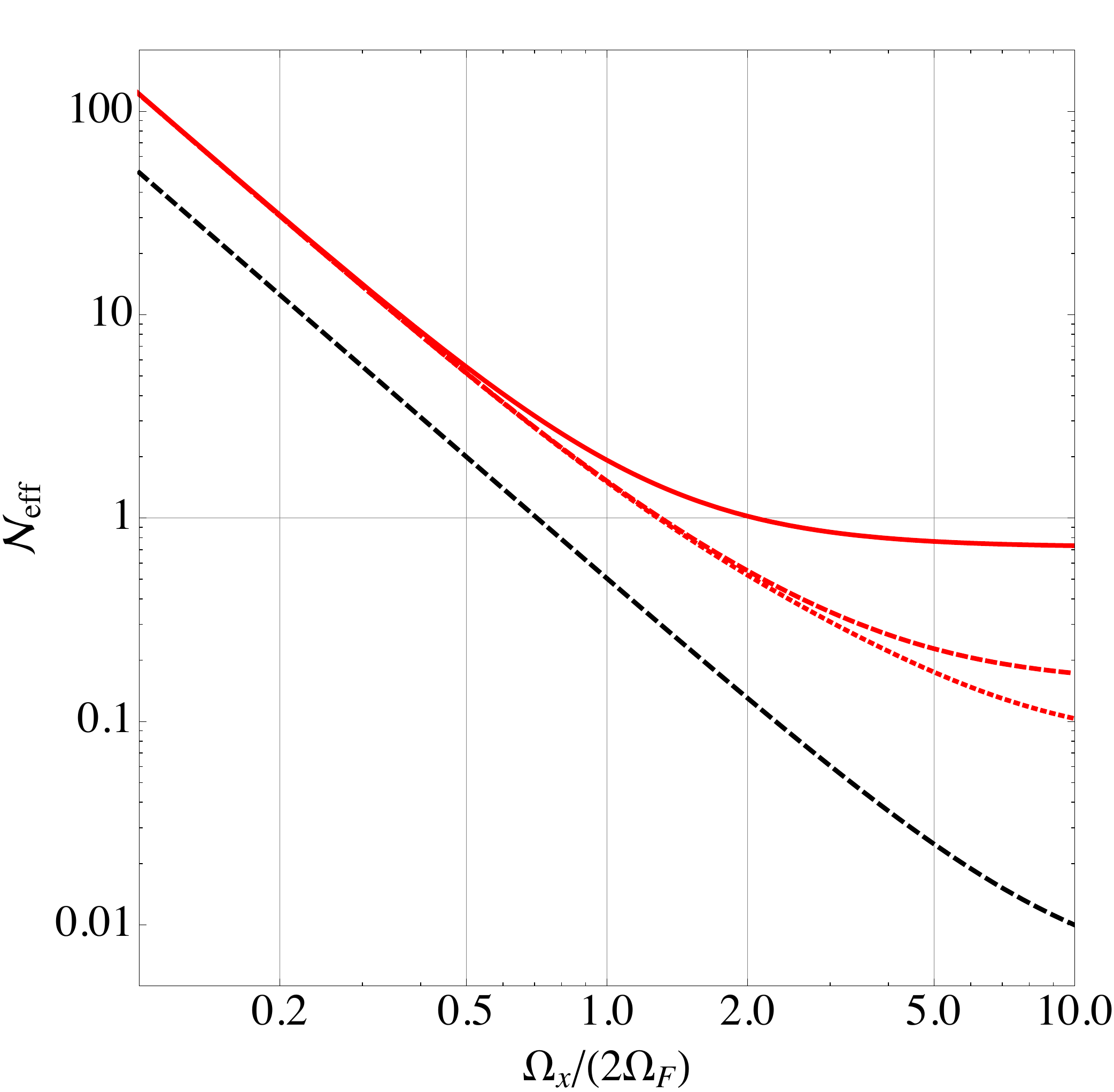}}
\caption{Effective occupation number $\mathcal{N}_{\rm eff}$ achievable as a function of $\Omega_x/(2\Omega_F)$ (which is also the ratio by which the device's classical noise beats the free-mass SQL, see Sec.~\ref{subsec:conditional} for details) by: (i) conditional state preparation (black dashed), (ii) steady state obtained by optimally feeding back  the out-going phase quadrature (red solid), (iii) steady state obtained by optimally feeding back an optimal, frequency independent output quadrature (red dashed) and (iv) steady state by feeding back phase quadrature, but with 10\,dB of frequency-independent input squeezing at an optimal quadrature (red dotted).  [Combination of detection quadrature optimization and squeezing injection did not turn out to improve much upon case (iv).] \label{fig:FOM}}
\end{figure}

\subsubsection{Figures of merit and comparisons with trapping-cooling and conditioning}
\label{subsec:cd:fm}

Because $V_{xp}^{\rm cond}\neq 0$, the controlled steady state of the mechanical oscillator is always imperfect.  As an example, if one starts from a free test mass, and measure the phase quadrature of the out-going light (a ``conventional measurement strategy''), the best one can do will be  $1/\sqrt{2}$.  This is because for a free mass, the frequency scale of the controlled oscillator is always set by the measurement process, and measurement decoherence can never be completely overcome. 

As shown by Danilishin et al.~\cite{Danilishin:2008}, for an oscillator starting with frequency $\omega_m$, quality factor $Q_m$ and temperature $T_m$, as we use phase-quadrature readout to measure the mirror's position  (as in Sec.~\ref{subsec:conditional}) and use optimal filter function for the controller to prepare its quantum state, there is a critical temperature 
\begin{equation}
T_c = \frac{\hbar \omega_m Q_m }{2\sqrt{2}k_B}
\end{equation}
if $T_m$ is  above this temperature, the optimal scheme is to use as much power as possible to measure this object, and use feedback to create an oscillator whose frequency is determined by the measurement and feedback process.  In this case, we are preparing an oscillator under strong measurement, and the best occupation number achievable is $1/\sqrt{2}$.  By contrast, below that critical temperature, the measurement scheme should be chosen at an optimal strength, and the achievable occupation number is proportional to
\begin{equation}
\mathcal{N}_{\rm eff} \approx 2^{-3/4} (T_m/T_c)^{1/2} \,, T_m \ll T_c\,.
\end{equation}
This illustrated in the left panel of Fig.~\ref{fig:FOM}.  Regardless of whether $T_m>T_c$ or $T_m<T_c$, feedback cooling either has a limited final occupation number, or has an optimal measurement strength --- This is the price we pay for having to make a measurement but not taking full advantage of the data.  By contrast, trapping-cooling schemes mentioned in Sec.~\ref{subsec:trapping} does not have such an issue, because there is in principle no limit for the optical dilution factor.  However, as we have mentioned in Sec.~\ref{subsec:trapping}, to obtain large dilution factor often comes with instabilities as well as excess radiation-pressure noise, and often requires more than one optical mode to be coupled to the mechanical oscillator.

For conditional-state preparation, we can avoid the above limitations posed by the requirement of a steady state.  In Fig.~\ref{fig:FOM}, for a free mass, we plot, as a function of the SQL-beating factor $\Omega_x/(2\Omega_F)$ of the device's classical noise budget (see Sec.~\ref{subsec:conditional} for definition),  the $\mathcal{N}_{\rm eff}$ achievable by a conditional state preparation measuring the out-going phase quadrature (without injecting squeezing), compared with steady-states prepared by feedback of  out-going phase quadrature, a general out-going quadrature, and  out-going  phase quadrature assuming frequency-independent squeezing of 10\,dB.  In addition to the classical sensing noise and force noise, characterized by $\Omega_x$ and $\Omega_F$, respectively, we have also added an optical loss of $2\%$.  A substantial gap exists between the quality of the steady state and the conditional state --- despite the use of squeezing and the optimization of readout quadrature.

\begin{figure*}
\centerline{\includegraphics[width=6in]{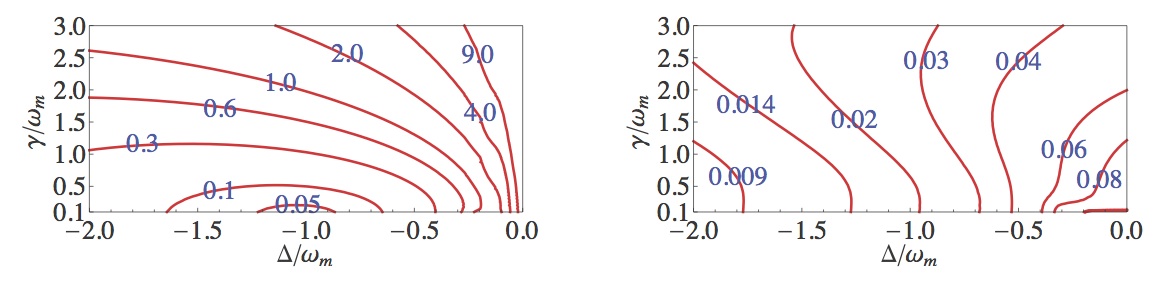}}
\caption{Recovering steady-state purity via quantum feedback control (in absence of classical noise).  [This figure is reproduced from Ref.~\cite{Miao:2010d}; its convention for $\Delta$ is opposite to ours.  Here $\Delta/\omega_m<0$ corresponds to light being red tuned.]  Left panel: effective occupation number achievable by radiation-pressure damping alone, as a function of detuning $\Delta/\omega_m$ and cavity bandwidth $\gamma/\omega_m$.  Here we require the resolved side-band limit, namely $\gamma \ll\Delta$ to obtain efficient cooling. (The case $\Delta/\omega_m = -1$ is extensively discussed in Sec.~\ref{subsec:radcooling}.)  Right panel: as an appropriate out-going quadrature field is detected and optimally fed back onto the oscillator, effective occupation number is shown to decrease significantly, and resolved side-band limit is not necessary.  \label{fig:recover}}
\end{figure*}

\subsubsection{Further developments}

If cooling-trapping is already possible in the {\it resolved side-band limit}, it should be more powerful than feedback cooling.  However, if a device uses radiation-pressure damping, but cannot achieve resolved side-band cooling, feedback cooling can be applied as a remedy.  Using the relation between the conditional and the optimally controlled state developed in Sec.~\ref{subsubsec:optctrl}, it is straightforward to re-evaluate radiation-pressure cooling experiments that are not at the resolved side-band limit. Ideally, the additional quantum uncertainty of the oscillator could be due to: (i) entanglement between oscillator and cavity optical mode, and (ii) information that leave from the out-going light.  In Ref.~\cite{Miao:2010d}, Miao et al.\ has carried out this study, and in Fig.~\ref{fig:recover}, we compare the effective occupation number achieved by a generic radiation-pressure cooling (left panel) and the one one achievable by measuring and optimally feeding back an output quadrature (right panel).   As in Sec.~\ref{subsec:radcooling}, here we have limited ourselves in the regime with power low enough not to shift the real part of the oscillator's eigenfrequency, but high enough such that the damping it creates dominates over the oscillator's original damping. In this regime, the additional occupation number $\mathcal{N}_{\rm eff}$ caused by light does not depend on optical power. As we can see from Fig.~\ref{fig:recover}, feedback is able to strongly suppress $\mathcal{N}_{\rm eff}$.  This indicates that, not only is the cavity mode not much entangled with the motion of the oscillator, the information that comes with the out-going light in such a way that $V_{\rm xp}^{\rm cond}\ll \hbar$ [See Sec.~\ref{subsubsec:optctrl}]. Recalling discussions at the end of Sec.~\ref{subsec:wiener}, this suggests that information for $\hat x(t)$ does not appear in the out-going light right at the ``last moment before $t$'', but instead had to be accumulated over a timescale much longer than $1/\omega_m$.

As a more advanced strategy of feedback cooling, as shown by Szorkovszky et al.~\cite{Szorkovszky:2011}, if the spring constant is modulated at twice the mechanical oscillation frequency, $2\omega_m$, the parametric amplification effect of this modulation, plus the effect of measurement and feedback, and allow substantial improvement towards preparing near pure quantum states.  Moreover, because of parametric amplification, the states they prepare can  be substantially squeezed, and has its $\Delta x$ and $\Delta p$ each oscillate at a frequency of $\omega_m$.  Such a ``breathing''  arises from the non-stationarity of the setup, and cannot be achieved by a purely stationary scheme.   This strategy is similar to previous proposals for electromechanical systems~\cite{Ruskov:2005,Ruskov:2005b}. 

\subsubsection{Measurement-based control versus coherent quantum control}

More broadly speaking, control theory deals with the dynamics of a composite system made up from a ``plant'' and a ``controller''.   The controller collects information about the plant (via observation), performs a calculation, and acts back onto the plant (via actuation). In classical physics, both the plant and controller satisfy classical equations of motion, the ``observation'' can in principle be perfect, but is usually noisy in practical situations. The dynamics of any one such composite system can be analyzed simply by solving the differential equations they satisfy.  However, control theory provides us with a language to describe the behaviors of such composite systems, and then a set of tools to: (i) estimate the behaviors of the composite systems for large classes of controllers without having to solve individually the joint equations of motion, and (ii) qualitatively design the controllers in order to achieve a certain set of behaviors.  

We have so far discussed feedback control scenarios in which the plant is quantum, but the controller is classical --- for this reason, the observation process has to convert quantum information into classical information via quantum measurement, thereby causing decoherence.  This is an important class of control systems to study, because one might imagine classical computers with complex algorithms being used for controlling more complex quantum systems.  

However, one can also imagine using a quantum system for the controller --- and in this way the observer will not have to convert quantum information into classical.  Such control systems are referred to ``coherent quantum control systems''~\cite{Lloyd:2000,Nurdin:2009}.   One can already consider optical cavities as quantum controllers, and one can consider trapping-cooling as coherent quantum control, while feedback cooling  as  quantum control.  Hamerly et al.~\cite{Hamerly:2012a,Hamerly:2012b} recently made an explicit comparison between measurement-based feedback control and coherent quantum control for cooling mechanical oscillators, and demonstrated the advantage of the latter.  Coherent feedback of a mechanical oscillator has been experimentally demonstrated by Kerckhoff et al.~\cite{Kerckhoff:2012}.   Recently, Jacobs and Wang  discussed the advantage of coherent control for driving the plant into a specific pure state within minimum time~\cite{Jacobs:2012}.

\subsection{State Tomography}
\label{sec:tomography}

A conditional state preparation cannot be a stand-alone scheme, because the out-going light only provides information about the expectation value of the position and momentum, which undergo random walk under measurement --- the remaining uncertainty of position and momentum are the true quantum uncertainty, and these can only be extracted by a series of additional experiments, each repeated many times.  We shall review the most straightforward process to carry out a verification process, which reconstructs the Wigner function of the test mass at a given moment~\cite{Miao:2010b}. The scheme we discuss here demonstrates such a possibility of state verification; depending on one's aim, the verification procedure can be much simplified~\cite{Mari:2011}.

\subsubsection{Error of Tomography}
More specifically, the goal of our state verification scheme is to obtain a map (or tomography) of the mechanical object's Wigner function with an error much less than Heisenberg uncertainty.  Experimentally,  for each quadrature $x_\theta$, we will measure the distribution of a new observable $x_\theta^{\rm meas}$,
\begin{equation}
 x_\theta^{\rm meas} = x_\theta + n_{\theta}^{\rm add}
\end{equation}
which is $x_\theta$, the oscillator quadrature we need to measure, superimposed with an additional error $n_\theta^{\rm add}$ --- noise contained in this error is assumed to be statistically independent from quantum fluctuations of $x_\theta$.
The distribution we measure, and attribute to $x_\theta$ is actually the distribution of $x^{\rm meas}$, which is the distribution of $x_\theta$ convolved with the distribution of $n_\theta^{\rm add}$:
\begin{equation}
\label{convolve}
p_{ x_\theta^{\rm meas}}(y) = \int  p_{x_\theta}(y-z) \exp\left({-\frac{z^2}{2 V_{\theta\theta}^{\rm add}}}\right)dz
\end{equation}
Here $p_{x_\theta}$ is the distribution of $x_\theta$, obtainable from its true Wigner function,  
\begin{equation}
V_{\theta\theta}^{\rm add}\equiv \langle n_{\theta}^{\rm add} n_{\theta}^{\rm add}\rangle
\end{equation}
is the variance of the $\theta$-quadrature measurement error $n_\theta^{\rm add}$. The verification procedure will obtain the unique Wigner function that is compatible with $p_{x_\theta^{\rm meas}}$, through a Radon transformation~\cite{Lvovsky:2009}.   We shall not assume the state preparation process would yield a Gaussian state (See Secs.~\ref{subsec:nongaussian} below), but we do assume that the verification process to be linear and Gaussian.  We will even find that  the additional noise is describable with a ``noise ellipse'':
\begin{eqnarray}
\langle n_{\theta_1}^{\rm add} n_{\theta_2}^{\rm add} \rangle &=& (\cos\theta \;\sin\theta)
\left[
\begin{array}{cc} 
V_{xx}^{\rm add} & V_{xp}^{\rm add} \\
V_{xp}^{\rm add} & V_{pp}^{\rm add} \\
\end{array}\right]
\left(
\begin{array}{c}
\cos\theta \\
\sin\theta
\end{array}\right) \nonumber\\
&=&  (\cos\theta \;\sin\theta)\,
\mathbf{V}^{\rm add}\left(
\begin{array}{c}
\cos\theta \\
\sin\theta
\end{array}\right)
\end{eqnarray}

In this case, the Wigner function we construct from the sequences of measurements  is simply the convolution of the true Wigner function with a Gaussian with noise ellpse specified by $\mathbf{V}^{\rm add}$:
\begin{eqnarray}
&& W^{\rm recon}(x,p) \nonumber\\
&=& \int du dv \,\frac{e^{-\frac{1}{2} (u\;v) \,\left(\mathbf{V}^{\rm add}\right)^{-1}\, \left(\scriptsize\begin{array}{c} u \\ v \end{array}\right)}}{2\pi \sqrt{\det \mathbf{V}_{\rm add}}}\,W(x-u,p-v) \nonumber\\
\end{eqnarray}
One can first check this $W^{\rm meas}$ does give Eq.~\eref{convolve}, and then use the uniqueness in the correspondence between Wigner function and the distribution of all quadratures.   In the case the prepared state is Gaussian, we simply return to gaining and additional piece in the covariance matrix:
\begin{equation}
\mathbf{V}_{\rm recon} = \mathbf{V} + \mathbf{V}_{\rm add}
\end{equation}
We can then define a verification process with {\it sub-Heisenberg} error as one with
\begin{equation}
\label{subheis}
\det \mathbf{V}^{\rm add} <\frac{\hbar^2}{4}\,.
\end{equation}
Unless this is satisfied, the measured Wigner function $W^{\rm recon}$ cannot have any negative regions.   In addition, verification of Gaussian entanglement by calculating negativity of the covariance matrix~\cite{Duan:2000,Simon:2000} also requires \eref{subheis}.  We shall refer to verification processes that takes equal sign in Eq.~\eref{subheis} as {\it Heisenberg-limited tomography}. 

Note that there is no fundamental limit to the accuracy for state tomography --- by doing a tomography we are not trying to convert $\hat x$ and $\hat p$ into classical numbers (the extent one can do that is limited by the Heisenberg Uncertainty), but instead  trying to map out the Wigner function, which is related to the  {\it distribution} we obtain when trying to convert linear combinations of $\hat x$ and $\hat p$ into classical numbers.

\subsubsection{Achieving Tomography and the use of Back-Action Evasion}
\label{subsec:bae}

 Now suppose we would like to carry out tomography for the Wigner function of the mechanical object at $t=0$, we must be able to measure the distribution for a large number of quadratures $x_\theta(t=0)$.  In order to do so with non-zero signal-to-noise ratio, we must collect data for a non-zero duration. Note that during the time after $t=0$, we have
\begin{equation}
\hat x(t) = \hat x(0) \cos\omega_m t + \frac{\hat p(0)}{m\omega_m}\sin\omega_m t = \hat x_{\omega_m t} 
\end{equation} 
This means, if we can accumulate the out-going light in an appropriate way, we shall we able to measure
\begin{equation}
\int_0^{+\infty} f(t) \hat x(t) dt
\end{equation}
which will yield a particular initial quadrature of the oscillator.  The complications are: (i) there will be measurement back action and classical force noise 
\begin{eqnarray}
\hat x(t) &=&\hat x(0) \cos\omega_m t + \frac{\hat p(0)}{m\omega_m}\sin\omega_m  t  \nonumber\\ &+&\int_0^t \frac{\sin\omega_m(t-t')}{m\omega_m} \left[\alpha \hat a_1(t') + \hat n_F(t')\right]
\end{eqnarray} 
and (ii) sensing noise:
\begin{eqnarray}
\hat b_1 (t) &=& \hat a_1(t) \\
\hat b_2(t) &=& \hat a_2(t) + \alpha \left[\hat n_x(t) +\hat  x(t)\right]
\end{eqnarray}
As it turns out, even in absence of classical noise $\hat n_F$ and $\hat n_x$, if we always measure the out-going $\hat b_2$ quadrature, and carry out filtering of 
\begin{equation}
\label{eqb2readout}
\int_0^{+\infty}g_2(t) \hat b_2(t) dt
\end{equation}
then by optimizing $g_2$, for particular quadratures $x_\theta$, the back-action and sensing-noise together imposes an exactly {\it Heisenberg-limited} tomography~\cite{Miao:2010b}.  One way to get out of this is to inject squeezing, so that for different injected squeezed vacuum, subject to the limit of $S_{a_1}S_{a_2}-S_{a_1 a_2}^2 =1$, we will obtain different Heisenberg-limited tomographies --- combining these information will lead to a sub-Heisenberg tomography. 

However, a remarkably  elegant scheme that relies on {\it back-action evasion} can be inspired by the {\it variational measurement} invented by Vyatchanin et al.~\cite{Vyatchanin:1993,Vyatchanin:1995}, who observed that if signal waveform is known, back-action evasion can be achieved if we measure a time-dependent optical quadrature.  (For us, each mechanical oscillator quadrature basically corresponds to the signal having a particular phase.)  The key is, in absence of classical noise, if we decide to use the particular filtering of $b_2$ shown in Eq.~\eref{eqb2readout}, to evade back-action, we simply need to add the right amount of $b_1(t)$, by detecting the combination of  
\begin{equation}
\label{eqb2}
\int_0^{+\infty}\left[g_1(t) b_1(t) + g_2(t) b_2(t) \right]dt
\end{equation}
in such a way that the $a_1$ content of the two terms cancel with each other.  In our case, the unique way to completely evade back action is to choose:
\begin{equation}
g_1(t) +\frac{\alpha^2}{m\omega_m} \int_t^{+\infty} dt' \sin\omega_m(t'-t) g_2(t') =0 
\end{equation}

\subsubsection{Figures of merit and experimental prospects}
\label{subsec:fom:verify}

In presence of classical noise, the filters $g_1$ and $g_2$ will have to be chosen such that the measurement time is short enough for force noise not to matter, and long enough for sensing noise not to matter. The result of the optimization for white force and sensing noise,  assuming $\Omega_q \gg \omega_m$, is analytical and 
\begin{eqnarray}
\label{vxxadd}
V_{xx}^{\rm add} &=& \frac{\sqrt{2}\hbar}{M\Omega_q}\cdot  \Lambda_x^{3/2} \xi_F^{1/2} \,, \\
V_{xp}^{\rm add} &=& -\Lambda_x \xi_F {\hbar} \,,\\
V_{pp}^{\rm add} &=&  {\sqrt{2}\hbar M\Omega_q} \cdot  \Lambda_x^{1/2}\xi_F^{3/2} 
\label{vppadd}
\end{eqnarray}
with
\begin{equation}
\label{xilambda}
\xi_F = \frac{\Omega_F}{\Omega_q}\,,\quad \xi_x =\frac{\Omega_q}{\Omega_x}\,,\quad \Lambda_x = \sqrt{\xi_x^2 +\frac{e^{-2q}}{2}}
\end{equation}
where $\Omega_q$, $\Omega_x$ and $\Omega_F$ are as defined in Sec.~\ref{subsec:conditional}, and $e^{-2q}$ is the power squeezing factor (i.e., $e^{-2q}=0.1$ for ``10\,dB squeezing'').     From Eqs.~\eref{vxxadd} and \eref{xilambda}, we obtain
\begin{eqnarray}
\label{vadd_tom}
\frac{\det\left(\mathbf{V}_{\rm add}\right)}{\hbar^2/4} &=& \Lambda_x^2 \xi_F^2  = \left(\frac{\Omega_F}{\Omega_x}\right)^2 +\frac{e^{-2q}}{2} \left(\frac{\Omega_F}{\Omega_q}\right)^2 \nonumber\\
&=&\frac{1}{4}\min_\Omega\left[\frac{S_{\rm cl}(\Omega)}{S_{\rm SQL}(\Omega)}\right] +  \frac{e^{-2q}}{2} \left(\frac{\Omega_F}{\Omega_q}\right)^2\,.
\end{eqnarray}
Here the error decreases monotonically as we increase $\Omega_q$ --- even vanishingly small in absence of classical noise --- thanks to back-action evasion.   In presence of classical noise, the necessary condition for sub-Heisenberg verification is for the classical noise to beat the SQL.  Interestingly, as we increase $\omega_m$, the tomographic error ellipse will change shape, but {\it Eq.~\eref{vadd_tom} for area of the error ellipse remains universally valid for all values of $\omega_m$}. 

Up till now, we have seen that although conditional state preparation only requires classsical noise to be below the SQL --- and then the quantum noise will automatically enforce Heisenberg Uncertainty for the mirror, thereby creating a nearly pure conditional state, an accurate state tomography requires us to use advanced detection techniques, including back-action evasion (via time-dependent homodyne detection) and rather substantial levels of  input squeezing.  At this moment, squeezing has been reliably applied to gravitational-wave detectors, improving sensitivity to a level that had not been achieved before~\cite{Abadie:2011}, frequency-domain variational measurements have been demonstrated at high frequencies in the context of frequency-dependent homodyne detection~\cite{Chelkowski:2005} and in the context of variational coupling with a nanomechanical oscillator~\cite{Hertzberg:2009}. Time-domain variational measurements have not been demonstrated. [See Sec.~\ref{subsec:sqlbeating} for more details.]

\subsection{Example of a preparation-evolution-verification experiment}
\label{subsub:timeline}

Before we move on to discuss applications of the basic strategies listed in this section, let us illustrate their use by outlining an experiment that prepares a squeezed state of the test mass, lets the squeezed state evolve for different durations of time, and observes the ``breathing'' of position uncertainty.  The requirements on classical noise levels are challenging, but perhaps realizable within the near future; this time-domain operation mode may incur transients, which must be studied carefully. 

\begin{figure}
\centerline{\includegraphics[width=2.75in]{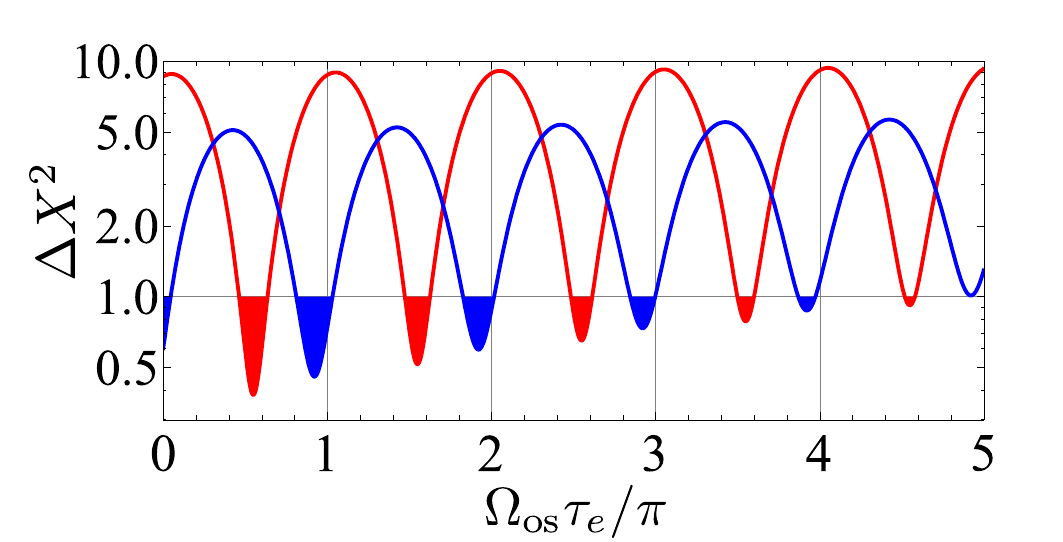}}
\caption{Breathing of $\Delta x$ of a squeezed state of a mechanical oscillator. \label{fig:breathe}}
\end{figure}

Suppose we have a device with $\Omega_x = 50 \Omega_F$. For example, a generation of gravitational-wave detectors beyond Advanced LIGO will have $\Omega_x \sim 10 \,$Hz, and $\Omega_F \sim 500\,$Hz.  This corresponds to an SQL-beating factor of 5 in amplitude [See Eq.~\eref{eq:sqlbeating}].  Let us also assume an optical spring can be imposed at will, and we can achieve 10\,dB of squeezing without loosing much of it.   Let us describe the timeline of a first scenario of the experiment, if we would like to obtain a full tompgraphy:
\begin{enumerate}
\item {\it State Preparation.}    In this process ($t'<0$), we assume a measurement strength that corresponds to  $\Omega_q = 0.23 \sqrt{\Omega_x\Omega_F}$, and there is no optical spring. We use Wiener Filters to obtain continuously the conditional expectations $\langle x\rangle$ and $\langle p\rangle$,  which undergo random walks.  The conditional variance should gradually approach a steady state --- but that is for the next stages of experiment to verify. 
\item  {\it State Evolution.}  At $t=0$, we  attach an optical spring to achieve $\Omega_{\rm opt} = 1.2 \sqrt{\Omega_x\Omega_F}$.  In order to remove the decoherence due to radiation pressure noise, we  monitor the out-going  $b_1$, which is proportional, up to sensing noise, to the radiation-pressure noise acting on the oscillator.   
\item {\it State Verification.}  At time $t=\tau_e$, we start the state-tomography process, we assume turning off the optical spring and starting verification with measurement strength of $\Omega_q'=3.0\sqrt{\Omega_x\Omega_F}$, we also assume 10\,dB squeezing is injected.  
\begin{enumerate}
\item At each run of the experiment for a fixed $\tau_e$, we can only verify one oscillator quadrature $\zeta$, obtaining one value.  We will have to subtract from this value the corresponding contributions from $\langle x\rangle$ and $\langle p \rangle$ obtained in Step (i), as well as the radiation-pressure force $b_1$ in Step (ii).
\item We will have to repeat (a) many times for the same $\zeta$, obtaining a distribution of $\hat x_\zeta(\tau)$.  Then for many values of $\zeta$, we will have to obtain their distributions. Assembling all these one-dimensional distributions, and applying a Radon transformation, we will finally be able to obtain the Wigner function at time $t=\tau_e$. 
\item We will have to repeat the above process for different values of $\tau_e$, to obtain the ``verified Wigner function'' of the oscillator as a function of time. 
\end{enumerate}
\end{enumerate}
If we were to only verify $\Delta x(\tau_e)$, we will be able to see the ``breathing effect'' due to mechanical squeezing, because Step (i) of the above process will be preparing  a momentum  squeezed state (relative to the ground state of a harmonic oscillator with $\Omega_{\rm opt}$) at the beginning. Given the noise budget, in Fig.~\ref{fig:breathe}, we show that, depending on $\tau_e$, the total verified $\Delta x$ oscillates as a function of time, following the red curve.  At the beginning we have an anti-squeezed state of the oscillator, with $\Delta x > \Delta x_{\rm vac}$, the zero-point standard deviation.  As $\tau_e$ increases, $\Delta x$ decreases, and eventually dips below vacuum fluctuation --- this indicates that the mechanical oscillator is truly squeezed.  Although thermal decoherence becomes more important in the longer term, in this setup, we shall see $\Delta x$ go below vacuum level five times.  Of course, each $\tau_e$ in the plot corresponds to many experimental runs, which are required to carry out tomography at $t=\tau_e$.  The turning on and off of optical spring is only there to ease the computation so that formulas in this paper can be used directly.   It should be possible to find experimental strategies in which one does not have to only turn on the optical spring during evolution, but instead can keep it there.

In the second scenario, we choose $\Omega_q = 2.5\sqrt{\Omega_x \Omega_F}$, $\Omega_{\rm opt} = 0.8\sqrt{\Omega_x\Omega_F}$ and $\Omega_q'=3.0\sqrt{\Omega_x\Omega_F}$.  In this case, we will be preparing a position-squeezed state at the beginning, therefore $\Delta x$ dips below vacuum level even at $\tau_e=0$.  We will be able to see $\Delta x$ dip below vacuum for 5 times. \here 

\section{Further Developments for Linear Systems}
\label{sec:further}

In this section, building on the basic experimental concepts discussed in  Sec.~\ref{sec:exp_linear}, we  discuss several more advanced experimental concepts for quantum optomechanical systems that aim at exposing features of quantum mechanics in macroscopic objects.

\subsection{Ponderomotive Squeezing and Entanglement between out-going optical modes}

\subsubsection{Ponderomotive Squeezing}

It was long known that a cavity with a moving mirror, when pumped with strong carrier field, can convert an input  coherent state into an output squeezed state~\cite{Pace:1993}; similar mechanism also applies to electromechanical devices~\cite{Woolley:2008}.  This is also referred to as {\it ponderomotive squeezing}, because ponderomotive force, i.e., low-frequency components of the force exerted onto the mirror by light, is involved in converting  quantum amplitude fluctuations of light into mirror motion, which in turn gets converted into quantum phase fluctuations of the out-going light.  The classical coupling mechanism has been demonstrated~\cite{Mow-Lowry:2004,Briant:2009,Marino:2010}, while the quantum squeezing has recently been observed by Brooks et al.\ using an atomic ensemble as the mechanical oscillator~\cite{Brooks:2011}.  

Another way of generating squeezing using moving mirrors is the Dynamical Casimir Effect~\cite{Nation:2012,Wilson:2011}, in which a mirror moving with high oscillatory velocity at $2\omega_0$ converts incoming vacuum into squeeze vacuum, but we shall not go into the details because the motion of the mirror here is highly classical.

The simplest experimental setup for generating ponderomotive squeezing is described in Sec.~\ref{subsec:elimination}, where the cavity's detuning and bandwidth are both much greater than our frequencies of interest.   The input-output relation of the system can be written as
\begin{equation}
b_1 =a_1\,,\quad 
b_2 = a_2 + \alpha x
\end{equation}
and 
\begin{equation}
- M(\Omega^2-\omega_{\rm opt}^2) x = \alpha a_1 + n_F\,,
\end{equation}
Combining the two equations, we realize that using a free mass without optical rigidity (i.e., $\omega_{\rm opt}^2 \ll \Omega^2$) will produce frequency-dependent squeezing~\cite{Kimble:2001}, with a frequency dependence that may not be most suitable for improving sensitivities of measuring devices~\cite{Kimble:2001,Purdue:2002b,Harms:2003,Chen:2003a,Danilishin:2004,Tsang:2010}.  

Nevertheless, as shown by Corbitt et al.~\cite{Corbitt:2006}, if we have a strong optical spring, then for frequencies below the optomechanical resonance ($\Omega \ll \omega_{\rm opt}$), the output squeezing is frequency independent (writing $\alpha^2=M\Omega_q^2$): 
\begin{equation}
b_1 = a_1\,,\quad b_2 = a_2 +\left(\frac{\Omega_q}{\omega_{\rm opt}}\right)^2 a_1 + \frac{n_F}{\sqrt{M}\omega_{\rm opt}^2/\Omega_q} \,.
\end{equation}
Remarkably, in absence of additional noise (e.g., due to optical losses),  we {\it always gets at least a mild squeezing} from the out-going field, even regardless of the level of thermal noise.  Suppose we measure 
\begin{equation}
b_\zeta = b_1 \cos\zeta + b_2\sin\zeta
\end{equation}
with $\zeta \ll 1$, we have
\begin{equation}
S_{b_\zeta} = 1+\frac{2\Omega_q^2}{\omega_{\rm opt}^2}\zeta + O(\zeta^2)
\end{equation}
this is less than unity if $\zeta$ is a negative number with a very small magnitude.  

However, in order to obtain a {\it substantial squeezing factor} of $e^{-2q}$, one must tune the optomechanics so that
\begin{equation}
\omega_{\rm opt} \approx e^{-q} \Omega_q
\end{equation}
and require a low enough thermal noise (assuming velocity damping),
\begin{equation}
\Omega_F \stackrel{<}{_\sim} e^{-2q} \omega_{\rm opt}\,.
\end{equation}
For a squeezing at level of $\sim $3\,dB, we need 
\begin{equation}
\Omega_F \stackrel{<}{_\sim} \omega_{\rm opt} \stackrel{<}{_\sim} \Omega_q\,.
\end{equation}
This is a similar requirement to having the spectrum of quantum back-action force comparable to that of thermal noise~\cite{Borkje:2010}, as has recently been achieved by Purdy et al.~\cite{Purdy:2012}.



\subsubsection{Entanglement between out-going fields}

Pirandola et al.~\cite{Pirandola:2003} suggested that several beams of light simultaneously incident on the same mirror will cause the returning beams to be entangled.  Wipf et al.~\cite{Wipf:2008} considered such an entanglement in the context of stable double optical springs~\cite{Corbitt:2007}, and  showed that the optical dilution effect makes the entanglement between out-going light also robust against thermal noise.

\subsubsection{Optomechanical generation of frequency-dependent squeezing.}

In sub-SQL quantum measurements it is often interesting to create frequency-dependent squeezing (See ~\ref{subsec:sqlbeating}).  One way to realize this is to use detuned Fabry-Perot Cavities as filters, yet for applications like LIGO, the long timescale of measurement means the requirement of narrow linewidth cavities ($\sim$100\,Hz), which must store light for a long time, and hence can be either very long or  very lossy, or both.  It was proposed that materials with Electromagnetically Induced Transparency (EIT), which makes light travel slowly, can be used to reduce the length of such filters~\cite{Mikhailov:2006}. Coupling to mechanical motion can also bring effects similar to EIT --- this has been referred to as Optomechanically-Induced Transparency (OMIT)~\cite{Weis:2010,Teufel:2011b,Safavi-Naeini:2011,Chang:2011}. This feature in principle allows us to build a optomechanically tunable frequency-dependent quadrature-rotation device in the audio frequency band.  However, the requirement on low thermal noise and low optical losses seem rather challenging~\cite{Zhao:2012}.


\begin{figure}
\centerline{\includegraphics[width=3.4 in]{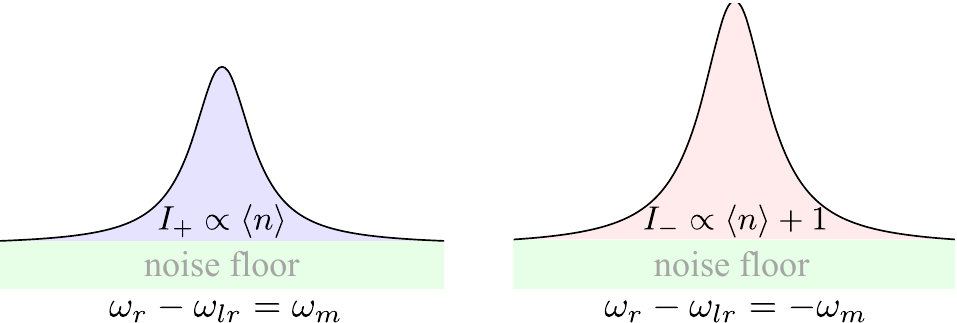}}
\caption{Amplitude-fluctuation spectra for transmitted, red-detuned (left) and blue-detuned (right) light.  The level of asymmetry measures how close the mechanical oscillator is to the ground state. In Sec.~\ref{subsec:asym}, Khalili et al.~\cite{Khalili:2012} offer two explanations for the asymmetry.}
\end{figure}

\subsection{Detecting Zero-Point Fluctuation of a Harmonic Oscillator}
\label{subsec:asym}

In this section, we discuss an alternative way to verify that a mechanical oscillator has been cooled to ground state --- without having to carry out state tomography.  The entire experimental strategy is carried out in steady state, and all measurements only involve output spectrum.  This strategy has been employed by several experimental groups to interprete their near-ground-state cooling experiments~\cite{Brahms:2012,Safavi-Naeini:2012,Safavi-Naeini:2012b}. We shall present two compatible yet very different ways of understanding this strategy.  

Here we consider a cavity with two partially transmissive mirrors, one movable and another fixed.  Suppose the movable mirror is already cooled to a nearly ground state, and we would like to probe the motion of the movable mirror by pumping the cavity  with light at angular frequency $\omega_0$, which is detuned from the cavity (the cavity resonates at a nearby frequency of $\omega_0 +\Delta$), and observe the amplitude fluctuations of the transmitted light. For simplicity, we shall assume the cavity is either resonant at $\omega_0+\omega_m$ (light is red detuned), or $\omega_0-\omega_m$ (light is blue detuned), and $\gamma \ll |\Delta|$ (i.e., resolved side-band limit). 

In both cases, because cavity is detuned, motion of the mirror would modulate the transmissivity of the cavity, thereby creating amplitude modulation to the out-going light.  One would therefore expect the amplitude fluctuation of the out-going light would have a peak near $\omega_m$.  However, a more detailed consideration indicates a difference between red- and blue-detuned cases. 

\subsubsection{Explanation in terms of transition between levels.}

The  first explanation notices that amplitude fluctuation of out-going light requires photon number fluctuations inside the cavity, which in turn requires the excitation of the cavity mode $(\hat A,\hat A^\dagger)$.  Under the rotating wave approximation~\cite{Schliesser:2006,Safavi-Naeini:2012,Safavi-Naeini:2012b}, in the red-detuned case, the interaction Hamiltonian between the cavity $(\hat A,\hat A^\dagger)$ and the mirror $(\hat B,\hat B^\dagger)$ can be written as
\begin{equation}
V_I = \hat A^\dagger \hat B +  \hat A \hat B^\dagger\,.
\end{equation}
This means, in order for mirror motion to affect photons' entrance into the cavity, the mirror has to go to a lower state, so that the photon can enter the cavity.  However, this is not going to happen if the mirror is already at ground state.    On other other hand, for the blue-detuned case, we have 
\begin{equation}
V_I = \hat A^\dagger \hat B^\dagger +  \hat A \hat B
\end{equation}
photon entering cavity requires mirror going up one level, which is possible.  In this case, motion of mirror affects photon's transmittance.  A more detailed calculation along these lines shows that 
\begin{equation}
\frac{1}{\bar n} =\frac{I_-}{I_+}-1
\end{equation}
where $I_-$ is the additional area below the power spectrum of the blue-detuned light, while $I_+$ is the additional area below the power spectrum of the red-detuned light.

\subsubsection{Explanation using linear quantum mechanics.}

Another point of view does not require concepts of the gound-state being the minimum-energy state, but simply looks at correlations between linear field operators. It is noticed that, in addition to the zero-point fluctuation of the mirror inducing an out-going amplitude fluctuation, the out-going amplitude also fluctuates due to the mirror motion driven by radiation pressure.  The out-going field is
\begin{equation}
\hat O =\hat Z + \alpha^2 \chi \hat F + \alpha \hat x^{(0)}
\end{equation}
where $\alpha$ is a coupling constant, and $\chi$ is the response of the mirror,  $\hat Z$ and $\hat F$ are our ``sensing noise'' and ``back-action noise'',  as defined in Sec.~\ref{subsec:SQL}, while $\hat x^{(0)}$ is the unperturbed position operator for the mirror.  As coupling is weak (keeping only leading effects in $\alpha$), we have
\begin{equation}
S_O = S_Z + \alpha^2 S_x + \alpha^2 \left[S_{ZF} \chi^* +S_{FZ} \chi\right]
\end{equation}
In our case, the back-action force
\begin{equation}
\hat F_\Omega = G \left[\hat A_\Omega^\dagger + \hat A_\Omega\right]
\end{equation}
has an important feature.  For $\Omega>0$,  in  the red detuned case ($\Delta > 0$), we have $\hat A^\dagger_\Omega \sim 0$, and
\begin{equation}
\hat F \propto G \hat a_{\omega_0-\Omega}\,,\quad 
\hat F (\Omega)|0\rangle  \approx 0
\end{equation}
We also have  [Cf.~\eref{commzf}]
\begin{equation}
\left[ \hat Z(\Omega) ,\hat F^\dagger(\Omega')\right] = -i \delta(\Omega-\Omega')
\end{equation}
This, and the definition of $S_{ZF}$ (which is symmetrized) leads to 
\begin{equation}
S_{ZF}^- = -i\hbar
\end{equation}
In the blue-detuned case, a similar argument leads to 
\begin{equation}
S_{ZF}^+ = + i\hbar
\end{equation}
In our two cases, this gives
\begin{equation}
S_O^\pm = S_Z + \alpha^2 S_x \pm 2\alpha^2\hbar \mathrm{Im}\chi 
\end{equation}
Interestingly, for zero-point fluctuation of an oscillator we have
\begin{equation}
S_x = 2\hbar \mathrm{Im} \chi
\end{equation}
This means for the two different detunings, we have 
\begin{eqnarray}
\mbox{red detuned:} & & \quad S_O^{-} = S_Z \,,  \\
\mbox{blue detuned:} & & \quad S_O^{+} =  S_Z + 2\alpha^2 S_x 
\end{eqnarray}
This is the same asymmetry as derived above~\cite{Miao:2012}.

\begin{figure}
\centerline{\includegraphics[width=3.25in]{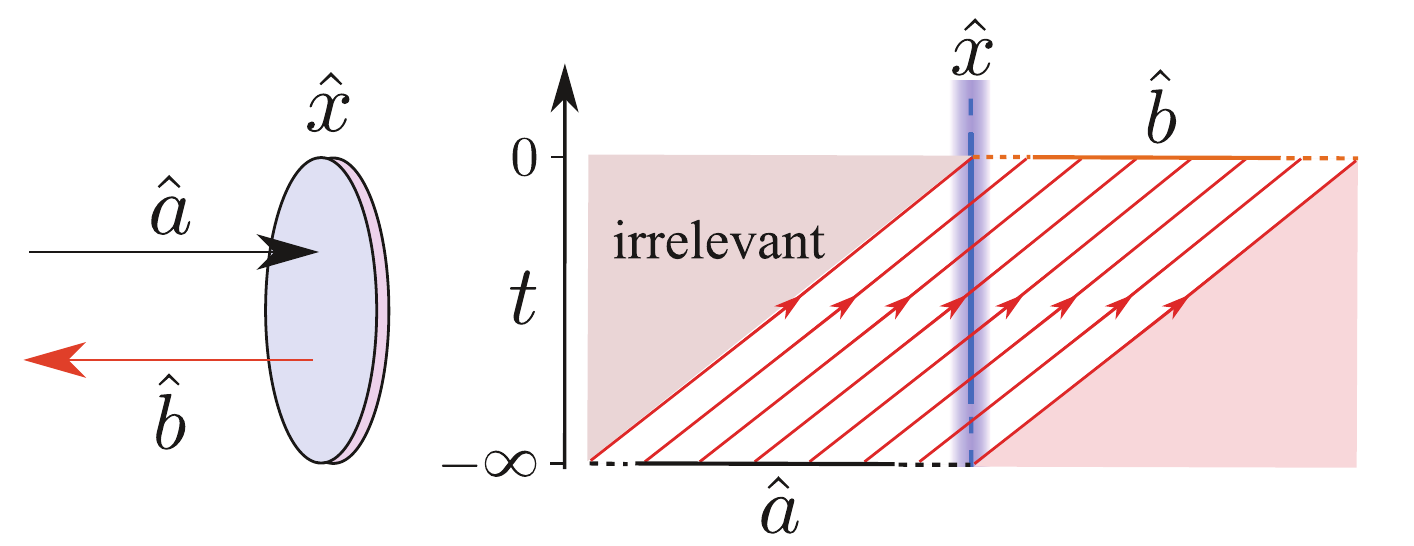}}
\caption{Entanglement between mechanical object and out-going light. Left panel shows the setup of our thought experiment: light measuring the position of a mirror. Right panel is a spacetime diagram that depicts all in-going fields up to $t=0$ as having originated as independent degrees of freedom at $t=-\infty$, which, after interacting with the mirror during $-\infty<t'<0$, emerges once more as independent degrees of freedom at  $t=0$.  In Sec.~\ref{subsec:massfield}, we consider the entanglement between the mirror's state and the out-going field's state at $t=0$. [Figure reproduced from Ref.~\cite{Miao:2010}] }
\end{figure}

\subsection{Entanglement between the mechanical oscillator and the out-going light}
\label{subsec:massfield}

Entanglement is the hallmark for quantum coherence, and often perceived to be difficult to achieve when one of the parties is a  macroscopic object.  As mechanical oscillators interact with light, entanglement may build up as a consequence.  Paternostro et al.~\cite{Paternostro:2007} and Vitali et al.~\cite{Vitali:2007} showed that in optomechanical systems, if temperature is low enough, stationary entanglement can build up between the mirror and the cavity's optical mode --- and that such entanglement can be verified by characterizing the out-going field.  On the other hand, it was also shown that if coupling is strong enough, entanglement may be present in limited even for highly ``classical objects''~\cite{Galve:2010}.

In terms of out-going light from an optomechanical system, 
Giovannetti et al.~\cite{Giovannetti:2007} and 
Bhattacharya et al.~\cite{Bhattacharya:2008d} both considered the entanglement between a mirror being measured and one spatial mode of out-going light (which predominantly couples to the mirror) --- both pointed out experimental regimes in which such entanglement can exist.   

\subsubsection{Universal entanglement.} Here we discuss a more general treatment given by Miao et al.~\cite{Miao:2010}: for a mechanical oscillator at temperature with thermal noise characterized by $\Omega_F$, being measured at time scale $\Omega_q$, if we  consider, at time $t=0$,   the joint quantum state of the  continuum of out-going light [i.e., $\{b_1(t'), b_2(t')|t'<0\}$]
and the mirror [i.e., $(x(0),p(0))$], we find that {\it they are always entangled, no matter how high  $\Omega_F$ is}.  For a high-$Q$ oscillator, the {\it logarithmic negativity} of this joint quantum state~\cite{Duan:2000,Simon:2000} is given by 
\begin{equation}
\mathcal{E} = \frac{1}{2}\log\left[1+\frac{25\Omega_q^2}{8\Omega_F^2}\right]
\end{equation}
which is small yet still non-zero when $\Omega_F \gg \Omega_q$, i.e., when the measurement is highly classical (i.e., highly dominated by thermal noise).   A subsequent analysis revealed that in the classical case, the spatial optical mode of the out-going field that is ``most entangled'' with the mirror leaves the mirror within a time scale of $\sim\Omega_F$. 


\subsubsection{Quantum Steering.} If we would like to take advantage of this ``universal entanglement'', we will have to return to the more stringent experimental conditions considered in Ref.~\cite{Paternostro:2007,Vitali:2007}.    One way would be to look at the effect of ``steering'', proposed first by Wiseman, Jones, and Doherty~\cite{Wiseman:2007,Wiseman:2012}.  The entanglement has to be strong enough in order for different choices of which quantity of the  the output light to measure to result in mutually incompatible conditional quantum states. 

In the case of a mechanical oscillator measured by light, steering can be demonstrated if there exist two quadratures $x_{\phi_1}$ and $x_{\phi_2}$, corresponding to two measurement strategies which measure time-dependent $\theta_1(t')$ and $\theta_2(t')$ quadratures of the out-field, respectively, during $-\infty<t'<0$, and the conditional variances satisfy 
\begin{equation}
\Delta x_{\phi_1}^{|\theta_1} \cdot \Delta x_{\phi_2}^{|\theta_2} < |\sin(\phi_1-\phi_2)|.
\end{equation}
In other words, choice made on the out-going field is able to modify the state of the mirror to an extent such that combination of information from both measurements would result in a state that is below Heisenberg Uncertainty --- which means that the quantum state must indeed be different depending on which measurement scheme is {\it chosen}. 

A closer examination reveals an intimate connection between steering and state tomography~\cite{Mueller-Ebhardt:2012}: in both cases, we are allowed to come up with multiple strategies (i.e., the out-going quadrature $\theta(t')$ to detect) to measure different mechanical quadratures as precisely as possible --- only that steering takes place during $-\infty<t'<0$, while tomography takes place during $0<t'<+\infty$ --- so they are time reversal to each other.  As a consequence, given an optomechanical device, covariance matrix that describes the minimum possible conditional variance of each quadrature, regardless of the choice of quadratures measured during $-\infty<t'<0$, is simply the time-reversed version of the error covariance matrix for tomography:
\begin{eqnarray}
V_{xx}^{\rm st} = V_{xx}^{\rm tm},\quad
V_{xp}^{\rm st} = -V_{xp}^{\rm tm},\quad
V_{pp}^{\rm st} = V_{pp}^{\rm tm}.
\end{eqnarray}
Here we have used the superscript ``st'' for steering, and ``tm'' for tomography.  The $\mathbf{V}^{\rm tm}$ here is the $\mathbf{V}^{\rm add}$ in Sec.~\ref{sec:tomography}, especially Sec.~\ref{subsec:fom:verify} and Eqs.~\eref{vxxadd}--\eref{vppadd}. 

As a consequence, in presence of classical noise,  condition for steerablity is the same as condition for achieving sub-Heisenberg state tomography.  If we follow Wiseman et al.~\cite{Wiseman:2012} and  define {\it steerability} as 
\begin{equation}
\mathcal{S} = -\log\left[2 \sqrt{\det\mathbf{V}^{\rm st}} /\hbar\right]\,,
\end{equation}
with $\mathcal{S} >0$ indicating the mirror's state is steerable by measuring out-going light, then 
\begin{equation}
\mathcal{S} = -\log\left[2 \sqrt{\det\mathbf{V}^{\rm tm}} /\hbar\right]
\end{equation}
If the above steerability is to be verified, we will have to carry out a tomography, which adds $\mathbf{V}^{\rm tm}$, and gives a {\it verifiable steerability} of 
\begin{equation}
\mathcal{S} = -\log\left[4 \sqrt{{V}_{xx}^{\rm tm}{V}_{pp}^{\rm tm}} /\hbar\right]
\end{equation}
As we can see from Eqs.~\eref{vxxadd}--\eref{vppadd}, steering can benefit significantly from squeezing.

\begin{figure*}
\centerline{\includegraphics[width=5.5in]{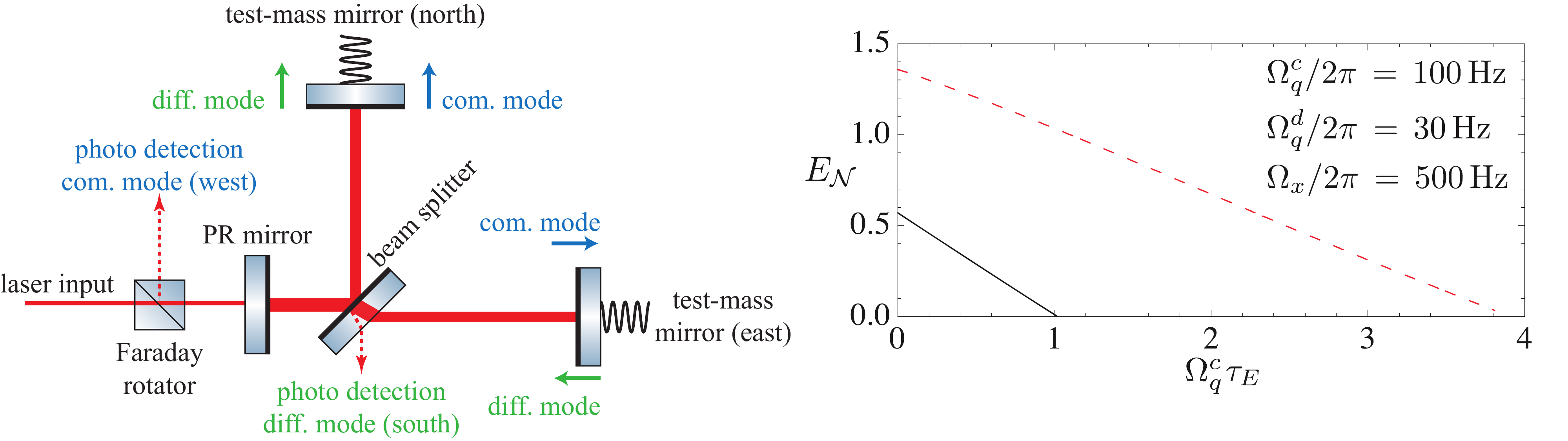}}
\caption{Left panel: sketch of a possible experimental setup to create entanglement between macroscopic test masses [taken from Ref.~\cite{Mueller-Ebhardt:2008b}]. This configuration requires laser noise not to be a significant contribution to the measurement.  More complex interferometer configurations will have to be used to eliminate laser noise.  Right panel: verifiable logarithmic negativity between two mechanical objects as a function of time elapsed after preparation [taken from Ref.~\cite{Miao:2010b}]. Dashed line assumes $\Omega_F=2\pi\times 10\,$Hz, while solid line assumes $\Omega_F=2\pi\times 20\,$Hz. See Sec.~\ref{subsec:massmass} for details. \label{fig:EN}}
\end{figure*}

\subsection{Entanglement between mechanical modes}
\label{subsec:massmass}

Using radiation pressure to create entanglement between mechanical modes has been discussed by many authors.  For example, Mancini et al.~\cite{Mancini:2002} discussed an effective entanglement between two moving mirrors of a driven cavity --- in terms of entanglement between the Fourier components of their motion.   Pirandola et al.~\cite{Pirandola:2003b}, based on work of Mancini et al.~\cite{Mancini:2003}, discussed how entanglement between two Fourier sidebands of light can be transferred to the entanglement of two mechanical oscillators.  The main theme of both works would be to prepare common and differential modes of the two mechanical oscillators to different nearly pure quantum states, and therefore the two oscillators would then be entangled. Such a strategy was also proposed by Zhang et al.~\cite{Zhang:2003}, and have been applied by other authors to propose generating this type of entanglement~\cite{Pinard:2007,Bhattacharya:2008c}.

One approximation commonly employed by the above pioneering work is to use Fourier components of the mechanical oscillators' Heisenberg operators, as well as Fourier components of sideband light, as effective degrees of freedom.  While such an approximation could be valid for scenarios involving oscillators with high quality factors and optical cavities with detuning matched to the frequencies of the oscillators, they do not directly apply to experiments with free masses, or oscillators whose restoring force are created by optical rigidity, and may not have very high quality factors.  The additional subtlety in quoting Fourier components is that since they are not defined at the same moment of time, entangling them may not allow testing of the more fundamental aspects of quantum mechanics.

Here we discuss in more details the work of M\"uller-Ebhardt et al., who not only  treated the quantum states of the mirrors in the time domain, but also incorporated continuous measurement processes.  Having gravitational-wave detectors and prototype experiments in mind, these authors also connected the possibility of entanglement with the noise budget of the system were it to be used as a measuring device~\cite{Mueller-Ebhardt:2008b}.  They further considered a three-stage experiment with preparation, evolution and verification of entangled states, as described in Ref.~\cite{Miao:2010b}.   For these authors, the two mechanical objects can be the two mirrors of a Michelson interferometer, and entanglement is realized as we measure the common and differential mode of motion using readout from the two ports of the Michelson, with different measurement strengths, as discussed by~\cite{Mueller-Ebhardt:2008b}. A possible experimental setup is shown on the left panel of Fig.~\ref{fig:EN}. Two identical movable end mirrors of a Michelson interferometer have their common and  different modes measured with different  strengths --- due to the existence of the power recycling mirror.  One obstacle towards implementation of this scheme is that classical laser noise directly affects our sensitivity for the common mechanical  mode. 

In case laser noise must be suppressed by interferometry, we could consider a Michelson interferometer with arm cavities, each with movable input and end mirrors.  If we inject two carrier fields to the bright port of the Michelson interferometer, one enters the arm cavities and senses the cavity length, while the other does not enter the cavity and only senses the input mirrors' locations (e.g., like the configuration suggested in Ref.~\cite{Rehbein:2007}, but for other purposes), we can create conditional entanglement if parameters of the two fields are adjusted appropriately.

Using notations of Sec.~\ref{subsec:conditional}, if we assume  $\Omega_x/\Omega_F = 25\sim 50$ (which correspond to a SQL-beating factor of $\sim$3.5 -- 5), and 10\,dB squeezing, appropriate choice of $\Omega_q$ for common and differential modes of two mechanical objects would lead to ``verifiable entanglement'' that survives for a time scale comparable to $1/\Omega_q$.  
An example is shown in Fig.~\ref{fig:EN}, where frequency scales are chosen to be suitable for GW detectors. Here the  logarithmic negativity $\mathcal{E}_{\mathcal{N}}$ is calculated taking into account thermal decoherence during the evolution stage, as well as verification error.   In Sec.~\ref{subsubsec:gravdecoh}, we shall discuss an application of measuring the survival time of entanglement.

More generally, Ludwig et al.~\cite{Ludwig:2010} analyzed the (unconditional) entanglement between two coupled oscillators at the same instant, for a general common bath to which they are both coupled to.

\subsection{Quantum Teleportation Mechanical States}
\label{subsec:teleport}

\begin{figure}
\centerline{\includegraphics[width=2.5in]{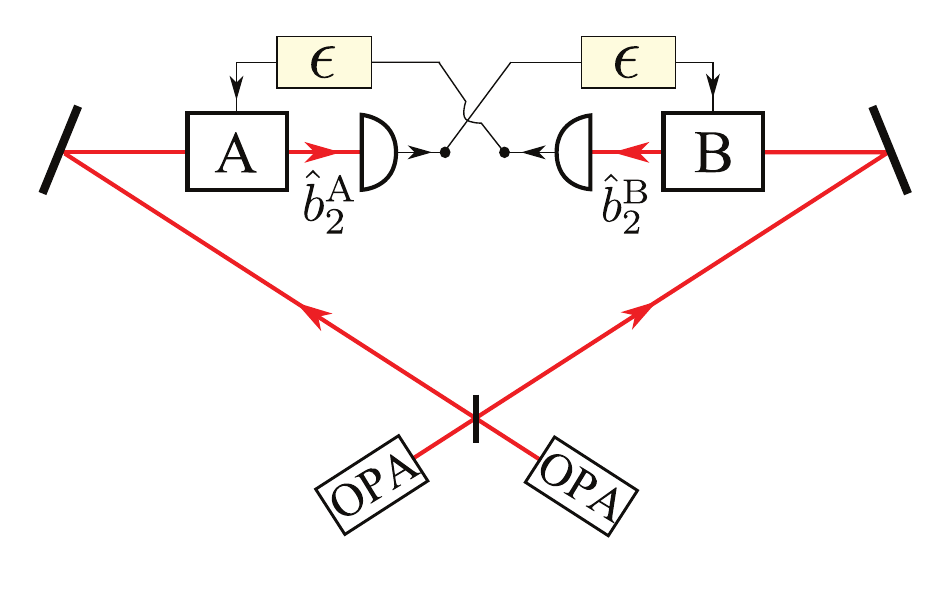}}
\caption{Sketch of an experimental scheme that teleports the  quantum state of a mechanical oscillator, as discussed in Sec.~\ref{subsec:teleport}. Two mechanical oscillators are driven by entangled beams; out-going light from each oscillator is fed back as a force to act on the other. 
\label{fig:teleport}
}
\end{figure}

Analogous to continuous-variable teleportation of optical states~\cite{Loock:2000}, one can teleport the quantum state of one mechanical oscillator to the other, if two entangled squeezed beams are used to drive them, each of their positions are measured --- and with results fed back to the other one (as shown in Fig.~\ref{fig:teleport}). 

If we label the two oscillators and their optical fields by $A$ and $B$, assuming that we measure the output phase quadrature in each case (i.e., $b_2^A$ and $b_2^B$) and feed it back as a force to the other object with a gain of $\epsilon$, the equations of motion will  be (setting $\hbar=M=1$) 
\begin{eqnarray}
b_2^A = a_2^A + \Omega_q x_A, \;\;
b_2^B = a_2^B + \Omega_q x_B.
\end{eqnarray}
and
\begin{eqnarray}
\dot x_A = p_A\,, \quad \dot p_A =-\omega_{\rm opt}^2 x_A +\Omega_q a_1^A -\epsilon b_2^B \\
\dot x_B = p_B\,, \quad \dot p_B =-\omega_{\rm opt}^2 x_B +\Omega_q a_1^B -\epsilon b_2^A
\end{eqnarray}
Here $\omega_{\rm opt}$ is the optomechanical resonant frequency of the mechanical oscillators, which could either be due to elastic or optical rigidity (the subscript we used suggests the use of an optical spring).  This means we shall be adopting the straw-man mode of Sec.~\ref{subsec:elimination}. 

Combining the above, we obtain the following equations of motion for the oscillators $(x_A,p_A,x_B,p_B)$: 
\begin{eqnarray}
\label{eqteleport1}
\dot x_A = p_A\,, \quad \dot p_A =-\omega_{\rm opt}^2 x_A -\Omega_q \epsilon x_B + n_A \\
\label{eqteleport2}
\dot x_B = p_B\,, \quad \dot p_B =-\omega_{\rm opt}^2 x_B -\Omega_q \epsilon x_A + n_B
\end{eqnarray}
In absence of classical noise, $n_A$ and $n_B$ are given by
\begin{equation}
n_A =  \Omega_q a_1^A -\epsilon a_2^B \,,\quad 
n_B =  \Omega_q a_1^B -\epsilon a_2^A \,.
\end{equation}
Equations~\eref{eqteleport1} and \eref{eqteleport2} describe two coupled oscillators driven with additional quantum noise --- but the coupling is only due to classical communication and feedback.  If we can make the noise level arbitrarily small, then the two oscillators will swap their quantum states back and forth --- just like two coupled pendulums. 
The ``sloshing frequency'' is determined by the difference in the two new eigenfrequencies of the coupled system:
\begin{equation}
\Omega_{\rm slosh} =\Omega_+ -\Omega_-, \;\;
\Omega_{\pm} = \sqrt{\omega_{\rm opt}^2 \pm \epsilon\Omega_q}
\end{equation}
The first exchange of quantum takes place at
\begin{equation}
\tau_{\rm ex} =\frac{\pi}{\Omega_{\rm slosh}}
\end{equation}

If we inject vacuum states to each oscillator, $n_A$ and $n_B$ will both be at the vacuum level --- and  we will add at least a noise ellipse limited by Heisenberg Uncertainty after one sloshing period. Fortunately, we note that $n_A$ and $n_B$ do commute, therefore we can make them simultaneously small --- of course, that only happens when we drive the two devices with highly entangled beams, each in turn is generated by interfering two highly squeezed beams --- as shown in Fig.~\ref{fig:teleport}.  

In presence of classical noise, the teleportation process has the following additional noise at $\tau_{\rm ex}$,
\begin{equation}
\mathbf{V}_{\rm add} =\frac{\pi}{8}\frac{\zeta_F \Omega_q^2 + \zeta_x \epsilon^2}{\Omega_{\rm slosh}}\left[
\begin{array}{cc}
\Omega_+^{-2}+\Omega_-^{-2} & 0  \\
0 & 2
\end{array}
 \right]
\end{equation}
where
\begin{equation}
\zeta_x =\sqrt{e^{-2q} +2 \xi_x^2},\quad
\zeta_F =\sqrt{e^{-2q} +2 \xi_F^2}
\end{equation}
and $\xi_x$ and $\xi_F$ are same as defined in Sec.~\ref{subsec:conditional} [Cf.~Eq.~\eref{eq:xiFxix}].  This means the teleportation will be quantum if $\xi_x$ and $\xi_F$ --- as well as $e^{-2q}$ are sufficiently small.

In presence of classical noise, we first of all must make $\xi_x$ and $\xi_F$ small, which requires $\Omega_q$ to be within the region $\Omega_F <\Omega_q <\Omega_x$.  For weak feedback or small $\epsilon$, $\Omega_{\rm slosh}$ will be small, which means it takes long for the two oscillators to exchange state, allowing more thermal decoherence.  However, if feedback is too strong, the system becomes unstable.  Given fixed values of $(\Omega_F,\Omega_x)$ and $e^{-2q}$, an asymptotic optimal value for  $\mathrm{det}(\mathbf{V}_{\rm add})$ is 
\begin{equation}
\mathrm{det}(\mathbf{V}_{\rm add}) =\frac{\hbar^2}{4}\pi^2\left(e^{-2q}+\frac{2\Omega_F}{\Omega_x}\right)
\end{equation}
which is achieved when $\Omega_q =\sqrt{\Omega_x\Omega_F}$ and $\omega_{\rm opt} \rightarrow +\infty$.  In order for the additional noise to be sub-Heisenberg, $\mathbf{V}_{\rm add}$, which is necessary for preserving any negative regions of the original state's Wigner function, we will need a substantial squeezing factor (recall that $e^{-2q}$ is the squeeze factor in power) as well as classical noise budget substantially below the free-mass SQL [recall that $\Omega_x/(2\Omega_F)$ is the factor that our total classical noise beat the SQL in power].

\subsection{Non-Gaussianity and Single Photons in Linear Systems}

\label{subsec:nongaussian}

Up till now, we have restrained ourselves to Gaussian states, which have well-behaved and non-negative Wigner functions.  If we also perform only measurements on observables that are linear in position and momentum of the mechanical oscillators, and the optical field operators, all our experimental results will be explainable in terms of classical random processes.  Even though we may still use such experiments to perform certain test quantum mechanics, as discussed in Sec.~\ref{sec:test}, we will be limited in the extent to which we are demonstrating quantum mechanics on macroscopic objects, and in the scope of tests we will be able to perform.  In this section, we shall discuss how to create non-Gaussian mechanical states, taking advantage of the fact that a single photon, being the Fock state $|1\rangle$,  is a highly non-Gaussian quantum state. 

\subsubsection{Strong versus weak coupling.}

At this stage, it is difficult to emit a significant number of photons with strong mutual coherence with each other.  On the other hand, the total effect of a large number of uncorrelated photons on a mechanical oscillator will usually be Gaussian. This seems we should send order $\sim 1$ photon to interact with the mechanical oscillator --- and the condition for this photon to  significantly influence the mechanical oscillator's quantum state is 
\begin{equation}
\mathcal{F}\frac{\hbar\omega_0}{c} \sim \sqrt{\hbar M\omega_m},
\end{equation}
namely, the cumulated momentum transfer on the left-hand side ($\mathcal{F}$ is finesse of the cavity) must be comparable to the momentum quantum uncertainty of the mechanical object on the right-hand side.  This is also equivalent to 
\begin{equation}
\frac{\lambda}{\mathcal{F}} \sim \sqrt{\frac{\hbar}{M\omega_m}}
\end{equation}
which means the linear dynamical range of the cavity has to be less than the position uncertainty of the mechanical object.    We shall postpone that possibility to Sec.~\ref{sec:nonlinear}.

\begin{figure*}
\centerline{\includegraphics[width=6.5in]{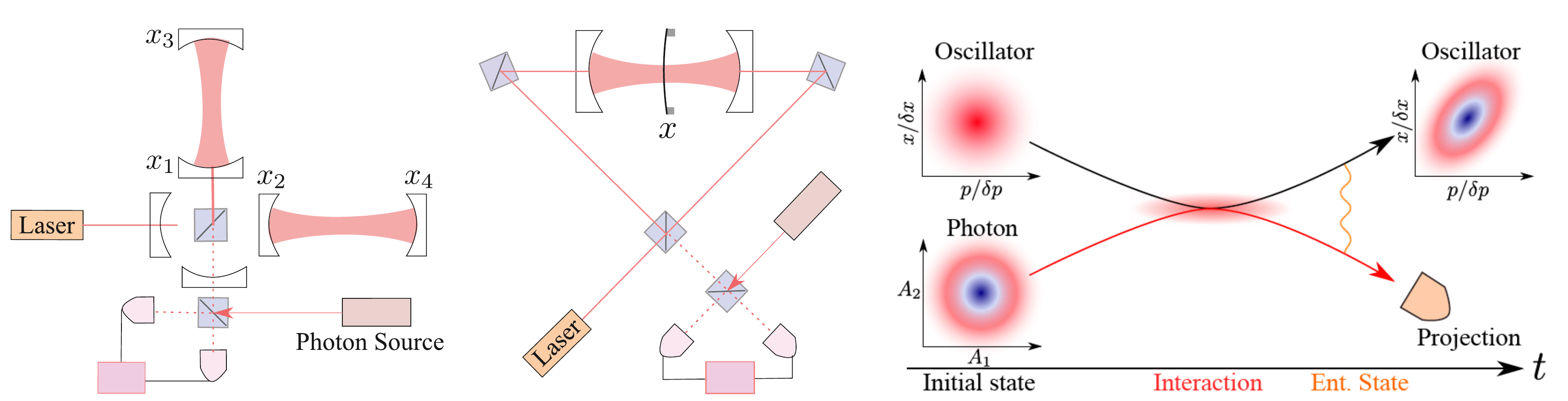}}
\caption{Illustration of non-Gaussian state-preparation using a linear system driven by a single photon, as discussed in Sec.~\ref{subsec:nongaussian}.   The left two panels show possible experimental setups, while the right panel illustrates the process of conditional preparation of non-Gaussian mechanical states.\label{fig:nongaussian}
}
\end{figure*}

\subsubsection{Coherent amplification due to pumping.}

In this section, we shall consider an amplification effect due to the ``beating'' between a single photon and a strong beam of light.  This idea was first discussed for optomechanical systems by Mancini et al.~\cite{Mancini:2003}, who noticed that force on a mirror is the square of the electric field, $E^2$, then if $E$ contains a large classical component $E_c$ and a quantum component  $\hat E_q$, then the beating of them, $E_c \hat E_q$ appears in the radiation-pressure force acting on the mirror, thereby may drive the mirror into a highly non-Gaussian state.   Mancini et al.\ restricted themselves to a scenario in which simplification using the rotation-wave approximation is possible: the mirror is a high-$Q$ oscillator, and coupled only to a single sideband of an optical mode --- which allowed them to consider a finite-mode Hamiltonian in which the mirror is linearly coupled to the optical mode, with a coupling coefficient that is highly amplified by the classical pumping.   In this idealized case, it is often possible to simply consider direct transfers between optical and mechanical states. 

Very similar to Mancini's proposal, but in a very different regime: coupling between high-frequency phonons of diamond and single photons, assisted by pulsed pumping, has recently been implemented experimentally, demonstrating entanglement of phonon modes of two pieces of diamond~\cite{Lee:2011}, as well as implementation of quantum memory using one piece of diamond~\cite{Lee:2011b}.  This was in turn motivated by proposals involving atomic ensembles~\cite{Duan:2001}.

More recently, Vanner et al.\ building upon the state-transfer mechanism, and proposed protocols for non-Gaussian quantum-state engineering of mechanical oscillators~\cite{Vanner:2013}.

\subsubsection{Treatment for a broadband (non-resolved-sideband) device.}

A more rigorous treatment of Mancini et al.'s strategy in more general  optomechanical systems was later carried out by Khalili et al.~\cite{khalili:2010}.   This does not require a simplified situation of state transfer between several known modes of oscillation, but instead considers the interaction between a mechanical mode and an infinite continuum of optical field.   Khalili et al.\  considered a Michelson interferometer, in which the force acting on the differential mode of the two mirrors is proportional to $\hat E_c \hat E_d$, where $\hat E_c$ is common-mode amplitude and $\hat E_d$ is differential-mode amplitude.  Suppose the common optical  mode is highly pumped, while differential mode only has vacuum fluctuations or single photon, the force on the differential mode of motion of the mirrors is approximately $\langle E_c\rangle \hat E_d$ --- which would be highly non-Gaussian if a single photon is to be injected from the dark port.   An illustration of the experimental strategy is shown in Fig.~\ref{fig:nongaussian}.

Unlike Mancini et al.~\cite{Mancini:2003}, the treatment of Khalili et al.~\cite{khalili:2010} did not use the rotating wave approximation, nor did they idealize the optical field into one single degree of freedom.  
%
%
%
%
Since the technical treatment itself is quite interesting, we will describer it here briefly. 
%
Now, suppose we have a vacuum for the optical field for a long time, until the system reaches a steady state, and then inject a single photon with known wavefunction; this corresponds to the following initial state for the optical field
\begin{eqnarray}
|\rm ini\rangle &=& \Gamma^\dagger |0\rangle =  \int\frac{d\Omega}{2\pi}\tilde\Phi(\Omega) a_{\omega_0 +\Omega}^\dagger |0\rangle  \nonumber\\
&=&\int_{-\infty}^0 \Phi(t) [a_1(t)+i a_2(t)]|0\rangle 
\end{eqnarray}
Here $\Phi(t)$ is the inverse-Fourier transform of $\tilde\Phi(\Omega)$,
\begin{equation}
\Phi(t)=  \int
\frac{d\Omega}{2\pi} \tilde\Phi(\Omega)e^{ - i \Omega t}\,,
\end{equation}
it is the profile of the spatial wavefunction of the photon, and we  impose the normalization condition of
\begin{equation}
\label{eqgammanorm}
\left[\Gamma,\Gamma^\dagger\right] = \int_{-\infty}^0  dt \left|\Phi^2(t)\right|  =\int_{-\infty}^{+\infty}\frac{d\Omega}{2\pi}  \left|\tilde\Phi^2(\Omega)\right|  = 1\,.
\end{equation}

For a preparation that targets a state at $t=0$, $\Phi(t)$ should only be non-zero for $t<0$.  With this initial state, we have, at $t=0$, 
\begin{eqnarray}
\label{eqJc}
&&J_c(\mu,\nu)\nonumber\\
&=&
\int D[k(t')] \nonumber\\
&&\;\langle 0| \hat\Gamma e^{
i\mu\hat x(0) +i\nu \frac{\hat p(0)}{M\omega_m}+
i \int_{-\infty}^0 k(t') \left[\hat b_\zeta(t') - \xi(t')\right]} \hat\Gamma^\dagger |0\rangle. \quad\qquad
\end{eqnarray}
If we denote
\begin{equation}
\hat O \equiv \mu\hat x(0) +\nu\frac{\hat p(0)}{M\omega_m}+
 \int_{-\infty}^0 k(t') \left[\hat b_\zeta(t') - \xi(t')\right],
\end{equation}
Then
\begin{equation}
J_c(\mu,\nu)  =\left(1 - \left|\left[\hat O,\hat \Gamma^\dagger\right]\right|^2\right) \langle 0|e^{i\hat O} |0\rangle\,.
\end{equation}
Note that: (i) the factor  $\langle 0|e^{i\hat O} |0\rangle$ is the generating function of the conditional state when there is only vacuum input, (ii) the commutator  $\left[\hat O,\hat \Gamma^\dagger\right]$ is a complex number, linear in $\mu$ and $\nu$, and causes the conditional state to be non-Gaussian because it contains $\mu$ and $\nu$.

Due to the normalization condition~\eref{eqgammanorm}, non-Gaussianity is only significant if the photon has a duration that matches the timescale of state preparation, $1/\Omega_q$.  Note that  the commutator $\left[\hat O,\hat \Gamma^\dagger\right]$ only yields $\mu$ and $\nu$ dependence (which is required for non-Gaussianity) when $\hat x(0)$ and $\hat p(0)$ have significant non-zero commutator with $\hat\Gamma^\dagger$, which only happens when the optical mode associated with $\hat\Gamma$ ``makes up a large portion of'' the Heisenberg operators of $\hat x(0)$ and $\hat p(0)$ --- and therefore this optical mode should have a duration comparable to $1/\Omega_q$.




\section{Examples of Non-Linear Optomechanical Systems}

More dramatic effects of quantum mechanics appear for nonlinear systems, or when nonlinear observables (e.g., those that are not linear combinations of position, momentum and optical-field operators) are measured.   The most famous example being the Quantum Zeno effect~\cite{Misra:1977}: as an observable with discrete spectrum is being measured, state reduction tends to keep the system from jumping between eigenstates of the operator being measured.  

However, for optomechanical systems, it is rather difficult to enter a regime in which motion of the mechanical object nonlinear: while achieving nonlinearity requires the mechanical object to move as much as possible, a quantum state with a larger spread in position tends to get destroyed by decoherence much faster. Nevertheless, nonlinearity has been observed in the effective motion of atomic ensembles~\cite{Gupta:2007,Purdy:2010}; they may become possible in  future photonic/phononic crystals~\cite{Painter:private}

Finally, while linear systems are all like, every nonlinear system is nonlinear in its own way. The rich variety of nonlinear systems means that it will be very difficult to provide a comprehensive discussions of nonlinear optomechanics.  In this paper, we will only review two aspects: Quantum Zeno effect (Sec.~\ref{subsec:zeno}) and few-photon-driven optical cavity with movable mirror (Sec.~\ref{subsec:fewphoton}). As it will turn out, both tend to require the strong-coupling condition
\begin{equation}
\label{eq:strong}
\frac{\lambda}{\mathcal{F}} \sim \sqrt{\frac{\hbar}{M\omega_m}}
\end{equation}

\label{sec:nonlinear}

\subsection{Phonon Counting and Signatures of Energy Quantization}
\label{subsec:zeno}

If we measure the occupation number of a harmonic oscillator continuously, quantum state reduction will eventually localize the oscillator into a Fock state with a definite energy and occupation number.  In presence of thermal noise, quantum jumps across different levels will be suppressed by the effect of the measurement.  This is called the quantum Zeno Effect~\cite{Misra:1977}.  
Santamore et al.~\cite{Santamore:2004} and Martin and Zurek~\cite{Martin:2007} proposed measuring the Zeno effect in mechanical oscillators, achieved by first (effective) coupling the phonon excitation number of the oscillator to the optical field (mode of a cavity), via
\begin{equation}
V_I \propto \hat n_{m} \hat n_{\rm cav}
\end{equation}
and then readout the optical field.  This process and its variants have been theoretically studied by Gangat et al.~\cite{Gangat:2011} and Ludwig et al.~\cite{Ludwig:2012}.

A more concrete experimental strategy was proposed by Thompson et al.~\cite{Thompson:2008} and Jayich et al.~\cite{Jayich:2008}, who considered  a membrane inside a cavity~\cite{Jayich:2008,Thompson:2008}, where an appropriate choice of the location of the membrane and the pumping frequency will allow the 
\begin{equation} 
\frac{d\omega_c}{dx}=0\,,\quad\frac{d^2\omega_c}{dx^2}\neq 0
\end{equation}
therefore the parametric coupling between the cavity and the mirror becomes approximately
\begin{equation}
\label{eq:quadraticH}
V_I = \hbar\left(\frac{d^2\omega_c}{dx^2}\frac{x^2}{2}\right) \hat A^\dagger \hat A
\end{equation}
In certain regimes, such quadratic coupling can be used to manipulate trapping potentials for an optically levitated test mass~\cite{Romero-Isart:2011c,Chang:2012}, or give rise to two-photon cooling and squeezing of the mechanical oscillator~\cite{Nunnenkamp:2010}.  Quadratic coupling between position and a qubit has also been shown to be able to prepare mechanical oscillators into highly non-Gaussian states and their superpositions~\cite{Jacobs:2009}. 

Perhaps the most intriguing  regime is when we make sure that low-frequency components of $(\hat A, \hat A^\dagger)$ have much higher coupling to the membrane than high-frequency components (e.g., due to cavity bandwidth), we will then be making a ``slow measurement'' on $\hat x^2$, which averages to be proportion to the phonon number of the membrane.  As it turns out, in the membrane-in-cavity strategy proposed by Thompson et al.~\cite{Thompson:2008}, one has to worry about the residual linear coupling~\cite{Miao:2009}, which only allows detection of discrete energy quantization if
\begin{equation}
\frac{\lambda}{\mathcal{F}} < \Delta x_q
\end{equation}
where $\lambda$ is the wavelength of light, $\mathcal{F}$ the cavity finesse (which is limited by optical loss), and $\Delta x_q$ the quantum uncertainty in the position of the membrane.  This is the same as the strong-coupling condition discussed before at the beginning of Sec.~\ref{subsec:nongaussian}.

As a much easier strategy to observe the signature of energy quantization, Clerk et al.\ proposed observing a pumped oscillator with a non-zero (rather high) occupation number~\cite{Clerk:2010,Clerk:2011} --- through detecting non-Gaussianity of the out-going current imparted by the non-Gaussianity in the position of the mirror.  This strategy allows the detection of certain signatures of energy quantization as long as the oscillator is cool enough:
\begin{equation}
k_B T < \hbar\omega_m
\end{equation}

\subsection{Few-photon-driven strongly coupled cavity}
\label{subsec:fewphoton}

In this section, we shall consider the injection of a few photons into a cavity with a movable mirror --- in the parameter regime of ``strong coupling'', namely that the momentum transfer from the photon to the mirror is comparable to the total quantum uncertainty of the mirror's momentum.  The key to the simplification we shall receive in this section is that the interaction Hamiltonian
\begin{equation}
V_I \propto \hat x\hat A^\dagger\hat A = \hat x \hat n_{\rm cav} 
\end{equation}
does not later the linearity of the mirror's Hamiltonian, but instead simply shifts the equilibrium position of the harmonic oscillator ---  if we keep a fixed number of photons in the cavity.   This has allowed Bose, Jacobs and Knight~\cite{Bose:1997} to  calculate quantum state evolution of the mirror under radiation pressure when the cavity is closed and prepared into a non-classical initial optical state.  They showed that the mirror can be prepared into highly non-classical quantum states this way. 

\subsubsection{Cavity driven by weak steady beam.} Motivated by progress in optomechanics, Rabl~\cite{Rabl:2011}, Nunnenkamp et al.~\cite{Nunnenkamp:2011} discussed the output photon statistics of such a strongly coupled cavity when weak light is injected.  They found the so-called ``photon blockade'' effect, namely the fact that once a photon is already in the cavity, a second one would not be able to enter very easily because the mirror is oscillating around a new equilibrium point. For pumping at a low power, Qian et al.\ further found that the mirror can be prepared into a highly non-Gaussian steady state~\cite{Qian:2012}. More recently, Nunnenkamp et al.\ considered the cooling of the mirror in such a few-photon regime~\cite{Nunnenkamp:2012}.

\begin{figure}
\centerline{\includegraphics[width=3in]{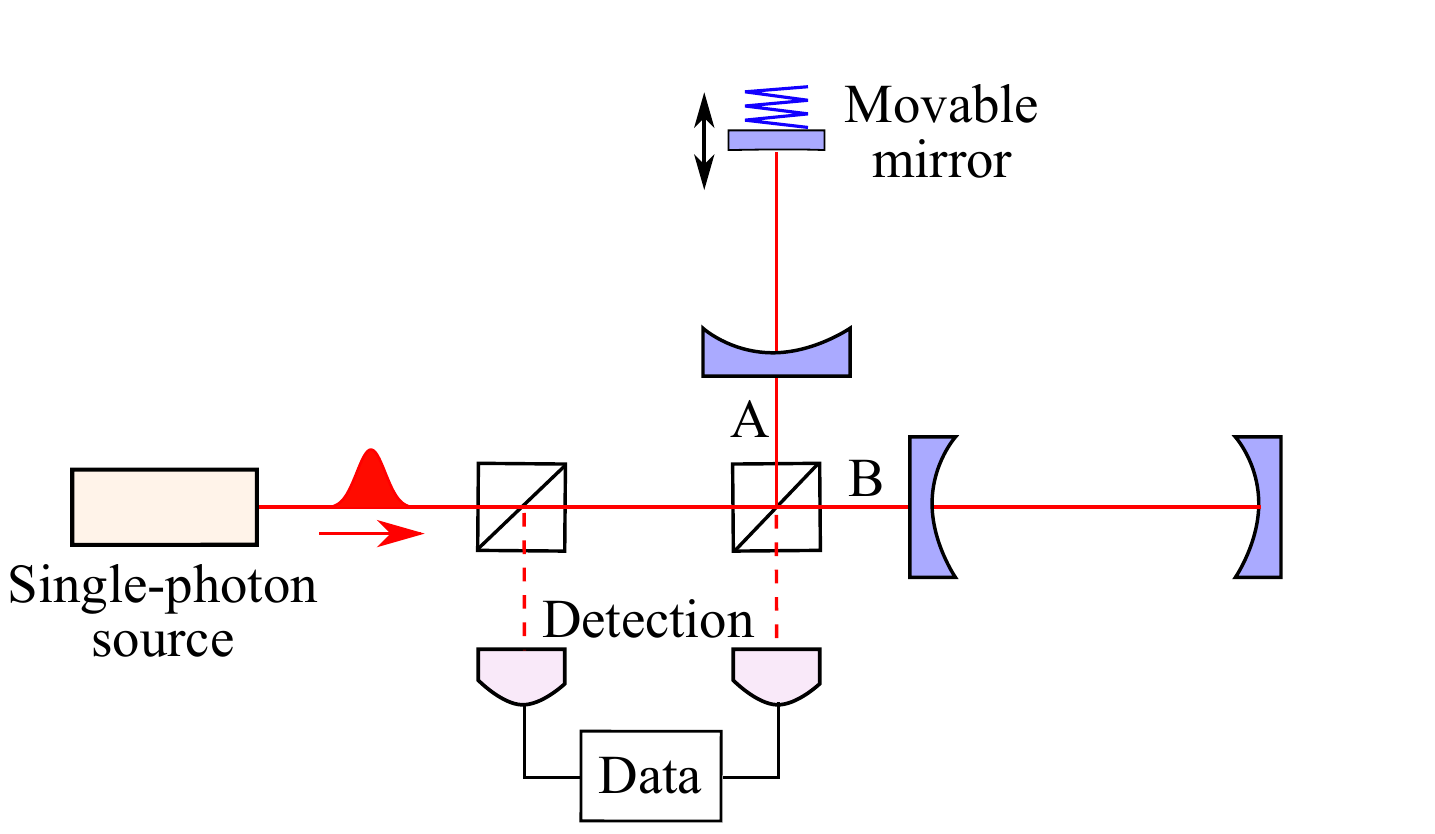}}
\caption{Single photon injected into a Michelson interferometer with one arm cavity having a movable mirror.  The cavities are originally empty, and we assume the ``strong-coupling regime'' defined by Eq.~\eref{eq:strong}. \label{fig:single}}
\end{figure}

\subsubsection{Cavity driven by a few photons with known arrival times.}
Hong et al.~\cite{hong:2011} and Liao et al.~\cite{Liao:2012} focused on the case of a single photon, and obtained exact solution of the interaction between a single injected photon and a cavity with a movable mirror, in the strong-coupling regime.  Moreover, Hong et al.\ assumed the out-going photon to be detected by a photodetector, and considered the state of the mirror conditioned on the arrival time of the photon.   Hong et al.'s work was motivated by Marshall et al.~\cite{Marshall:2003}'s proposal of using this strongly coupled experiment to probe gravity decoherence: in the original proposal, as well subsequent analyses of this experimental setup~\cite{Bassi:2005}, the shape of the injected photon has always been ignored --- yet as Hong et al.\ has shown, that shape does affect the interpretation of the experimental result.  Hong et al~\cite{hong:2011} also showed that, in fact, by adjusting the wavefunction of the incoming photon, the quantum state of the mirror can be prepared into a wide range of possible quantum states.  Liao et al.\ later extended their calculation to incorporate two correlated incoming photons~\cite{Liao:2012b}.

\subsubsection{Cavity driven by short pulses.} Finally, another interesting regime that triggers nonlinearities would be a cavity driven by a number of (non-entangled) photons that arrive together as a pulse with duration much shorter than the oscillation period of the oscillator. In the non-linear regime of such pulsed optomechanics (Cf.~\cite{Vanner:2011,Machnes:2012,Hofer:2011}), as Vanner has shown, non-Gaussian states can be prepared for the mechanical oscillator~\cite{Vanner:2011b} due to optical nonlinearity in the Hamiltonian~\eref{eq:h12}. Vanner has also shown that non-Gaussianity is more easily achievable here than when the Hamiltonian~\eref{eq:quadraticH}, which is quadratic in $x$. 



\section{Tests of Quantum Mechanics and Fundamental Physics}

\label{sec:test}

In this section, we will venture beyond the standard quantum mechanics.  First, in Sec.~\ref{sec:QMG}, we will discuss how optomechanics may provide new opportunities for testing gravity's modification to quantum mechanics.   Then in Sec.~\ref{subsec:testQM}, we shall explore  the possibility of a framework that allows us to systematically test for modifications to quantum mechanics for macroscopic mechanical objects --- regardless of motivations of the modifications. 

\subsection{Quantum Mechanics and Gravity}
\label{sec:QMG}

\subsubsection{Gravity Decoherence}
\label{subsubsec:gravdecoh}

Penrose and others have speculated that gravity may destroy macroscopic quantum superpositions~\cite{Penrose:1998,Penrose:2006,Diosi:2007}.  In particular, if a macroscopic object is in a superposition state $|\psi\rangle  = |\psi_1\rangle +|\psi_2\rangle$, with $|\psi_1\rangle$ and $|\psi_2\rangle$ corresponding to very different mass distributions, then that difference, $\delta\rho$, would gradually make the superposition a classical one, at a time scale 
\begin{equation}
\tau \sim \hbar/E_G
\end{equation}
and
\begin{equation}
\label{eq:EG}
E_G = G\int d^3 \mathbf{x} d^3 \mathbf{x}' \frac{\delta\rho(\mathbf{x})\delta \rho(\mathbf{x}')}{\left|\mathbf{x}-\mathbf{x}'\right|}
\end{equation}
is the self gravitational field of the difference in mass distributions.  Experiments have been proposed to look for such decoherence~\cite{Marshall:2003,vanWezel:2008,vanWezel:2012}. 

An order-of-magnitude estimate for the time scale of gravity decoherence shows that it is rather difficult to detect.  In fact, for a mechanical object assumed to have a uniform density $\rho_0$, if we prepare a superposition at a length scale close to the vacuum state of an oscillator with frequency $\Omega_q$, and write the decoherence time scale as $\tau$, then
\begin{equation}
\label{gravitytime}
\Omega_q\tau = \left(\frac{\Omega_q}{\sqrt{G\rho_0}}\right)^2
\end{equation}
This is the number of mechanical oscillation cycles it takes for gravity decoherence to completely destroy the quantum superposition.
For a typical material, Si, with $\rho_0 =2.3\times 10^3\,\mathrm{kg}/\mathrm{m}^3$, we have
\begin{equation}
\sqrt{G\rho_0} \sim 4\times 10^{-4}\,\mathrm{s}^{-1}\,,
\end{equation}
this 
requires the measurement time scale to be 
\begin{equation}
 1/\Omega_q \approx 2.5\times 10^3\,\mathrm{s}\,,
 \end{equation}
  which is rather difficult to achieve on the ground.  If we focus on Eq.~\eref{gravitytime}, the size or mass of the object does not directly enter.  However, because experiments with heavier test masses normally operate at low frequencies, and therefore makes Eq.~\eref{gravitytime} more achievable.  We note an alternative speculation of gravity decoherence is more approachable, as discussed by Miao et al.~\cite{Miao:2010b}, where instead of using Eq.~\eref{eq:EG}, the scale of gravity decoherence is given by the {\it difference} of gravitational self energy between two components of the quantum state which are superimposed. 

However, if we take literally Eq.~\eref{eq:EG}, we will have to consider the high concentration of matter near lattice points, if we are using a crystal that is cooled much below its Debye temperature.  In this way, we will have to use a much higher density 
\begin{equation}
\rho = \Lambda \rho_0
\end{equation}
with
\begin{equation}
\Lambda = {m}/({12\sqrt{\pi}\rho_0\Delta x_{\rm zp}^3 })\,.
\end{equation}
Here $m$ is the mass of the individual atom in the lattice, and $\Delta x_{\rm zp}$ is the zero point position uncertainty of that atom along each direction.  This distance can be measured rather accurately using X-ray diffraction~\cite{Housley:1966}.  It has been estimated that $\Lambda \approx 8.3 \times 10^3$ for Si crystal~\cite{Yang:2012}.  We should therefore require 
\begin{equation}
\Omega_q\tau = \frac{1}{\Lambda}\left(\frac{\Omega_q}{\sqrt{G\rho_0}}\right)^2
\end{equation} 
which leads to 
\begin{equation}
1/\Omega_q \approx 28\,\mathrm{s}
\end{equation}
which might be achievable on the ground with the help of torsional bars at the quantum regime~\cite{Ando:2010}.   However, due to the lack  of a mathematical formulation for gravity docoherence, it is not easy to work out the exact density to use from first principle.

In Ref.~\cite{Romero-Isart:2011}, Romero-Isart discussed more types of decoherence models that are motivated by the interplay between gravity and quantum mechanics, and outlined an ambitious research program not using nearly Gaussian states, but states that have wavefunction spread comparable to or bigger than the physical size of the mechanical objects.  These experiments are intended to be performed in space, where decoherence from a suspension system will no longer be an issue~\cite{Kaltenbaek:2012}.

\subsubsection{Semiclassical gravity}

\label{subsec:semiclassical}
Theoretical physicists have long argued that gravity should be quantum, just like the electroweak and strong interactions~\cite{Feynman:1995}. Major theoretical effort are being made to elaborate how gravity can be reconciled with quantum mechanics.  From an experimental point of view, however, it might still be good to consider ruling out gravity being classical~\cite{Carlip:2008}. 

The most straightforward way to construct a classical theory of spacetime that supports quantum matter is to write the Einstein's equation as:
\begin{equation}
\label{semiclass}
G_{\mu\nu} = \langle 8\pi T_{\mu\nu}\rangle\,,
\end{equation}
where the expectation value is evaluated at the quantum state of the entire universe.  This was proposed by M{\o}ller~\cite{moller:1962} and Rosenfeld~\cite{Rosenfeld:1963}, and often referred to as {\it semiclassical gravity}.  The most fatal flaw of semiclassical gravity is that it does not allow state reduction --- because that makes the right-hand side not divergence free, violating the Bianchi Identity on the left-hand side (see Chapter 14 of Ref.~\cite{Wald:1984}).  On the other hand, the most obvious state-reduction-free interpretation of quantum mechanics, the ``Many-World Interpretation''~\cite{Everett:1957} when combined with Eq.~\eref{semiclass}, gives rather absurd predictions~\cite{Page:1981}.  For this reason, in order for Eq.~\eref{semiclass} to work literally, there has to exist an interpretation of quantum mechanics that does not require state reduction yet still explains the phenomenology of quantum measurements.

\begin{figure}
\centerline{\includegraphics[width=3.in]{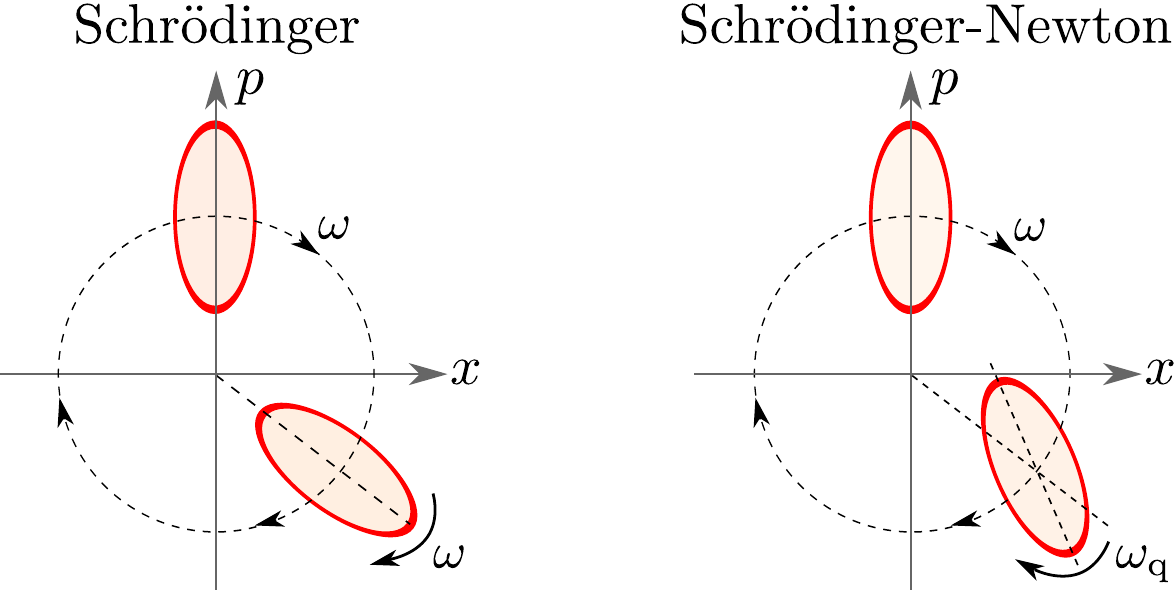}}
\caption{Evolution of an object's Wigner function under standard Schr\"odinger equation and the Schr\"odinger-Newton equation.  See Sec.~\ref{subsec:semiclassical} for details.}
\end{figure}

Given Eq.~\eref{semiclass}, one can derive a Schr\"odinger-Newton (SN) equation for an $n$-particle system in the non-relativistic regime, which reads:
\begin{eqnarray}
i\hbar\frac{\partial \varphi(t,\mathbf{X})}{\partial t} &=&\sum_k\left[
 -\frac{\hbar^2\nabla_k^2}{2m_k} + \frac{m_k U(t,\mathbf{x}_k)}{2}\right]\varphi(t,\mathbf{X})\nonumber\\ &+& V(\mathbf{X})\varphi(t,\mathbf{X})\,,
\end{eqnarray}
here  $\phi(t,\mathbf{X})$ is the joint wavefuntion of the particles, $\mathbf{X}=(\mathbf{x}_1, \ldots, \mathbf{x}_n)$,  $U$ is the Schr\"odinger-Newton potential, given by
\begin{equation}
\nabla^2 U(t,\mathbf{x}) =4\pi   \int d^{3n} \mathbf{X} \sum_k m_k \delta(\mathbf{x}-\mathbf{x}_k) |\varphi(t,\mathbf{X})|^2
\end{equation}
Because $U$ depends on $\varphi$, the SN equation is a non-linear equation --- although one can show that this equation still preserves total probability.

The SN equation has often been used in the regime in which the additional $U$ term is significant --- often when attempts were made to connect this equation to gravity decoherence~\cite{Moroz:1999} or to replace state reduction~\cite{Salzman:2006}.  However, if we restrict ourselves in the perturbative regime, the SN equation has very simple predictions.  First of all, for an $n$-particle system that is a solid material, one can derive a Center-of-Mass (CM) SN equation --- during the derivation, one finds that concentration of mass around equilibrium positions of atoms does increase the effect of the SN term.  The  CM SN equation, for a mechanical oscillator with mass $M$ (made up from a piece of crystal cooled to a temperature much below its Debye temperature) and resonant frequency $\omega_c$,  can be written as~\cite{Yang:2012}:
\begin{equation}
\label{SNCM}
i\hbar\frac{\partial\Psi}{\partial t} = \left[\frac{\hbar^2\nabla^2}{2M}+\frac{M\omega_c^2 x^2}{2} +\frac{1}{2}\mathcal{C} \left(x-\langle \hat x\rangle\right)^2 \right] \Psi\,.
\end{equation}
Here the constant $\mathcal{C}$ arises from the Schr\"odinger-Newton term, and give by 
\begin{equation}
\label{eqCSN}
\mathcal{C}=- \frac{1}{2}\frac{\partial^2}{\partial z^2}\left[ \int \frac{G\tilde \rho_{\rm int}(\mathbf{y})\tilde\rho_{\rm int}(\mathbf{y}')}{|\mathbf{z} +\mathbf{y} -\mathbf{y}'|}d\mathbf{y} d\mathbf{y}' \right]_{\mathbf{z}=0}\,,
\end{equation}
where $\tilde \rho_{\rm int}(\mathbf{y})$ is the mass distribution in the center-of-mass frame (the subscript ``int'' indicates internal motion). Further calculation assuming Gaussian distribution of mass near lattice sites gives
\begin{equation}
 \mathcal{C} = \frac{GmM}{12\sqrt{\pi}\Delta x_{\rm zp}^3}\,,
\end{equation}
where $M$ is the object's total mass, $m$ is the mass of each individual atom, and $\Delta x_{\rm zp}$ is the zero point position uncertainty of each atom along each direction.


It is straightfoward to show that Eq.~\eref{SNCM} describes a separate motion for the expectation values $\langle x\rangle$, $\langle p\rangle$ and their covariance matrix, $(V_{xx}, V_{xp},V_{pp})$, which describes quantum uncertainty. In the phase space, $(\langle x\rangle,\langle p\rangle)$ rotates with frequency $\omega_c$, while the ellipse that represents quantum uncertainty rotates at a slightly faster frequency
\begin{equation}
\omega_{q} =\sqrt{\omega_c^2 +\frac{\mathcal{C}}{M} }
\end{equation}
For a mechanical object probed by light, one can then show that the classical thermal noise peaks at $\omega_c$, while quantum radiation-pressure noise should peak at $\omega_q$.  One can resolve the two peaks if
\begin{equation}
Q \stackrel{>}{_\sim} \frac{\mathcal{C}}{M\omega_c^2} =\frac{\omega_{\rm SN}^2}{\omega_c^2}
\end{equation}
For silicon crystal at temperature around 10\,K, we have
\begin{equation}
\omega_{\rm SN} =0.036\,\mathrm{s}^{-1}
\end{equation}
and we have the requirement of
\begin{equation}
Q  > 3\times 10^6 \cdot \left(\frac{10\,{\rm Hz}}{f_c}\right)^2 
\end{equation}

When we have two mechanical objects interacting through gravity --- assuming both of them to only move within a small range ---  the SN formulation provides an ``interaction Hamiltonian'' of 
\begin{equation}
\label{eq:h12}
H_{12} = \frac{\mathcal{C}_{12}}{2} \left[ \left(x_1-\langle x_2\rangle\right)^2 +
\left(x_2-\langle x_1\rangle\right)^2\right]\,,
\end{equation}
which generates canonical equations by treating expectation values as numbers.
This correctly recovers the mutual gravitational attraction at the level of expectation value.  However, because the coupling is only through one expectation value multiplying the other operator, i.e., $\langle x_1\rangle x_2$, quantum uncertainty cannot be communicated from one object to the other through gravity.  This is therefore a mathematical model for the notion that ``classical gravity cannot be used to transfer quantum information''.   Unfortunately, the term $\mathcal{C}_{12}$ does not have any amplification due to mass concentration --- it is at the scale of $GM/L^3$, and therefore quite weak. ($M$ is the mass of the objects, and $L$ their distance.)  Testing the distinction between classical and quantum gravity through information transfer is therefore much more difficult than through the effect of self-gravity.   Nevertheless, the fact that the CM SN equation can provide a description of such an distinction between classical- and quantum-information transfer makes it perhaps more worthwhile to test than its derivation might suggest. 

\subsubsection{Existence of a minimum length scale} 

\label{subsec:length}

 It was speculated that, due to the existence of a minimum length scale in quantum gravity, the Planck length, Heisenberg Uncertainty should be modified, so that the minimum-uncertainty state would not have a position uncertainty less than the Planck Length.   It was also suggested that this modification for a point particle, could be~\cite{Maggiore:1994,Das:2008,Ali:2011} 
\begin{equation}
\left[\hat x,\hat p\right] = i \hbar \left[1+\beta_0\left(\frac{\hat p}{M_P c}\right)^2\right]
\end{equation}
or a similar form, where $M_P =\sqrt{\hbar c/G} =2\times 10^{-8}\,$kg is the Planck mass, with the general feature that
\begin{equation}
\label{modifyxp}
\left[\hat x,\hat p\right] = i \hbar \left[1+f(\hat p)\right]
\end{equation}

Recently, Pikovsky et al.~\cite{Pikovski:2012}, motivated by this modification, proposed an elegant scheme to probe such an uncertainty principle --- a sequence of four optical pulses, separated by one quarter of the oscillator's period of oscillation,  are applied on a movable mirror to create an evolution operator that explicitly contains the commutator $\left[\hat x,\hat p\right]$.  An alternative scheme to probe the commutator, using a  Sagnac interferometer with movable mirrors, was proposed by Ran Yang et al.~\cite{RYang:2010} --- although they did not provide much motivation for testing the commutation relation.

As Pikovski et al.~\cite{Pikovski:2012} has shown, the modification to the commutation relation, if small, can be perturbatively absorbed into the definition of momentum, in such a way that the commutation relation returns to a canonical one, but the Hamiltonian gains a new term for any free mass.  This poses a problem for this program of modified quantum mechanics: classical mechanics (i.e., evolution equations for expectation values) is also modified by this quantum-gravity-motivated modification to quantum mechanics.  In fact, classical contribution in experimental tests (i.e., from $\langle \hat p \rangle^2$) may easily be much larger than quantum contribution (i.e., from $\langle \hat p^2 \rangle -\langle \hat p\rangle^2$), as is the case in Pikovski et al.'s experimental proposal~\cite{Pikovski:2012}.  Although classical mechanics may not have been specifically tested against the specific gravity model in the mass regime proposed here, the mere fact that classical mechanics is widely accepted casts doubt on the program of modifying commutators in the fashion of Eq.~\eref{modifyxp}.  We will further discuss  this issue below in Sec.~\ref{subsec:testQM}.

\subsection{Towards a more systematic approach}
\label{subsec:testQM}

It is not clear with which chance models mentioned in Sec.~\ref{sec:QMG} will turn out to be true.  Nevertheless, they provided motivations for testing quantum mechanics for macroscopic objects. In this section, I will comment on how this might proceed, if we aim at testing whether macroscopic objects, when isolated well enough from the environment, do follow the Schr\"odinger equation.  Here by ``macroscopic objects'', we could also mean composite ones, e.g., the differential mode of motion of end mirrors of a Michelson interferometer. 

In doing so, one might take two different experimental strategies.  The first one would be an extension of the work of Romero-Isart et al.~\cite{Romero-Isart:2011}: one can create wavefunctions of mechanical objects that spread to a coherence distance much larger than the size of the mechanical object, let them evolve for a finite amount of time, and directly measure deviations from predictions of Schr\"odinger equation.   This kind of test will be model independent, and tell us rather directly in which ways Schr\"odinger equation is violated.   However, experimental challenges for setting up this kind of experiments will be high~\cite{Kaltenbaek:2012}.

A second approach would be to keep measuring small-scale motions of macroscopic mechanical oscillators near their equilibrium positions, using experimental techniques that are already used for precision measurements~\cite{testqm:2012}.  To make this work, one has to {\it parametrize} how the true dynamics of the object deviates from standard quantum mechanics, and how such deviations may show up in measurements with mechanical oscillators.  The test of semiclassical gravity discussed in Sec.~\ref{subsec:semiclassical} is a good illustration of one incidence of such a program: from a {\it particular} modification to the Schr\"odinger Equation (motivated by semiclassical gravity), we obtained a particular observational signature mechanical oscillator.  The next steps would be to (i) formulate the semiclassical test in a parametrized way~\cite{tobar:2012}, and (ii)  systematically collect a more general set of parametrized modifications to quantum mechanics, obtain their signatures in mechanical-oscillator experiments, and put experimental upper limits on all of them.  This would be analogous to the precision test of General Relativity (GR), formulated in the 1970s~\cite{Will:2006}.  

Here we note that our semiclassical gravity model can be easily incorporated into the {\it nonlinear quantum mechanics} model of Weinberg~\cite{Weinberg:1989,Weinberg:1989b}, which he proposed as a framework for precision tests of quantum mechanics.  It is interesting to note that none of the experiments that set upper limits on the Weinberg formalism were able to rule out semiclassical gravity --- scenario in Sec.~\ref{subsec:semiclassical} is much more efficient in probing semiclassical gravity, because of: (i) high mass concentration near the lattice sites and (ii) long coherence time. For this reason,  macroscopic quantum mechanics experiments is likely going to push further the test of other types of nonlinearities that fall within Weinberg's framework~\cite{testqm2:2012}.



\section{Summary and outlook}
\label{sec:conclusion}

\subsection{Summary of this paper }

In this paper, after reviewing theoretical tools for analyzing optomechanical systems in Secs.~\ref{sec:linear}, \ref{subsec:stochastic}, and \ref{sec:unified}, we have outlined experimental concepts that can allow quantum optomechanical systems to demonstrate and test the quantum mechanics of macroscopic objects --- a new regime that had not been accessed before.

In Sec.~\ref{sec:exp_linear}, we have organized the basic experimental concepts, namely quantum-state preparation and quantum-state tomography, into a {\it preparation-evolution-verification} strategy for macroscopic quantum mechanics experiments.  It is the human being's inability to interact quantum mechanically with experimental apparatus has predestined us to such a cumbersome paradigm:  a large number of repetitive and identical experiments --- each starting from quantum measurement and ending at quantum measurement --- with data that eventually collected and analyzed to be compared with quantum mechanics. In Sec.~\ref{subsub:timeline}, we have designed a sample timeline for such a three-stage experiment. If we only need to test certain interesting aspects of quantum mechanics or constrained by experimental possibilities, these idealized concepts here can be simplified and/or modified to fit our purposes. In the subsequent sections~\ref{sec:further} and \ref{sec:nonlinear}, we discussed how these basic experimental concepts can form more complex experiments that demonstrate different aspects of quantum mechanics on macroscopic objects.

\begin{table*}
\centerline{
\begin{tabular}{ccc}
Techniques & Force Measurement & Macroscopic Quantum Mechanics \\
\hline
\hline
\begin{tabular}{c} Classical  noise \\
reduction
\end{tabular}
 &  \begin{tabular}{c}
 obtain sensitivity  
 reaching SQL
 \end{tabular}
 &
\begin{tabular}{c}
 bring detector  
 to quantum regime
 \end{tabular}
 \\
\hline 
\begin{tabular}{c} Optical Spring  \\ and Damping \\
{\small [\ref{rigidity}]} \end{tabular} &  \begin{tabular}{c} resonant signal  enhancement \end{tabular} & 
\begin{tabular}{c} Trapping$^{P,E}$ {\small [\ref{subsec:cooling}, \ref{subsub:timeline}]}   \end{tabular} \\  
\hline
\begin{tabular}{c}
Back-Action Evasion \\
(Variational) \\
{\small [\ref{subsec:SQL},\ref{subsec:bae}]}
\end{tabular}
& 
\begin{tabular}{c} (frequency\ domain) \\ 
avoid radiation pressure noise \end{tabular} 
 & \begin{tabular}{c} time domain, sub-Heisenberg \\
state  tomography$^V$  \\
{\small [\ref{sec:tomography}]}
 \end{tabular} \\ 
\hline 
\begin{tabular}{c} 
Back-Action Evasion \\
{(Stroboscopic)}
\end{tabular}
& 
\begin{tabular}{c} 
QND readout  
for oscillators 
\end{tabular}
&
\begin{tabular}{c} 
Pulsed Optomechanics  \\ 
Ref.~\cite{Vanner:2011}
\end{tabular}
\\
\hline
\begin{tabular}{c} optical  squeezing \\
{\small [\ref{app:twomode}]}
  \end{tabular} & \begin{tabular}{c} overall improvement, \\
  crucial for LIGO-3/ET
  \end{tabular} & \begin{tabular}{c} overall improvement$^{P,E,V}$ \\
important for tomography$^{V}$
{\small [\ref{sec:tomography}]}\\
required by teleportation$^P$ {\small [\ref{subsec:teleport}] }
\end{tabular}
\\
\hline
single photon {\small [\ref{subsec:nongaussian}]}
 & (not yet useful) & 
\begin{tabular}{c} non-Gaussian   state$^P$ 
{\small [\ref{subsec:nongaussian},\ref{subsec:fewphoton}]}
 \end{tabular}  \\
\hline
\begin{tabular}{c}
signal processing
\end{tabular}  &
\begin{tabular}{c} extract GW 
signal\end{tabular}
 & 
\begin{tabular}{c} compute 
conditional state$^P$  {\small [\ref{subsec:conditional}]} \end{tabular}
\\ 
\hline  
\begin{tabular}{c} measurement-based \\ control  {\small [\ref{subsec:colddamping}]} \end{tabular}& \begin{tabular}{c} 
crucial; but should \\not affect noise level
\end{tabular} & 
\begin{tabular}{c} feedback cooling$^P$ {\small [\ref{subsec:colddamping}]} \\ teleportation$^P$ {\small [\ref{subsec:teleport}]}\end{tabular} \\
\hline 
\begin{tabular}{c} coherent quantum \\ control {\small [\ref{subsec:colddamping}]}
\end{tabular} &
\begin{tabular}{c} possible guidance to new \\
designs with higher sensitivity
\end{tabular}  &
Trapping$^{P,E}$ {\small [\ref{subsec:cooling}, \ref{subsub:timeline}]}
\\
\hline
\end{tabular}}
\caption{Techniques of GW detectors and their application to exploring MQM. Superscripts of $P$, $E$ and $V$ stand for preparation, evolution and verification, respectively. Numbers in square brackets in the first column indicate sections in which the technique is first introduced in this paper, while those in the third column indicate the sections in which the technique is applied.   \label{tab:techniques}} 
\end{table*}

In Sec.~\ref{sec:exp_linear}, we have also connected the feasibility of these concepts to the device's noise levels if they were to be used as device that measure classical forces.  In order to draw more connections between techniques of force measurement (e.g., under development in the gravitational-wave detection community) and experiments in macroscopic quantum mechanics, I have listed, in Table~\ref{tab:techniques}, several key techniques and the roles they may play in these two different efforts.

In Sec.~\ref{sec:test}, we showed how certain modifications to quantum mechanics, due to considerations on the interplay between gravity and quantum mechanics, can be tested using optomechanical systems.  This serves as a motivation for a program of ``precision test of macroscopic quantum mechanics''. 




\subsection{Optomechanics as part of quantum technology and many-body optomechanics}

In this paper, we have not emphasized on the tremendous technological impact of optomechanics: a mechanical component can be inserted into previously existing quantum system to achieve noval capabilities.   This often requires the construction of {\it hybrid systems}, in which mechanical degrees of freedom are strongly coupled with optical or atomic degrees of freedoms; this involves: (i)  the transfer of quantum information between a mechanical mode and an optical/microwave mode~\cite{Wang:2012,Verhagen:2012,Palokaki:2012}, between mechanical mode and a single atom~\cite{Aoki:2006,Hammerer:2009,Glaetzle:2010,Wallquist:2010,Singh:2008} or an atomic ensemble~\cite{Hammerer:2008,Muschik:2011,Vasilyev:2012,Hammerer:2010,Steinke:2011}, (ii) the use of mechanical motion as transducers that mediate interactions between optical/atomic or other mechanical modes~\cite{Lin:2010,Stannigel:2010,Stannigel:2011,Hill:2012,Winger:2011,Hafezi:2012,Dong:2012}, (iii) the use of mechanical structure with internal dynamics to confine atoms~\cite{Hammerer:2010b,Bhattacharya:2010}. Quantum information processing protocols using optomechanics alone have also been proposed~\cite{Schmidt:2012,Vanner:2012,Habraken:2012,Stannigel:2012}. Another exciting direction is to use optomechanics as a tool to study complex quantum dynamics~\cite{Seok:2012,Buchmann:2012,Bhattacharya:2008b}, especially those of many-body systems~\cite{Ludwig:2012b,Jacobs:2012b,Holmes:2012,ChenW:2009,Zhang:2010}.

The reader is referred to these other review articles for a more complete picture of optomechanics~\cite{Kippenberg:2008,Marquardt:2009,Wallquist:2009,Aspelmeyer:2010,Meystre:2012}


\subsection{Optomechanics and the nature of quantum state reduction}

Finally, concerning the nature of quantum measurement: does optomechanics also provide a new opportunity for probing further in into the nature of quantum-state reduction, and the statistical nature of the wavefunction?  On the one hand, we will be continuously measuring a macroscopic object for which quantum mechanics has never been tested before, while on the other hand, internal consistency of quantum mechanics allows us to formulate the measurement as being applied onto the out-going optical field, for which quantum mechanics has been well tested.  This means: (i) testing the deterministic quantum dynamics of a system consisting a macroscopic mechanical object, as we have discussed in this paper is a crucial foundation for the correctness of quantum measurement theory involving macroscopic objects, and that (ii) if quantum mechanics turns out to hold precisely for such systems, then the omnipresence of quantum measurement processes in many physics phenomenons will likely indicate that optomechanical systems do not offer unique insight.



\ack

The author would like to thank Professor Kip S.\ Thorne for introducing him to research in quantum measurement theory; he would also like to thank members of the AEI-Caltech-MIT-MSU macroscopic quantum mechanics group for discussions and collaborations over the past several years. For more recent discussions on testing quantum mechanics,  he would like to thank: Rana Adhikari, Markus Aspelmeyer, Thomas Corbitt, Stefan L.\ Danilishin, Bassam Helou, Bei-Lok Hu, Poghos Kazarian, Farid Ya.\ Khalili, Da-Shin Lee, Nergis Mavalvala, Haixing Miao, David McClelland, Larry Price, Oriol Romero-Isart, Nick Smith, Roman Schnabel, Kentaro Somiya, Kip Thorne, Mike Tobar, and Huan Yang.  He acknowledges support from the Keck Institute for Space Studies (KISS), which made possible some of the above discussions. 
The author's research is supported by NSF Grant PHY-1068881 and CAREER Grant PHY-0956189, the David and Barbara Groce Startup Fund at Caltech,  and the Institute of Quantum Information and Matter (IQIM), a NSF Physics Frontier Center with support from the Moore Foundation.  His research in this direction  has also been supported in the past by NSF Grants PHY-0653653 and PHY-0601659, as well as the Sofja Kovalevskaja Program of the Alexander von Humboldt Foundation (funded by the Ministry of Education of Germany).

\appendix 

\begin{appendices}

\section{Classical Random Processes and Spectral Density}
\label{app:classicalrandom}

In this section, we review key facts of classical stochastic processes, and establish our conventions and notations. 
\subsection{Gaussian Processes, Correlation Functions and Linear Regression.}
\label{app:subsec:gauss}

We first recall the feature of Gaussian random vectors, that the joint distribution $p$ of any set of such random variables, $x_{1,\ldots,n}$, if jointly Gaussian, can already be determined from the covariance matrix $V_{ij} = \langle x_i x_j\rangle$:
\begin{equation}
p_{x_1,\ldots,x_n}(y_1,\ldots,y_n) =\frac{1}{\sqrt{(2\pi)^{n} | \mbox{\boldmath$\Gamma$}|}}\exp\left(-\frac{y_i \Gamma^{ij} y_j}{2} \right)\,,
\end{equation}
where
\begin{equation}
 \mbox{\boldmath$\Gamma$} = \mathbf{V}^{-1}
 \end{equation}
 is sometimes referred to as the information matrix.

A Gaussian random process is a sequence of Gaussian random variables, $x(t)$.  
 For simplicity, let us consider processes that have zero mean values at all times.  All statistical characters of such a Gaussian random process will be determined by its two-time correlation function 
\begin{equation}
C_{xx}(t,t') = \langle x(t) x(t')\rangle
\end{equation}
The system is also {\it stationary} if  correlation functions only depend on the relative time  difference:
\begin{equation}
C_{xx}(t,t') =C_{xx}(t-t')
\end{equation}
Given another random process $y(t)$, suppose they are correlated, with cross correlation function
\begin{equation}
C_{xy}(t,t') =\langle x(t) y(t')\rangle\,.
\end{equation} 
If they are jointly stationary, we have
\begin{equation}
C_{xy}(t,t') =C_{xy}(t-t')\,.
\end{equation} 

A question often asked in statistics is: at time $t$, suppose we have already known $\{x(t'): 0<t'<t\}$, what is our best estimate for $y(t)$, and what would be the error?
 To this end, we first find the {\it conditional expectation} of $y(t)$, which can be obtained through the conditional probability density of $y$:
\begin{equation}
E[y(t)|x(t'):0<t'<t] =\int   \xi p_{y(t)|x(t'):0<t'<t}(\xi)  d\xi\,.
\end{equation}
This is a function of $\{x(t'): 0<t'<t\}$, and hence a random variable.  But it has a definite value once $\{x(t'): 0<t'<t\}$ is known.  In the Gaussian case, we can show that,
\begin{equation}
\label{eqG}
E[y(t)|x(t'):0<t'<t] = \int_0^t G(t,t') x(t') dt'\,,
\end{equation}
and that this is always a non-biased estimation for $x$; we can further show that among all possible linear functions of $\{x(t'):0<t'<t\}$, this one has the least mean-square error. 

In order to prove the above, including obtaining  $G$, let us define
\begin{equation}
R(t) \equiv y(t) -\int_0^t G(t,t') x(t') dt'
\end{equation}
and  set up the equation
\begin{equation}
\label{econd}
E\left[R(t) x(t'')\right]=0,\quad\forall 0<t''<t\,.
\end{equation}
First of all, Eq.~\eref{econd} determines $G$ uniquely. Defining 
\begin{equation}
u(t') = E[y(t) x(t')]\,,
\end{equation}
we can write
\begin{equation}
G(t')  = \int_0^t C^{-1}_{xx} (t',t'') u(t'') dt''\,.
\end{equation} 
where  we have defined $C_{xx}^{-1}$ as the inverse of $C_{xx}$ or
\begin{equation}
\int_{0}^t  C_{xx}^{-1}(t_1,t_2)C_{xx}(t_2,t_3) dt_2 =\delta(t_1-t_3)\,.
\end{equation}
Secondly, because of Eq.~\eref{econd}, and because expectation values of $R(t)$ and $x(t'')$ themselves vanish, $R(t)$  is independent from all of $x(t'')$, and 
\begin{equation}
E[R | x(t'):0<t'<t]= 0\,.
\end{equation}
on the other hand, it is obvious that 
\begin{eqnarray}
&& E\left[\int_0^t G(t,t') x(t') dt' \Big| x(t'):0<t'<t\right] \nonumber\\
&=& \int_0^t G(t,t') x(t') dt'\,.
\end{eqnarray}
we therefore have proven that our $G$ makes Eq.~\eref{eqG} satisfied.  Finally, minimum least-square error can be proved noting that since $R$ is independent from all $x(t')$ adding any additional term in the predictor for $y$ will yield a error variance that is greater than the variance of $R$\,.

\subsection{Spectral Density}

In this paper, we use the single-sided cross spectral density, for real-valued classical stationary random processes (with zero expectation value):
\begin{equation}
\frac{1}{2}S_{xy}(\Omega) \delta(\Omega-\Omega') =\langle \tilde x(\Omega) \tilde y^*(\Omega') \rangle\,.
\end{equation}
This lead to the Wiener-Khintchin relation of
\begin{equation}
\langle x(t) y(t')\rangle = \int_0^{+\infty}\frac{d\Omega}{2\pi} S_{xy}(\Omega) e^{-i\Omega(t-t')}
\end{equation}
In particular, 
\begin{equation}
\mathrm{Var}[x(t)] = \int_0^{+\infty}\frac{d\Omega}{2\pi}S_{xx}(\Omega)
\end{equation}
which means the variance of $x(t)$, which measures the fluctuation in $x$, is made up from fluctuations at all frequencies, with $S_{xx}(\Omega)$ measuring the noise-power contribution from a unit frequency band around $\Omega$.  

In this paper, we will often discuss spectral density of a quantum operator that is a function of time, or cross spectral density of two quantum operators. In order to do so, we must be careful about ordering of operators.  For two fields $\hat A(t)$ and $\hat B(t)$, which Fourier transforms into $\hat A(\Omega)$ and $\hat A(\Omega)$, for a state $|\psi\rangle$, we will {\it define} spectral density by symmetrization:
\begin{equation}
\frac{1}{2}\delta(\Omega-\Omega') S_{AB}(\Omega) =\frac{1}{2} \langle \psi| A(\Omega)B^\dagger(\Omega')+B^\dagger(\Omega')A(\Omega)|\psi\rangle
\end{equation}
We should be careful that not for all states will the right-hand side evaluate into a $\delta(\Omega-\Omega')$ --- but only for those can we define the cross spectrum of $A$ and $B$.

\subsection{Wiener Process and Ito Calculus}
\label{app:wiener}

If we define a process $v$ with 
\begin{equation}
S_v (\Omega) =2\,,
\end{equation}
it will then be a {\it white noise}, because noise power is constant over all frequencies. We will have, in the time domain,
\begin{equation}
\langle v(t) v(t') \rangle =\delta (t-t')
\end{equation}
and in particular
\begin{equation}
\mathrm{Var}[v] =\langle v^2(t)\rangle = +\infty
\end{equation}
because fluctuations arise from all possible frequencies until infinity.  If we define $\bar v$ as the average of $v$ over a time interval of $\Delta t$, we obtain
\begin{equation}
\langle \bar v^2 \rangle  =\frac{1}{\Delta t^2}\int_t^{t+\Delta t} \langle v(t')v(t'') \rangle dt'dt'' =\frac{1}{\Delta t} \,,
\end{equation}
the longer we average, the less the level of fluctuation.

If we further define
\begin{equation}
W (t) \equiv \int_0^t v(t')dt,
\end{equation}
we will obtain
\begin{equation}
\langle W(t)W(t') \rangle = \min(t,t')
\end{equation}
The process $W$ is called the Wiener Process. Physically, $W(t)$ describes the displacement of a particle undergoing a Brownian motion.  For a finite increment of time, $\Delta t$, we can also define \begin{equation}
\Delta W \equiv  W(t+\Delta t)-W(t) = \int_{t}^{t+\Delta t} v(t') dt'
\end{equation}
which is often called the {\it Wiener Increment} if we take $\Delta t \rightarrow dt$. 
 It would not be difficult to calculate
\begin{equation}
\langle \left(\Delta W\right)^{2n-1}\rangle =0 \,,
\;\;
\langle\Delta W^2\rangle =   \Delta t\,,\;\;
\langle\Delta W^4\rangle =3 \Delta t^2\,,\ldots
\end{equation}
which intuitively means that during $\Delta t$, $\Delta W$ fluctuates with magnitude of $\sim \sqrt{\Delta t}$.  This means $W(t)$ will be almost everywhere differentiable. In general, if a quantity $u$ has a finite spectrum $S_u(\Omega)$ everywhere, and $S_u(\Omega) \sim \Omega^{-2}$ as $\Omega\rightarrow+\infty$, then $u$ should be everywhere continuous but not differentiable. If $S_u(\Omega) \sim \Omega^{-2n}$, $n\ge 1$, then $u$ will have finite derivatives up to the $n$-th order.

In physics, we often consider processes driven by white noise.  The simplest case would be to consider a Langevin type equation of 
\begin{equation}
\dot x(t) = f(t) + g(t) v(t)
\end{equation}
which can be rewritten as
\begin{equation}
dx =f dt + g dW\,,
\end{equation} 
whose solution can be written as
\begin{equation}
\label{itoproc}
x(t) = \int_0^t f(t') dt' + \int_0^t g(t') dW(t')
\end{equation}
Processes like $x$ here are, driven by an integral of Wiener Increments, are called Ito processes.  The latter integral in Eq.~\ref{itoproc} can be constructed in the same way as a Riemann-Stieltjes integral, noting that as we subdivide the integral fine enough, the error introduced within each subinterval by a grid with size $\Delta t$ will have a standard deviation of $\Delta t^{3/2}$ --- and all these are statistically independent from each other. 

One can show that for any smooth enough function $y(x)$, $y(x(t))$ is also an Ito process.   However, if $y$ is a nonlinear function,  because $dW$ fluctuates with scale $\sqrt{\Delta t}$, we will have to expand to second order in $dW$:
\begin{equation}
\label{dy}
\Delta  y = y'(x) (f \Delta t + g \Delta W) + \frac{y'' (x)}{2} g^2 \Delta W^2 + o(\Delta t)\,.
\end{equation}
Equation~\ref{dy} only serves as a reminder that the second-order term, which is proportional to $(\Delta W)^2$ and has an expectation value and standard deviation at the order of $\Delta t$, must not be ignored. The obstacle against using Eq.~\ref{dy} directly for further deduction is that $\Delta W^2$ and $\Delta W$ are highly correlated to each other (because the former is the square of the latter).
The more informative way to deal with the problem is to divide 
$\Delta t$ into $N+1$ sections $t=t_0 <t_1 <\ldots<t_{N+1} = t+\Delta t$, we can write
\begin{eqnarray}
\Delta y &= & \left[y'_0 (f_0 \Delta t_0 + g_0 \Delta W_0) + \ldots  +y'_N (f_{N} \Delta t_{N} + g_N \Delta W_N)\right]\nonumber\\
&+& \frac{1}{2}\left[y''_0 g_0^2 (\Delta W_0)^2+ \ldots + y''_N g_N^2 (\Delta W_N)^2\right] +o(\Delta t),\qquad
\end{eqnarray}
with subscript indicating the time at which the functions are evaluated. 
As $N \rightarrow +\infty$, the first term remains proportional to $\Delta t$ and $\Delta W$, while the second term approaches its expectation value with unit probability, and therefore we can write
\begin{equation}
dy = y'(x) (f dt + g dW) + \frac{1}{2}y''(x) g^2 dt
\end{equation}
This means, as we apply the chain rule for Ito processes, we need to use the Ito rule:
\begin{equation}
dW^2 = dt
\end{equation}

In this discussion, we have adopted $dW(t) = W(t+dt)-W(t)$. In this way, the increment we receive at time $t$ is always statistically independent from anything we have already known up to this time.  This is the fundamental feature of Ito calculus.

\section{Classical Kalman Filtering for Linear Systems}
\label{kalman}

Suppose we have a linear system (with dynamical variables $\vec x$ and evolution matrix $\mathbf{M}$) driven by white force noise
\begin{equation}
d \vec {x} = \mathbf{M} \vec{x} + d\vec{W}_1\,, 
\end{equation}
while at the same time being observed with white sensing noise
\begin{equation}
d\vec{y} =\mathbf{A}  \vec{x} dt + d\vec{W}_2
\end{equation}
Here $d \vec y$ represents the measurement outcome during $dt$ and $\mathbf{A}$ is the transfer matrix.  Let us build a set of evolution equations for our knowledge about the system, i.e., $\vec x$.  At time $t$, suppose we have prior knowledge about $\vec x$, obtained by measuring $\vec y(t'<t)$, in the form of a conditional probability density, 
$p[\vec x|\vec y(t'<t)]$.  Let us assume we gather data for additional $\Delta t$, and find out our new conditional density for $\vec x$:
\begin{eqnarray}
&&p\left[\vec x(t+\Delta t)=\vec \xi\big|\vec y(t'<t),\Delta\vec{y}=\vec\eta\right] \nonumber\\
&\propto& 
\int p\left[\vec x(t+\Delta t)=\vec \xi,\vec x(t)= \vec\zeta,\vec y(t'<t),\Delta\vec y=\vec \eta\right] d\vec \zeta  \nonumber\\
&\propto&
\int  p\left[\vec x(t+\Delta t)=\vec \xi, \Delta\vec y=\vec \eta \big| \vec x(t)=\vec  \zeta, \vec y(t'<t)\right] 
\nonumber\\ && \,\qquad \qquad \qquad\qquad \times p\left[x(t)=\vec \zeta \big|\vec y(t'<t)\right] d\vec \zeta   
\end{eqnarray}
The normalization factor here can be determined by imposing integral over $\xi$ to go to unity.  Note that this is of the form of updating from an earlier time to a later time.  Denoting 
\begin{equation}
\mathbf{N}_{11} dt = d\vec W_1 \, d \vec W_1^t \,,\quad  \mathbf{N}_{22} dt = d\vec{W}_2 \, d\vec{W}_2^t\,,
\end{equation}
and writing the conditional covariance matrix of $\vec x$ as $\mathbf{V}$, the conditional expectation as $\vec x_c$, we obtain
\begin{equation}
\label{kalmanmean}
d\vec x_c =\mathbf{ M } \vec{x_c} dt +\mathbf{V} \mathbf{A}^t \mathbf{N}_{22}^{-1}\left(dy - \mathbf{A} \vec x_c dt\right)
\end{equation}
and
\begin{equation}
\label{kalmanvar}
\dot{\mathbf{V}} = \left(\mathbf{V M} +\mathbf{ M V}\right)  + \mathbf{N}_{11} -\mathbf{V}\mathbf{A}^t \mathbf{N}_{22}^{-1} \mathbf{A V} \,.
\end{equation}
These are called the linear {\it Kalman Filter equations}.  Note that the first terms on the right-hand side of both equations arise from free dynamics, while the additional terms from measurement. In Eq.~\eref{kalmanmean}, the additional term is actually a stochastic term proportional to $d\vec W_2$.  In Eq.~\eref{kalmanvar}, the first additional term is due to force noise, so when $\mathbf{N}_{11}$ grows, the covariance matrix increases; the second additional term is due to information gained during measurement --- as information becomes more accurate, or  $\mathbf{N}_{22}\rightarrow 0$, $V$ decreases very fast, while as information becomes less accurate, $\mathbf{N}_{22}\rightarrow +\infty$, this term ceases to be important.  Note that we always require $\mathbf{N}_{22}$ to be invertible, which means no measurement channel is noise free.

\section{Two-mode quantum optics}
\label{app:twomode}
\subsection{One-mode quantum optics}

In constructing a quantum field theory, we often start from annihilation and creation operators of a free field, which satisfy the canonical commutation relation:
\begin{equation}
\label{commaad}
\left[a_\omega,a_{\omega'}\right] =\left[a^\dagger_\omega,a_{\omega'}^\dagger\right]=0\,,\quad
\left[a_\omega,a_{\omega'}^\dagger\right] =2\pi\delta(\omega-\omega')
\end{equation}
This, together with $a_\omega |0\rangle=0$, leads to:
\begin{equation}
\langle 0 | a_\omega a^\dagger_{\omega'} |0\rangle = 2\pi\delta(\omega-\omega')\,,\quad 
\langle 0 | a^\dagger_\omega a_{\omega'} |0\rangle=0\,.
\end{equation}
This asymmetry in two-point correlation functions is directly related to the fact that $a$ and $a^\dagger$ do not commute.  Because $a$ and $a^\dagger$ are intimately related to positive- and negative-frequency Fourier components of the electric field operator in the Heisenberg picture, 
\begin{eqnarray}
{\hat E}^{(+)}(t,x) &=&\int_0^{+\infty} \frac{d\omega}{2\pi}\sqrt{\frac{\hbar\omega}{2}} \hat a^\dagger_\omega e^{-i\omega t +i\omega x}\,,\\
{\hat E}^{(-)}(t,x) &=&\int_0^{+\infty} \frac{d\omega}{2\pi}\sqrt{\frac{\hbar\omega}{2}} \hat  a_\omega e^{+i\omega t -i\omega x}\,,
\end{eqnarray}
with the total field operator the sum of the two:
\begin{eqnarray}
\hat E(t) ={\hat E}^{(+)}(t)+{\hat E}^{(-)}(t)
\end{eqnarray}
Our normalization is very simple because we relate the intensity of the propagating wave along this one dimension as
\begin{equation}
\hat I(t,x) = c \hat E^2(t,x)\,.
\end{equation}
In this way, since the energy density of the optical field is the energy flux divided by $c$, and we have
\begin{equation}
\hat H =\int  \hat I(t,x) dx = \int_0^{+\infty} \hbar \omega \left[ \hat a^\dagger_\omega \hat a_\omega +\frac{1}{2}\right]\,.
\end{equation}


\subsection{Two-mode quantum optics}

  Let us, we imagine a classical {\it carrier field} at $\omega_0$ frequency, and consider amplitude and phase modulations to this carrier.  This can either be described as having a coherent state of 
\begin{equation}
|\alpha_{\omega_0}\rangle = e^{\alpha a_{\omega_0}^\dagger+ \alpha^* a_{\omega_0}}|0\rangle
\end{equation}
or, though a unitary evolution, as having the annihilation and creation operators shifted by 
\begin{equation}
a_\omega = \alpha \delta(\omega-\omega_0) + a_\omega \,,\quad a_\omega^\dagger = \alpha^* \delta(\omega-\omega_0) +a_\omega^\dagger\,.
\end{equation}
Let us restrict ourselves to focus on field contents around a fixed frequency $\omega_0$ (e.g., within a bandwidth of $\Lambda$) and we have
\begin{eqnarray}
\hat E(t) &= & \alpha e^{-i\omega_0 t} + \alpha^* e^{i\omega_0 t} \nonumber\\
&+&\sqrt{\hbar\omega_0} \int\limits_{|\omega-\omega_0| < \Lambda}\frac{d\omega}{2\pi} \left[\hat a_\omega e^{-i\omega t} + \hat a_\omega^\dagger e^{i\omega t}\right]  \nonumber\\
&=&A\cos(\omega_0 t +\phi) \nonumber\\ &+&\sqrt{\hbar\omega_0} \left[\hat E_1(t) \cos\omega_0 t  +\hat E_2 (t)   \sin\omega_0 t \right]
\end{eqnarray}
with $\alpha = A e^{i\phi}$, and $\hat E_{1,2}(t)$ slowly varying fields defined as
\begin{equation}
\hat E_j(t) =\int_0^\Lambda  \frac{d\Omega}{2\pi}
\left[ a_j(\Omega) e^{-i\Omega t}+a_j^\dagger(\Omega) e^{i\Omega t}\right]\,,\quad
j=1,2,
\end{equation}
and
\begin{equation}
a_1(\Omega) =\frac{a_{\omega_0+\Omega} +a^\dagger_{\omega_0-\Omega} }{\sqrt{2}}\,,\quad 
a_2(\Omega) =\frac{a_{\omega_0+\Omega} - a^\dagger_{\omega_0-\Omega} }{\sqrt{2}i}\,.
\end{equation}
The fields $\hat E_1(t)$ and $\hat E_2(t)$ are called {\it quadrature fields}; their linear combinations, a general quadrature, is defined as
\begin{equation}
\hat E_\theta(t) = \hat E_1 (t) \cos\theta + \hat E_2 (t) \sin\theta\,.
\end{equation}
The quadrature $\hat E_\theta$ can  act as amplitude or  phase modulations to the carrier, when $\theta$ takes appropriate values. The formalism here is often referred to as {\it two-photon} quantum optics, because it deals with field fluctuations at $\omega_0\pm\Omega$ at the same time.   This formalism was first developed by Caves and Schumaker.  In vacuum state, we have
\begin{equation}
S_{a_1 a_1}= S_{a_2 a_2} =1\,,\quad S_{a_1 a_2}=0\,.
\end{equation}

\subsection{Two-Mode Squeezing}
\label{app:2modesqueezing}

A two-mode squeezed state normally refers to a Gaussian state of a continuum field in which $\omega_0 \pm\Omega$ sidebands are correlated. Formally, such a state can be represented by 
\begin{equation}
|\psi\rangle = \exp\left\{
\int\frac{d\Omega}{2\pi}
\left[\chi(\Omega) a_{\omega_0+\Omega}^\dagger a_{\omega_0-\Omega}^\dagger 
-\mathrm{ h.c.}
\right]\right\} |0\rangle 
\end{equation}
which contain pairs of correlated photons.  An alternative way is to transform the state back to vacuum, which results in a Bogolubov transformation involving $\hat a(\omega_0\pm \Omega)$ and $\hat a^\dagger(\omega_0\pm \Omega)$, which is easily represented in the quadrature picture.  At each sideband frequency $\Omega$, we need to relate the quadrature operators $a_{1,2}$ to two other quadrature operators $n_{1,2}$ which are at vacuum state, through
\begin{equation}
\left(\begin{array}{c} a_1 \\ a_2\end{array}\right) = \mathbf{R}_\phi  \mathbf{S}_r
\left(\begin{array}{c} n_1 \\ n_2\end{array}\right)
\end{equation}
with
\begin{eqnarray}
\mathbf{R}_\phi &=&
\left(\begin{array}{cc} \cos\phi & -\sin\phi \\ \sin\phi & \cos\phi\end{array}\right)\,,  \\
 \mathbf{S}_r &=& \left(
\begin{array}{cc}
e^r \\ & e^{-r}\end{array}\right)
\end{eqnarray}
where empty matrix entries are zeros. It is as if these quadrature operators are rotated and squeezed/stretched.  Here $\phi$ is usually referred to the squeeze angle (or phase), and $e^r$ is referred to as the amplitude squeeze factor ($e^{2r}$ is the power squeeze factor, and ofter quoted in dB).  In principle, $\phi$ and $r$ can be functions of sideband frequency $\Omega$. 

\section{Facts about Wigner Functions} 

In this appendix, we collect a set of facts involving the Wigner function, a quasi-probability distribution that exists for all quantum states of a mechanical degrees of freedom.  

\subsection{Definition}

The formal way to obtain the Wigner function from the density matrix is to first define the generating function
\begin{equation}
\label{eqJgen}
J(\mu,\nu) = \mathrm{tr} \left[ e^{i\mu\hat x + i\nu \hat p} \hat\rho\right] 
\end{equation}
and then write
\begin{equation}
\label{WfromJ}
\mathcal{W}(x,p) =\int \frac{d\mu}{2\pi} \frac{d\nu}{2\pi} e^{-i\mu x-i\nu p} J(\mu,\nu)
\end{equation} 
We can then combine these two and write
\begin{equation}
\mathcal{W}(x,p) =\int \frac{d\mu}{2\pi} \frac{d\nu}{2\pi} \langle  e^{-i\mu (x-\hat x) -i\nu (p-\hat p)} \rangle \,.
\end{equation}
This is almost like ``$\langle \delta (x-\hat x)\delta(p-\hat p)\rangle$'', and therefore a reasonable definition for a joint probability density of $x$ and $p$ --- except that $\hat x$ and $\hat p$ do not commute and we cannot write an expression this way.

However, if we first integrate out any linear combination of $x$ and $p$, obtaining a {\it marginal distribution} for the orthogonal distribution, we will have only one integral left and the above process works.  For example, integrating over $p$, we have
\begin{equation}
\int \mathcal{W}(x,p)dp =\frac{1}{2\pi} \int d\mu \langle e^{-i\mu(x-\hat x)}\rangle =\langle \delta (x -\hat x)\rangle
\end{equation}
This means any {\it marginal distribution} of $\mathcal{W}(x,p)$ along the $x$ direction agrees with the $x$ distribution of the quantum state.

\subsection{Stochastic Differential Equation (SDE) for the Conditional Wigner Function}
\label{app:wig}

Let us derive the continuous-measurement SDE for the Wigner function. 
%
Starting from the SME, 
\begin{eqnarray}
\label{smealt}
d\hat \rho &=& -i \left[\hat H,\hat\rho\right] dt + \frac{\alpha}{\sqrt{2}}
\left\{\hat x-\langle\hat x\rangle ,\hat \rho\right\} dW \nonumber\\
&-&\frac{\alpha^2}{4}\left[\hat x,\left[\hat x,\hat\rho\right]\right]dt
\end{eqnarray}
using the definition of the generating function $J$ [Cf.~Eq.~\eref{eqJgen}] and using $\hat H = p^2/(2m) + m\omega^2 x^2/2$ for concreteness, we obtain
\begin{eqnarray}
\label{SMEJ}
dJ  &=&\left(\frac{\mu}{m}\partial_\nu  J -m\omega^2\nu\partial_\mu J
\right) dt \nonumber\\
&+&\sqrt{2} \alpha\left(i\partial_\mu -\langle \hat x\rangle \right)J dW -\frac{\alpha^2}{4}\nu^2 J dt\,,
\end{eqnarray}
where the first term on the right-hand side comes from $H$, and the rest from the measurement process.  Here we have used the Baker-Campbell-Hausdorf formula,
\begin{equation}
e^{\hat A + \hat B} =e^{\hat A}e^{\hat B}e^{-\frac{1}{2}\left[\hat A,\hat B\right]}
\end{equation}
 and obtained
\begin{eqnarray}
\partial_\alpha e^{i\mu\hat x + i\nu \hat p}& =&e^{i\mu\hat x + i\nu \hat p} \left( i\hat x -\frac{i\nu}{2}\right)=\left( i\hat x +\frac{i\nu}{2}\right) e^{i\mu\hat x + i\nu \hat p}\nonumber\\ &=&\frac{i}{2}
\left\{\hat x, e^{i\mu\hat x + i\nu \hat p}\right\} 
\,, \\
\partial_\nu e^{i\mu\hat x + i\nu \hat p} &=&e^{i\mu\hat x + i\nu \hat p} \left( i\hat p +\frac{i\mu}{2}\right)=\left( i\hat p -\frac{i\mu}{2}\right) e^{i\mu\hat x + i\nu \hat p} \nonumber\\
&=&\frac{i}{2}
\left\{\hat p, e^{i\mu\hat x + i\nu \hat p}\right\}  \,.
\end{eqnarray}
As a next step, from Eq.~\eref{WfromJ}, we simply perform a Fourier transform  of Eq.~\eref{SMEJ}, and obtain
\begin{eqnarray}
d\mathcal{W}+\left[\frac{p}{m}\partial_x  -m\omega^2 x \partial_p \right]\mathcal{W} dt &=&\sqrt{2} \alpha(x-\langle \hat x\rangle) \mathcal{W} dW\nonumber\\ &+&\frac{\alpha^2}{4}
\partial_p^2 \mathcal{W} dt\,.\end{eqnarray}
Terms on the right-hand side originate from the measurement; the first term is a random drift driven by $dW$, while the second term is a diffusion.

\subsection{Path Integral and Causal Wiener Filtering}
\label{app:causal}   
   
In Sec.~\ref{subsec:path}, we encountered the following path integral:
\begin{eqnarray}
&& J_c\left[\mu,\nu|\xi(t')\right] \nonumber\\ &\propto  &\langle \psi_c| e^{i\left[\mu \hat x(t) +\nu \hat p(t)\right] }|\psi_c\rangle  \nonumber\\
&=&
\int D[k(t')] \mathrm{tr}  \left[ e^{
i\mu\hat x(t) +i\nu\hat p(t)+
i \int_0^t k(t') \left[\hat b_\zeta(t') - \xi(t')\right]} \hat\rho_{\rm ini}\right]
\end{eqnarray}
Let us show that this is as if all quantities involved are classical random processes.  First of all, let us assume we already have the Wiener filters $G_x$ and $G_p$, and the integrand can be changed into 
\begin{equation}
\left\langle e^{i\mu\hat x(t) +i\nu\hat p(t)+
i \int_0^t [k(t')- \mu G_x(t') - \nu G_p(t')] \left[\hat b_\zeta(t') - \xi(t')\right]}\right\rangle
\end{equation} 
because we are simply shifting each $k(t')$ by a constant, which does not change the value of the path integral.  This becomes
\begin{eqnarray}
 J_c\left[\mu,\nu|\xi(t')\right]
\propto \left\langle e^{i\hat  O }\right\rangle
\end{eqnarray} 
with
\begin{eqnarray}
\hat O &=&\mu\left[\hat x(t)-\int_0^t G_x(t') \left[\hat b_\zeta(t')-\xi(t')\right]\right] \nonumber\\
&+&\nu\left[\hat p(t)-\int_0^t G_p(t') \left[\hat b_\zeta(t')-\xi(t')\right]\right]\,.
\end{eqnarray}
Here we have discarded a factor that does not depend on $(\mu,\nu)$, and the path integral is contained within that factor.   This has been possible because all of $\{b_\zeta(t'):0<t'<t\}$ commute with $\mu \hat x(t)+\nu\hat p(t)$, 
\begin{eqnarray}
\left\langle b_\zeta(t')\left[\hat x(t)-\int_0^t G_x(t') \hat b_\zeta(t')\right] \right\rangle&=&0 \\
\left\langle b_\zeta(t')\left[\hat p(t)-\int_0^t G_p(t') \hat b_\zeta(t')\right] \right\rangle&=&0 
\end{eqnarray}
and if $[\hat A,\hat B]=0$ and $\langle \hat A\rangle =\langle \hat B\rangle =\langle \hat A \hat B\rangle=0$, we have
\begin{equation}
\left\langle e^{\hat A+\hat B}\right\rangle = e^{\Delta A^2/2} e^{\Delta B^2/2} =\left\langle e^{\hat A}\right\rangle\left\langle e^{\hat B}\right\rangle
\end{equation}
Here we have also used the fact that for linear operators on Gaussian states, 
\begin{equation}
\left\langle e^{\hat A} \right \rangle = e^{\left\langle \hat A  \right \rangle} e^{\Delta A^2/2}
\end{equation}
where
\begin{equation}
\Delta A^2 =\langle \hat A^2\rangle - \langle\hat A\rangle^2
\end{equation}

\end{appendices}


\bibliographystyle{iopart-num}	

\bibliography{references}		

\providecommand{\newblock}{}
\begin{thebibliography}{100}
\expandafter\ifx\csname url\endcsname\relax
  \def\url#1{{\tt #1}}\fi
\expandafter\ifx\csname urlprefix\endcsname\relax\def\urlprefix{URL }\fi
\providecommand{\eprint}[2][]{\url{#2}}

\bibitem{Zurek:2003}
Zurek W 2003 {\em Reviews of Modern Physics\/} {\bf 75} 715

\bibitem{Schwab:2005}
{Schwab} K~C and {Roukes} M~L 2005 {\em Physics Today\/} {\bf 58} 070000--42

\bibitem{Aspelmeyer:2012}
Aspelmeyer M, Meystre P and Schwab K 2012 {\em Physics Today\/} {\bf 65} 29--35
  \urlprefix\url{http://link.aip.org/link/?PTO/65/29/1}

\bibitem{Braginsky:1968}
{Braginski{\v i}} V~B 1968 {\em Soviet Journal of Experimental and Theoretical
  Physics\/} {\bf 26} 831

\bibitem{Braginsky:1986}
Braginsky V and Mitrofanov V 1986 {\em Systems with small dissipation\/}
  (University of Chicago Press)

\bibitem{Braginsky:1992a}
Braginsky V~B and Khalili F~Y 1992 {\em {Quantum Measurement}\/} (Cambridge
  University Press)

\bibitem{Caves:1980a}
Caves C~M, Thorne K~S, Drever R~W~P, Sandberg V~D and Zimmermann M 1980 {\em
  Rev. Mod. Phys.\/} {\bf 52}(2) 341--392
  \urlprefix\url{http://link.aps.org/doi/10.1103/RevModPhys.52.341}

\bibitem{Braginsky:1996}
Braginsky V~B and Khalili F~Y 1996 {\em Rev. Mod. Phys.\/} {\bf 68}(1) 1--11
  \urlprefix\url{http://link.aps.org/doi/10.1103/RevModPhys.68.1}

\bibitem{Braunstein:1994}
Braunstein S and Caves C 1994 {\em Physical Review Letters\/} {\bf 72}
  3439--3443

\bibitem{Tsang:2011}
Tsang M, Wiseman H~M and Caves C~M 2011 {\em Phys. Rev. Lett.\/} {\bf 106}(9)
  090401 \urlprefix\url{http://link.aps.org/doi/10.1103/PhysRevLett.106.090401}

\bibitem{Tsang:2012}
Tsang M 2012 {\em Phys. Rev. Lett.\/} {\bf 108}(17) 170502
  \urlprefix\url{http://link.aps.org/doi/10.1103/PhysRevLett.108.170502}

\bibitem{Tsang:2012c}
Tsang M and Caves C~M 2012 {\em Phys. Rev. X\/} {\bf 2}(3) 031016
  \urlprefix\url{http://link.aps.org/doi/10.1103/PhysRevX.2.031016}

\bibitem{Tsang:2012d}
Tsang M 2012 {\em Physical Review Letters\/} {\bf 108} 230401

\bibitem{Danilishin:2012a}
Danilishin S~L and Khalili F~Y 2012 {\em Living Reviews in Relativity\/} {\bf
  15} \urlprefix\url{http://www.livingreviews.org/lrr-2012-5}

\bibitem{Chen:2009}
Chen Y, Danilishin S, Khalili F and M{\"u}ller-Ebhardt H 2009 {\em arXiv
  preprint arXiv:0910.0319\/}

\bibitem{Punturo:2010}
Punturo M, Abernathy M, Acernese F, Allen B, Andersson N, Arun K, Barone F,
  Barr B, Barsuglia M, Beker M {\em et~al.\/} 2010 {\em Classical and Quantum
  Gravity\/} {\bf 27} 194002

\bibitem{Miao:2012c}
Miao H, Yang H, Adhikari R~X and Chen Y 2012 {\em LIGO Document T-1200008\/}

\bibitem{Feynman:1963}
Feynman R and Vernon F 1963 {\em Annals of Physics\/} {\bf 24} 118 -- 173 ISSN
  0003-4916
  \urlprefix\url{http://www.sciencedirect.com/science/article/pii/000349166390068X}

\bibitem{Keldysh:1964}
Keldysh L 1964 {\em Zh.\ Eksp.\ Teor.\ Fiz\/}

\bibitem{Caldeira:1983}
Caldeira A and Leggett A 1983 {\em Physica A: Statistical Mechanics and its
  Applications\/} {\bf 121} 587 -- 616 ISSN 0378-4371
  \urlprefix\url{http://www.sciencedirect.com/science/article/pii/0378437183900134}

\bibitem{Weiss:1999}
Weiss U 1999 {\em Quantum dissipative systems\/} vol~10 (World Scientific
  Publishing Company Incorporated)

\bibitem{Davies:1969}
{Davies} E~B 1969 {\em Communications in Mathematical Physics\/} {\bf 15}
  277--304

\bibitem{Davies:1970}
{Davies} E~B 1970 {\em Communications in Mathematical Physics\/} {\bf 19}
  83--105

\bibitem{Barchielli:1991}
Barchielli A and Belavkin V~P 1991 {\em Journal of Physics A: Mathematical and
  General\/} {\bf 24} 1495

\bibitem{Caves:1986}
Caves C~M 1986 {\em Phys. Rev. D\/} {\bf 33}(6) 1643--1665
  \urlprefix\url{http://link.aps.org/doi/10.1103/PhysRevD.33.1643}

\bibitem{Caves:1987a}
Caves C~M 1987 {\em Phys. Rev. D\/} {\bf 35}(6) 1815--1830
  \urlprefix\url{http://link.aps.org/doi/10.1103/PhysRevD.35.1815}

\bibitem{Caves:1987b}
Caves C~M and Milburn G~J 1987 {\em Phys. Rev. A\/} {\bf 36}(12) 5543--5555
  \urlprefix\url{http://link.aps.org/doi/10.1103/PhysRevA.36.5543}

\bibitem{Carmichael:1991}
Carmichael H~J 1991 {\em An Open Systems Approach to Quantum Optics\/}
  (Springer-Verlag)

\bibitem{Carmichael:1993}
Carmichael H 1993 {\em An open systems approach to Quantum Optics: lectures
  presented at the Universit{\'e} Libre de Bruxelles, October 28 to November 4,
  1991\/} vol~18 (Springer)

\bibitem{Strunz:1999}
Strunz W, Di{\'o}si L and Gisin N 1999 {\em Physical review letters\/} {\bf 82}
  1801--1805

\bibitem{Strunz:1999b}
Strunz W, Di{\'o}si L, Gisin N and Yu T 1999 {\em Physical Review Letters\/}
  {\bf 83} 4909--4913

\bibitem{Wiseman:2010}
Wiseman H~M and Milburn G~J 2010 {\em Quantum Measurement and Control\/}
  (Cambridge University Press)

\bibitem{Clerk:2010b}
Clerk A~A, Devoret M~H, Girvin S~M, Marquardt F and Schoelkopf R~J 2010 {\em
  Rev. Mod. Phys.\/} {\bf 82}(2) 1155--1208
  \urlprefix\url{http://link.aps.org/doi/10.1103/RevModPhys.82.1155}

\bibitem{Law:1995}
Law C~K 1995 {\em Phys. Rev. A\/} {\bf 51}(3) 2537--2541
  \urlprefix\url{http://link.aps.org/doi/10.1103/PhysRevA.51.2537}

\bibitem{Caves:1985}
Caves C~M and Schumaker B~L 1985 {\em Phys. Rev. A\/} {\bf 31}(5) 3068--3092
  \urlprefix\url{http://link.aps.org/doi/10.1103/PhysRevA.31.3068}

\bibitem{Schumaker:1985}
Schumaker B~L and Caves C~M 1985 {\em Phys. Rev. A\/} {\bf 31}(5) 3093--3111
  \urlprefix\url{http://link.aps.org/doi/10.1103/PhysRevA.31.3093}

\bibitem{Kimble:2001}
Kimble H~J, Levin Y, Matsko A~B, Thorne K~S and Vyatchanin S~P 2001 {\em Phys.
  Rev. D\/} {\bf 65}(2) 022002
  \urlprefix\url{http://link.aps.org/doi/10.1103/PhysRevD.65.022002}

\bibitem{Harry:2010}
Harry G {\em et~al.\/} 2010 {\em Classical and Quantum Gravity\/} {\bf 27}
  084006

\bibitem{Tsang:2012b}
Tsang M and Nair R 2012 {\em Phys. Rev. A\/} {\bf 86}(4) 042115
  \urlprefix\url{http://link.aps.org/doi/10.1103/PhysRevA.86.042115}

\bibitem{chen:2003}
Chen Y 2003 {\em Topics of LIGO Physics: Quantum Noise in Advanced
  Interferometers and Template Banks for Compact-Binary Inspirals\/} Ph.D.
  thesis California Institute of Technology
  \urlprefix\url{http://resolver.caltech.edu/CaltechETD:etd-05302003-044325}

\bibitem{Thorne:1978}
Thorne K~S, Drever R~W~P, Caves C~M, Zimmermann M and Sandberg V~D 1978 {\em
  Phys. Rev. Lett.\/} {\bf 40}(11) 667--671
  \urlprefix\url{http://link.aps.org/doi/10.1103/PhysRevLett.40.667}

\bibitem{Braginsky:1980}
Braginsky V, Vorontsov Y and Kip S 1980 {\em SCIENCE\/} {\bf 209} 1

\bibitem{Bocko:1996}
Bocko M and Onofrio R 1996 {\em Reviews of Modern Physics\/} {\bf 68} 755

\bibitem{Yuen:1983}
Yuen H 1983 {\em Physical Review Letters\/} {\bf 51} 719--722

\bibitem{Ozawa:1988}
Ozawa M 1988 {\em Physical review letters\/} {\bf 60} 385--388

\bibitem{Khalili:2012}
Khalili F~Y, Miao H, Yang H, Safavi-Naeini A~H, Painter O and Chen Y 2012 {\em
  Phys. Rev. A\/} {\bf 86}(3) 033840
  \urlprefix\url{http://link.aps.org/doi/10.1103/PhysRevA.86.033840}

\bibitem{Braginsky:2003}
Braginsky V, Gorodetsky M, Khalili F, Matsko A, Thorne K and Vyatchanin S 2003
  {\em Physical Review D\/} {\bf 67} 082001

\bibitem{Abramovici:1992}
Abramovici A, Althouse W, Drever R, G{\"u}rsel Y, Kawamura S, Raab F, Shoemaker
  D, Sievers L, Spero R, Thorne K {\em et~al.\/} 1992 {\em Science (New York,
  NY)\/} {\bf 256} 325

\bibitem{Accadia:2011}
Accadia T, Acernese F, Antonucci F, Astone P, Ballardin G, Barone F, Barsuglia
  M, Basti A, Bauer T, Bebronne M {\em et~al.\/} 2011 {\em Classical and
  Quantum Gravity\/} {\bf 28} 114002

\bibitem{Somiya:2012}
Somiya K 2012 {\em Classical and Quantum Gravity\/} {\bf 29} 124007

\bibitem{Abbott:2004}
Abbott R, Adhikari R, Allen G, Baglino D, Campbell C, Coyne D, Daw E, DeBra D,
  Faludi J, Fritschel P {\em et~al.\/} 2004 {\em Classical and Quantum
  Gravity\/} {\bf 21} S915

\bibitem{Driggers:2012}
Driggers J, Harms J and Adhikari R 2012 {\em arXiv preprint arXiv:1207.0275\/}

\bibitem{Gonzalez:1994}
Gonz{\'a}lez G and Saulson P 1994 {\em The Journal of the Acoustical Society of
  America\/} {\bf 96} 207

\bibitem{Nawrodt:2011}
Nawrodt R, Rowan S, Hough J, Punturo M, Ricci F and Vinet J 2011 {\em General
  Relativity and Gravitation\/} {\bf 43} 593--622

\bibitem{Levin:1998}
Levin Y 1998 {\em Physical Review D\/} {\bf 57} 659

\bibitem{Braginsky:1999b}
Braginsky V, Gorodetsky M and Vyatchanin S 1999 {\em Physics Letters A\/} {\bf
  264} 1--10

\bibitem{Kondratiev:2011}
Kondratiev N, Gurkovsky A and Gorodetsky M 2011 {\em Physical Review D\/} {\bf
  84} 022001

\bibitem{Hong:2012}
Hong T, Yang H, Gustafson E, Adhikari R and Chen Y 2012 {\em arXiv preprint
  arXiv:1207.6145\/}

\bibitem{Kwee:2012}
Kwee P, Bogan C, Danzmann K, Frede M, Kim H, King P, P{\"o}ld J, Puncken O,
  Savage R, Seifert F {\em et~al.\/} 2012 {\em Optics Express\/} {\bf 20}
  10617--10634

\bibitem{Evans:2010}
Evans M, Barsotti L and Fritschel P 2010 {\em Physics Letters A\/} {\bf 374}
  665--671

\bibitem{Adhikari:2012}
Adhikari R~X 2012 {\em Review of modern physics\/}

\bibitem{Caves:1981}
Caves C 1981 {\em Physical Review D\/} {\bf 23} 1693

\bibitem{Unruh:1982}
Unruh W {\em Experimental Gravitation, and Measurement Theory, {\'e}dit{\'e}
  par P. Meystre et MO Scully (Plenum, New-York, 1982)\/}  647

\bibitem{Jaekel:1990}
Jaekel M~T and Reynaud S 1990 {\em EPL (Europhysics Letters)\/} {\bf 13} 301
  \urlprefix\url{http://stacks.iop.org/0295-5075/13/i=4/a=003}

\bibitem{Vyatchanin:1993}
Vyatchanin S and Matsko A 1993 {\em Soviet Journal of Experimental and
  Theoretical Physics\/} {\bf 77} 218--221

\bibitem{Vyatchanin:1995}
Vyatchanin S and Zubova E 1995 {\em Physics letters A\/} {\bf 201} 269--274

\bibitem{Tsang:2010}
Tsang M and Caves C 2010 {\em Physical review letters\/} {\bf 105} 123601

\bibitem{Purdue:2002a}
Purdue P 2002 {\em Phys. Rev. D\/} {\bf 66}(2) 022001
  \urlprefix\url{http://link.aps.org/doi/10.1103/PhysRevD.66.022001}

\bibitem{Purdue:2002b}
Purdue P and Chen Y 2002 {\em Phys. Rev. D\/} {\bf 66}(12) 122004
  \urlprefix\url{http://link.aps.org/doi/10.1103/PhysRevD.66.122004}

\bibitem{Chen:2003a}
Chen Y 2003 {\em Phys. Rev. D\/} {\bf 67}(12) 122004
  \urlprefix\url{http://link.aps.org/doi/10.1103/PhysRevD.67.122004}

\bibitem{Danilishin:2004}
Danilishin S 2004 {\em Physical Review D\/} {\bf 69} 102003

\bibitem{Braginsky:1990}
Braginsky V and Khalili F 1990 {\em Physics Letters A\/} {\bf 147} 251 -- 256
  ISSN 0375-9601
  \urlprefix\url{http://www.sciencedirect.com/science/article/pii/037596019090442Q}

\bibitem{Xiao:1987}
Xiao M, Wu L and Kimble H 1987 {\em Physical review letters\/} {\bf 59}
  278--281

\bibitem{Cutler:2002}
Cutler C and Thorne K 2002 An overview of gravitational-wave sources {\em
  Proceedings of the GR16 Conference on General Relativity and Gravitation, ed.
  N. Bishop and SD Maharaj (World Scientific, 2002)\/} pp 72--111

\bibitem{McKenzie:2002}
McKenzie K, Shaddock D, McClelland D, Buchler B and Lam P 2002 {\em Physical
  review letters\/} {\bf 88} 231102

\bibitem{Vahlbruch:2006}
Vahlbruch H, Chelkowski S, Hage B, Franzen A, Danzmann K and Schnabel R 2006
  {\em Physical review letters\/} {\bf 97} 11101

\bibitem{Goda:2008}
Goda K, Miyakawa O, Mikhailov E, Saraf S, Adhikari R, McKenzie K, Ward R, Vass
  S, Weinstein A and Mavalvala N 2008 {\em Nature Physics\/} {\bf 4} 472--476

\bibitem{Abadie:2011}
Abadie J, Abbott B, Abbott R, Abbott T, Abernathy M, Adams C, Adhikari R,
  Affeldt C, Allen B, Allen G {\em et~al.\/} 2011 {\em Nature Physics\/} {\bf
  7} 962--965

\bibitem{Chelkowski:2005}
Chelkowski S, Vahlbruch H, Hage B, Franzen A, Lastzka N, Danzmann K and
  Schnabel R 2005 {\em Physical Review A\/} {\bf 71} 013806

\bibitem{Dwyer:2012}
Dwyer S 2012 {\em Private Communications\/}

\bibitem{Schnabel:2008}
Schnabel R 2008 {\em Nature Physics\/} {\bf 4} 440--441

\bibitem{Schnabel:2010}
Schnabel R, Mavalvala N, McClelland D and Lam P 2010 {\em Nature
  communications\/} {\bf 1} 121

\bibitem{Sheon:2011}
SHEON C, Stefszky M, Mmow-Lowry C, Buchler B, Mckenzie K, DANIEL A, PING K and
  DAVID E 2011 {\em International Journal of Modern Physics D\/} {\bf 20}
  2043--2049

\bibitem{McClelland:2011}
McClelland D, Mavalvala N, Chen Y and Schnabel R 2011 {\em Laser \& Photonics
  Reviews\/} {\bf 5} 677--696

\bibitem{Braginsky:1978}
Braginsky V, Vorontsov Y and Khalili F 1978 {\em Sov.\ Phys.\ JETP Lett.\/}
  {\bf 33}

\bibitem{Hertzberg:2009}
Hertzberg J, Rocheleau T, Ndukum T, Savva M, Clerk A and Schwab K 2009 {\em
  Nature Physics\/} {\bf 6} 213--217

\bibitem{Yang:2012b}
Yang H, Miao H and Chen Y 2012 {\em Phys. Rev. A\/} {\bf 85}(4) 040101
  \urlprefix\url{http://link.aps.org/doi/10.1103/PhysRevA.85.040101}

\bibitem{Buonanno:2001}
Buonanno A and Chen Y 2001 {\em Classical and Quantum Gravity\/} {\bf 18} L95

\bibitem{Buonanno:2001a}
Buonanno A and Chen Y 2001 {\em Phys. Rev. D\/} {\bf 64}(4) 042006
  \urlprefix\url{http://link.aps.org/doi/10.1103/PhysRevD.64.042006}

\bibitem{Buonanno:2002a}
Buonanno A and Chen Y 2002 {\em Phys. Rev. D\/} {\bf 65}(4) 042001
  \urlprefix\url{http://link.aps.org/doi/10.1103/PhysRevD.65.042001}

\bibitem{Buonanno:2003a}
Buonanno A and Chen Y 2003 {\em Phys. Rev. D\/} {\bf 67}(6) 062002
  \urlprefix\url{http://link.aps.org/doi/10.1103/PhysRevD.67.062002}

\bibitem{Tarabrin:2012}
Tarabrin S, Khalili F, Kaufer H, Schnabel R and Hammerer K 2012 {\em arXiv
  preprint arXiv:1212.6242\/}

\bibitem{Korth:2012}
Korth W, Miao H, Corbitt T, Cole G, Chen Y and Adhikari R 2012 {\em arXiv
  preprint arXiv:1210.0309\/}

\bibitem{Mueller-Ebhardt:2008a}
Mueller-Ebhardt H 2008 Ph.D. thesis University of Hannover

\bibitem{Khalili:2011a}
Khalili F, Danilishin S, M\"uller-Ebhardt H, Miao H, Chen Y and Zhao C 2011
  {\em Phys. Rev. D\/} {\bf 83}(6) 062003
  \urlprefix\url{http://link.aps.org/doi/10.1103/PhysRevD.83.062003}

\bibitem{Braginskii:1967}
Braginsky V~B and Manukin A~B 1967 {\em JETP\/} {\bf 25}

\bibitem{Braginskii:1970}
Braginsky V~B, Manukin A~B and Tikhonov M~Y 1970 {\em JETP\/} {\bf 31}

\bibitem{Hirakawa:1979}
Hirakawa H, Tsubono K and Oide K 1979 {\em Japanese Journal of Applied
  Physics\/} {\bf 18} 177--180
  \urlprefix\url{http://jjap.jsap.jp/link?JJAP/18/177/}

\bibitem{Linthorne:1990}
Linthorne N~P, Veitch P~J and Blair D~G 1990 {\em Journal of Physics D: Applied
  Physics\/} {\bf 23} 1
  \urlprefix\url{http://stacks.iop.org/0022-3727/23/i=1/a=001}

\bibitem{Tobar:1993}
Tobar M~E and Blair D~G 1993 {\em Journal of Physics D: Applied Physics\/} {\bf
  26} 2276 \urlprefix\url{http://stacks.iop.org/0022-3727/26/i=12/a=028}

\bibitem{Blair:1995}
Blair D~G, Ivanov E~N, Tobar M~E, Turner P~J, van Kann F and Heng I~S 1995 {\em
  Phys. Rev. Lett.\/} {\bf 74}(11) 1908--1911
  \urlprefix\url{http://link.aps.org/doi/10.1103/PhysRevLett.74.1908}

\bibitem{Cuthbertson:1996}
Cuthbertson B~D, Tobar M~E, Ivanov E~N and Blair D~G 1996 {\em Review of
  Scientific Instruments\/} {\bf 67} 2435--2442

\bibitem{Sheard:2004}
Sheard B, Gray M, Mow-Lowry C, McClelland D and Whitcomb S 2004 {\em Physical
  Review A\/} {\bf 69} 051801

\bibitem{Somiya:2005}
Somiya K, Beyersdorf P, Arai K, Sato S, Kawamura S, Miyakawa O, Kawazoe F,
  Sakata S, Sekido A and Mio N 2005 {\em Applied optics\/} {\bf 44} 3179--3191

\bibitem{Miyakawa:2006}
Miyakawa O, Ward R, Adhikari R, Evans M, Abbott B, Bork R, Busby D, Heefner J,
  Ivanov A, Smith M {\em et~al.\/} 2006 {\em Physical Review D\/} {\bf 74}
  022001

\bibitem{Hopkins:2003}
Hopkins A, Jacobs K, Habib S and Schwab K 2003 {\em Phys. Rev. B\/} {\bf
  68}(23) 235328
  \urlprefix\url{http://link.aps.org/doi/10.1103/PhysRevB.68.235328}

\bibitem{Vanner:2011}
Vanner M, Pikovski I, Cole G, Kim M, Brukner {\v{C}}, Hammerer K, Milburn G and
  Aspelmeyer M 2011 {\em Proceedings of the National Academy of Sciences\/}
  {\bf 108} 16182--16187

\bibitem{Machnes:2012}
Machnes S, Cerrillo J, Aspelmeyer M, Wieczorek W, Plenio M and Retzker A 2012
  {\em Physical Review Letters\/} {\bf 108} 153601

\bibitem{Hofer:2011}
Hofer S, Wieczorek W, Aspelmeyer M and Hammerer K 2011 {\em Physical Review
  A\/} {\bf 84} 052327

\bibitem{Phillips:1998}
Phillips W~D 1998 {\em Reviews of Modern Physics\/}

\bibitem{Vyatchanin:1977}
{Vyatchanin} S~P 1977 {\em Soviet Physics Doklady\/} {\bf 22} 321

\bibitem{Hirakawa:1977}
Hirakawa H, Hiramatsu S and Ogawa Y 1977 {\em Physics Letters A\/} {\bf 63} 199
  -- 202 ISSN 0375-9601
  \urlprefix\url{http://www.sciencedirect.com/science/article/pii/0375960177908738}

\bibitem{Hirakawa:1978}
Hirakawa H, Oide K and Ogawa Y 1978 {\em Journal of the Physical Society of
  Japan\/} {\bf 44} 337

\bibitem{Oide:1978}
Oide K, Ogawa Y and Hirakawa H 1978 {\em Japanese Journal of Applied Physics\/}
  {\bf 17} 429--432

\bibitem{Oide:1979}
Oide K, Hirakawa H and Fujimoto M~K 1979 {\em Phys. Rev. D\/} {\bf 20}(10)
  2480--2483 \urlprefix\url{http://link.aps.org/doi/10.1103/PhysRevD.20.2480}

\bibitem{Marquardt:2007}
Marquardt F, Chen J~P, Clerk A~A and Girvin S~M 2007 {\em Phys. Rev. Lett.\/}
  {\bf 99}(9) 093902
  \urlprefix\url{http://link.aps.org/doi/10.1103/PhysRevLett.99.093902}

\bibitem{Wilson-Rae:2007}
Wilson-Rae I, Nooshi N, Zwerger W and Kippenberg T~J 2007 {\em Phys. Rev.
  Lett.\/} {\bf 99}(9) 093901
  \urlprefix\url{http://link.aps.org/doi/10.1103/PhysRevLett.99.093901}

\bibitem{Schliesser:2006}
Schliesser A, Del'Haye P, Nooshi N, Vahala K~J and Kippenberg T~J 2006 {\em
  Phys. Rev. Lett.\/} {\bf 97}(24) 243905
  \urlprefix\url{http://link.aps.org/doi/10.1103/PhysRevLett.97.243905}

\bibitem{Arcizet:2006b}
Arcizet O, Cohadon P, Briant T, Pinard M and Heidmann A 2006 {\em Nature\/}
  {\bf 444} 71--74

\bibitem{Gigan:2006}
Gigan S, B{\"o}hm H, Paternostro M, Blaser F, Langer G, Hertzberg J, Schwab K,
  B{\"a}uerle D, Aspelmeyer M and Zeilinger A 2006 {\em Nature\/} {\bf 444}
  67--70

\bibitem{Teufel:2008}
Teufel J, Harlow J, Regal C and Lehnert K 2008 {\em Physical review letters\/}
  {\bf 101} 197203

\bibitem{Groblacher:2009}
Gr{\"o}blacher S, Hammerer K, Vanner M and Aspelmeyer M 2009 {\em Nature\/}
  {\bf 460} 724--727

\bibitem{Groblacher:2009b}
Gr{\"o}blacher S, Hertzberg J, Vanner M, Cole G, Gigan S, Schwab K and
  Aspelmeyer M 2009 {\em Nature Physics\/} {\bf 5} 485--488

\bibitem{Rocheleau:2009}
Rocheleau T, Ndukum T, Macklin C, Hertzberg J, Clerk A and Schwab K 2009 {\em
  Nature\/} {\bf 463} 72--75

\bibitem{Schliesser:2008}
Schliesser A, Rivi{\`e}re R, Anetsberger G, Arcizet O and Kippenberg T 2008
  {\em Nature Physics\/} {\bf 4} 415--419

\bibitem{Schliesser:2009}
Schliesser A, Arcizet O, Rivi{\`e}re R, Anetsberger G and Kippenberg T 2009
  {\em Nature Physics\/} {\bf 5} 509--514

\bibitem{Teufel:2011}
Teufel J, Donner T, Li D, Harlow J, Allman M, Cicak K, Sirois A, Whittaker J,
  Lehnert K and Simmonds R 2011 {\em Nature\/} {\bf 475} 359--363

\bibitem{Chan:2011}
Chan J, Alegre T, Safavi-Naeini A, Hill J, Krause A, Gr{\"o}blacher S,
  Aspelmeyer M and Painter O 2011 {\em Nature\/} {\bf 478} 89--92

\bibitem{Riviere:2011}
Riviere R, Deleglise S, Weis S, Gavartin E, Arcizet O, Schliesser A and
  Kippenberg T 2011 {\em Physical Review A\/} {\bf 83} 063835

\bibitem{Callen:1951}
Callen H and Welton T 1951 {\em Physical Review\/} {\bf 83} 34--40

\bibitem{Miao:2010d}
Miao H, Danilishin S, M{\"u}ller-Ebhardt H and Chen Y 2010 {\em New Journal of
  Physics\/} {\bf 12} 083032

\bibitem{Elste:2009}
Elste F, Girvin S and Clerk A 2009 {\em Physical review letters\/} {\bf 102}
  207209

\bibitem{Xuereb:2011}
Xuereb A, Schnabel R and Hammerer K 2011 {\em Physical Review Letters\/} {\bf
  107} 213604

\bibitem{Jahne:2009}
J{\"a}hne K, Genes C, Hammerer K, Wallquist M, Polzik E and Zoller P 2009 {\em
  Physical Review A\/} {\bf 79} 063819

\bibitem{Xuereb:2012}
Xuereb A, Usami K, Naesby A, Polzik E and Hammerer K 2012 {\em New Journal of
  Physics\/} {\bf 14} 085024

\bibitem{Usami:2012}
Usami K, Naesby A, Bagci T, Nielsen B, Liu J, Stobbe S, Lodahl P and Polzik E
  2012 {\em Nature Physics\/} {\bf 8} 168--172

\bibitem{Jahne:2008}
Jaehne K, Hammerer K and Wallquist M 2008 {\em New Journal of Physics\/} {\bf
  10} 095019

\bibitem{Braginsky:1999}
Braginsky V and Khalili F 1999 {\em Physics Letters A\/} {\bf 257} 241 -- 246
  ISSN 0375-9601
  \urlprefix\url{http://www.sciencedirect.com/science/article/pii/S0375960199003370}

\bibitem{Bhattacharya:2007}
Bhattacharya M and Meystre P 2007 {\em Physical review letters\/} {\bf 99}
  73601

\bibitem{Corbitt:2007}
Corbitt T, Chen Y, Innerhofer E, M\"uller-Ebhardt H, Ottaway D, Rehbein H, Sigg
  D, Whitcomb S, Wipf C and Mavalvala N 2007 {\em Phys. Rev. Lett.\/} {\bf
  98}(15) 150802
  \urlprefix\url{http://link.aps.org/doi/10.1103/PhysRevLett.98.150802}

\bibitem{Bhattacharya:2007b}
Bhattacharya M and Meystre P 2007 {\em Physical review letters\/} {\bf 99}
  153603

\bibitem{Bhattacharya:2008}
Bhattacharya M, Uys H and Meystre P 2008 {\em Physical Review A\/} {\bf 77}
  033819

\bibitem{Corbitt:2007b}
Corbitt T, Wipf C, Bodiya T, Ottaway D, Sigg D, Smith N, Whitcomb S and
  Mavalvala N 2007 {\em Phys. Rev. Lett.\/} {\bf 99}(16) 160801
  \urlprefix\url{http://link.aps.org/doi/10.1103/PhysRevLett.99.160801}

\bibitem{Romero-Isart:2010}
Romero-Isart O, Juan M, Quidant R and Cirac J 2010 {\em New Journal of
  Physics\/} {\bf 12} 033015

\bibitem{Pflanzer:2012}
Pflanzer A, Romero-Isart O and Cirac J 2012 {\em Physical Review A\/} {\bf 86}
  013802

\bibitem{Chang:2010}
Chang D, Regal C, Papp S, Wilson D, Ye J, Painter O, Kimble H and Zoller P 2010
  {\em Proceedings of the National Academy of Sciences\/} {\bf 107} 1005--1010

\bibitem{Singh:2010}
Singh S, Phelps G, Goldbaum D, Wright E and Meystre P 2010 {\em Physical review
  letters\/} {\bf 105} 213602

\bibitem{Pender:2012}
Pender G, Barker P, Marquardt F, Millen J and Monteiro T 2012 {\em Physical
  Review A\/} {\bf 85} 021802

\bibitem{Monteiro:2012}
Monteiro T, Millen J, Pender G, Marquardt F, Chang D and Barker P 2012 {\em
  arXiv preprint arXiv:1207.1567\/}

\bibitem{Chang:2012}
Chang D, Ni K, Painter O and Kimble H 2012 {\em New Journal of Physics\/} {\bf
  14} 045002

\bibitem{Lechner:2012}
Lechner W, Habraken S, Kiesel N, Aspelmeyer M and Zoller P 2012 {\em arXiv
  preprint arXiv:1212.4691\/}

\bibitem{Rehbein:2008}
Rehbein H, M\"uller-Ebhardt H, Somiya K, Danilishin S~L, Schnabel R, Danzmann K
  and Chen Y 2008 {\em Phys. Rev. D\/} {\bf 78}(6) 062003
  \urlprefix\url{http://link.aps.org/doi/10.1103/PhysRevD.78.062003}

\bibitem{Clerk:2008}
Clerk A, Marquardt F and Jacobs K 2008 {\em New Journal of Physics\/} {\bf 10}
  095010

\bibitem{Abbott:2009}
Abbott B, Abbott R, Adhikari R, Ajith P, Allen B, Allen G, Amin R, Anderson S,
  Anderson W, Arain M {\em et~al.\/} 2009 {\em New Journal of Physics\/} {\bf
  11} 073032

\bibitem{Mueller-Ebhardt:2009}
M\"uller-Ebhardt H, Rehbein H, Li C, Mino Y, Somiya K, Schnabel R, Danzmann K
  and Chen Y 2009 {\em Phys. Rev. A\/} {\bf 80}(4) 043802
  \urlprefix\url{http://link.aps.org/doi/10.1103/PhysRevA.80.043802}

\bibitem{Mueller-Ebhardt:2008b}
M\"uller-Ebhardt H, Rehbein H, Schnabel R, Danzmann K and Chen Y 2008 {\em
  Phys. Rev. Lett.\/} {\bf 100}(1) 013601
  \urlprefix\url{http://link.aps.org/doi/10.1103/PhysRevLett.100.013601}

\bibitem{Vitali:1998}
Mancini S, Vitali D and Tombesi P 1998 {\em Phys. Rev. Lett.\/} {\bf 80}(4)
  688--691 \urlprefix\url{http://link.aps.org/doi/10.1103/PhysRevLett.80.688}

\bibitem{Courty:2001}
Courty J~M, Heidmann A and Pinard M 2001 {\em The European Physical Journal D -
  Atomic, Molecular, Optical and Plasma Physics\/} {\bf 17}(3) 399--408 ISSN
  1434-6060 10.1007/s100530170014
  \urlprefix\url{http://dx.doi.org/10.1007/s100530170014}

\bibitem{Genes:2008}
Genes C, Vitali D, Tombesi P, Gigan S and Aspelmeyer M 2008 {\em Phys. Rev.
  A\/} {\bf 77}(3) 033804
  \urlprefix\url{http://link.aps.org/doi/10.1103/PhysRevA.77.033804}

\bibitem{Mow-Lowry:2008}
Mow-Lowry C, Mullavey A, Gossler S, Gray M and McClelland D 2008 {\em Physical
  review letters\/} {\bf 100} 10801

\bibitem{Arcizet:2006}
Arcizet O, Cohadon P~F, Briant T, Pinard M, Heidmann A, Mackowski J~M, Michel
  C, Pinard L, Fran\ifmmode~\mbox{\c{c}}\else \c{c}\fi{}ais O and Rousseau L
  2006 {\em Phys. Rev. Lett.\/} {\bf 97}(13) 133601
  \urlprefix\url{http://link.aps.org/doi/10.1103/PhysRevLett.97.133601}

\bibitem{Doherty:2000}
Doherty A~C, Habib S, Jacobs K, Mabuchi H and Tan S~M 2000 {\em Phys. Rev. A\/}
  {\bf 62}(1) 012105
  \urlprefix\url{http://link.aps.org/doi/10.1103/PhysRevA.62.012105}

\bibitem{Danilishin:2008}
Danilishin S~L, Mueller-Ebhardt H, Rehbein H, Somiya K, Schnabel R, Danzmann K,
  Corbitt T, Wipf C, Mavalvala N and Chen Y 2008 {\em arXiv:0809.2024\/}

\bibitem{Doherty:2012}
Doherty A~C, Szorkovszky A, Harris G~I and Bowen W~P 2012 {\em Philosophical
  Transactions of the Royal Society A: Mathematical, Physical and Engineering
  Sciences\/} {\bf 370} 5338--5353 (\textit{Preprint}
  \eprint{http://rsta.royalsocietypublishing.org/content/370/1979/5338.full.pdf+html})
  \urlprefix\url{http://rsta.royalsocietypublishing.org/content/370/1979/5338.abstract}

\bibitem{Szorkovszky:2011}
Szorkovszky A, Doherty A, Harris G and Bowen W 2011 {\em Physical review
  letters\/} {\bf 107} 213603

\bibitem{Ruskov:2005}
Ruskov R, Schwab K and Korotkov A 2005 {\em Physical Review B\/} {\bf 71}
  235407

\bibitem{Ruskov:2005b}
Ruskov R, Schwab K and Korotkov A 2005 {\em Nanotechnology, IEEE Transactions
  on\/} {\bf 4} 132--140

\bibitem{Lloyd:2000}
Lloyd S 2000 {\em Phys. Rev. A\/} {\bf 62}(2) 022108
  \urlprefix\url{http://link.aps.org/doi/10.1103/PhysRevA.62.022108}

\bibitem{Nurdin:2009}
Nurdin H, James M and Petersen I 2009 {\em Automatica\/} {\bf 45} 1837--1846

\bibitem{Hamerly:2012a}
Hamerly R and Mabuchi H 2012 {\em arXiv preprint arXiv:1206.0829\/}

\bibitem{Hamerly:2012b}
Hamerly R and Mabuchi H 2012 {\em arXiv preprint arXiv:1206.2688\/}

\bibitem{Kerckhoff:2012}
Kerckhoff J, Andrews R, Ku H, Kindel W, Cicak K, Simmonds R and Lehnert K 2012
  {\em arXiv preprint arXiv:1211.1950\/}

\bibitem{Jacobs:2012}
Jacobs K and Wang X 2012 {\em arXiv preprint arXiv:1211.1724\/}

\bibitem{Miao:2010b}
Miao H, Danilishin S, M{\"u}ller-Ebhardt H, Rehbein H, Somiya K and Chen Y 2010
  {\em Physical Review A\/} {\bf 81} 012114

\bibitem{Mari:2011}
Mari A, Kieling K, Nielsen B, Polzik E and Eisert J 2011 {\em Physical review
  letters\/} {\bf 106} 10403

\bibitem{Lvovsky:2009}
Lvovsky A and Raymer M 2009 {\em Reviews of Modern Physics\/} {\bf 81} 299

\bibitem{Duan:2000}
Duan L, Giedke G, Cirac J and Zoller P 2000 {\em Physical Review Letters\/}
  {\bf 84} 2722--2725

\bibitem{Simon:2000}
Simon R 2000 {\em Physical Review Letters\/} {\bf 84} 2726--2729

\bibitem{Pace:1993}
Pace A, Collett M and Walls D 1993 {\em Physical Review A\/} {\bf 47} 3173

\bibitem{Woolley:2008}
Woolley M, Doherty A, Milburn G and Schwab K 2008 {\em Physical Review A\/}
  {\bf 78} 062303

\bibitem{Mow-Lowry:2004}
Mow-Lowry C, Sheard B, Gray M, McClelland D and Whitcomb S 2004 {\em Physical
  review letters\/} {\bf 92} 161102

\bibitem{Briant:2009}
Briant T, Verlot P, Tavernarakis A, Cohadon P and Heidmann A 2009 Quantum
  optomechanical correlations induced by radiation pressure between light and
  mirrors {\em Proc. of SPIE Vol\/} vol 7225 pp 72250M--1

\bibitem{Marino:2010}
Marino F, Cataliotti F~S, Farsi A, de~Cumis M~S and Marin F 2010 {\em Phys.
  Rev. Lett.\/} {\bf 104}(7) 073601
  \urlprefix\url{http://link.aps.org/doi/10.1103/PhysRevLett.104.073601}

\bibitem{Brooks:2011}
Brooks D, Botter T, Brahms N, Purdy T, Schreppler S and Stamper-Kurn D 2011
  {\em arXiv preprint arXiv:1107.5609\/}

\bibitem{Nation:2012}
Nation P, Johansson J, Blencowe M and Nori F 2012 {\em Reviews of Modern
  Physics\/} {\bf 84} 1

\bibitem{Wilson:2011}
Wilson C, Johansson G, Pourkabirian A, Simoen M, Johansson J, Duty T, Nori F
  and Delsing P 2011 {\em Nature\/} {\bf 479} 376--379

\bibitem{Harms:2003}
Harms J, Chen Y, Chelkowski S, Franzen A, Vahlbruch H, Danzmann K and Schnabel
  R 2003 {\em Physical Review D\/} {\bf 68} 042001

\bibitem{Corbitt:2006}
Corbitt T, Chen Y, Khalili F, Ottaway D, Vyatchanin S, Whitcomb S and Mavalvala
  N 2006 {\em Phys. Rev. A\/} {\bf 73}(2) 023801
  \urlprefix\url{http://link.aps.org/doi/10.1103/PhysRevA.73.023801}

\bibitem{Borkje:2010}
B{\o}rkje K, Nunnenkamp A, Zwickl B, Yang C, Harris J and Girvin S 2010 {\em
  Physical Review A\/} {\bf 82} 013818

\bibitem{Purdy:2012}
Purdy T, Peterson R and Regal C 2012 {\em arXiv preprint arXiv:1209.6334\/}

\bibitem{Pirandola:2003}
Pirandola S, Vitali D, Tombesi P and Lloyd S 2006 {\em Phys. Rev. Lett.\/} {\bf
  97}(15) 150403
  \urlprefix\url{http://link.aps.org/doi/10.1103/PhysRevLett.97.150403}

\bibitem{Wipf:2008}
Wipf C, Corbitt T, Chen Y and Mavalvala N 2008 {\em New Journal of Physics\/}
  {\bf 10} 095017
  \urlprefix\url{http://stacks.iop.org/1367-2630/10/i=9/a=095017}

\bibitem{Mikhailov:2006}
Mikhailov E, Goda K, Corbitt T and Mavalvala N 2006 {\em Physical Review A\/}
  {\bf 73} 053810

\bibitem{Weis:2010}
Weis S, Rivi\`ere R, Del\'eglise S, Gavartin E, Arcizet O, Schliesser A and
  Kippenberg T~J 2010 {\em Science\/} {\bf 330} 1520--1523 (\textit{Preprint}
  \eprint{http://www.sciencemag.org/content/330/6010/1520.full.pdf})
  \urlprefix\url{http://www.sciencemag.org/content/330/6010/1520.abstract}

\bibitem{Teufel:2011b}
Teufel J, Li D, Allman M, Cicak K, Sirois A, Whittaker J and Simmonds R 2011
  {\em Nature\/} {\bf 471} 204--208

\bibitem{Safavi-Naeini:2011}
Safavi-Naeini A, Alegre T, Chan J, Eichenfield M, Winger M, Lin Q, Hill J,
  Chang D and Painter O 2011 {\em Nature\/} {\bf 472} 69--73

\bibitem{Chang:2011}
Chang D, Safavi-Naeini A, Hafezi M and Painter O 2011 {\em New Journal of
  Physics\/} {\bf 13} 023003

\bibitem{Zhao:2012}
Zhao C and Ward R {\em Private communications.\/}

\bibitem{Brahms:2012}
Brahms N, Botter T, Schreppler S, Brooks D~W~C and Stamper-Kurn D~M 2012 {\em
  Phys. Rev. Lett.\/} {\bf 108}(13) 133601
  \urlprefix\url{http://link.aps.org/doi/10.1103/PhysRevLett.108.133601}

\bibitem{Safavi-Naeini:2012}
Safavi-Naeini A~H, Chan J, Hill J~T, Alegre T~P~M, Krause A and Painter O 2012
  {\em Phys. Rev. Lett.\/} {\bf 108}(3) 033602
  \urlprefix\url{http://link.aps.org/doi/10.1103/PhysRevLett.108.033602}

\bibitem{Safavi-Naeini:2012b}
Safavi-Naeini A, Chan J, Hill J, Groeblacher S, Miao H, Chen Y, Aspelmeyer M
  and Painter O 2012 {\em arXiv preprint arXiv:1210.2671\/}

\bibitem{Miao:2012}
Khalili F~Y, Miao H, Yang H, Safavi-Naeini A~H, Painter O and Chen Y 2012 {\em
  Phys. Rev. A\/} {\bf 86}(3) 033840
  \urlprefix\url{http://link.aps.org/doi/10.1103/PhysRevA.86.033840}

\bibitem{Miao:2010}
Miao H, Danilishin S and Chen Y 2010 {\em Phys. Rev. A\/} {\bf 81}(5) 052307
  \urlprefix\url{http://link.aps.org/doi/10.1103/PhysRevA.81.052307}

\bibitem{Paternostro:2007}
Paternostro M, Vitali D, Gigan S, Kim M~S, Brukner C, Eisert J and Aspelmeyer M
  2007 {\em Phys. Rev. Lett.\/} {\bf 99}(25) 250401
  \urlprefix\url{http://link.aps.org/doi/10.1103/PhysRevLett.99.250401}

\bibitem{Vitali:2007}
Vitali D, Gigan S, Ferreira A, B{\"o}hm H, Tombesi P, Guerreiro A, Vedral V,
  Zeilinger A and Aspelmeyer M 2007 {\em Physical review letters\/} {\bf 98}
  30405

\bibitem{Galve:2010}
Galve F, Pach{\'o}n L and Zueco D 2010 {\em Physical review letters\/} {\bf
  105} 180501

\bibitem{Giovannetti:2007}
Giovannetti V, Mancini S and Tombesi P 2007 {\em EPL (Europhysics Letters)\/}
  {\bf 54} 559

\bibitem{Bhattacharya:2008d}
Bhattacharya M, Giscard P and Meystre P 2008 {\em Physical Review A\/} {\bf 77}
  013827

\bibitem{Wiseman:2007}
Wiseman H~M, Jones S~J and Doherty A~C 2007 {\em Phys. Rev. Lett.\/} {\bf
  98}(14) 140402
  \urlprefix\url{http://link.aps.org/doi/10.1103/PhysRevLett.98.140402}

\bibitem{Wiseman:2012}
Wiseman H~M and Gambetta J~M 2012 {\em Phys. Rev. Lett.\/} {\bf 108}(22) 220402
  \urlprefix\url{http://link.aps.org/doi/10.1103/PhysRevLett.108.220402}

\bibitem{Mueller-Ebhardt:2012}
Mueller-Ebhardt H, Miao H, Danilishin S and Chen Y 2012 {\em arXiv preprint
  arXiv:1211.4315\/}

\bibitem{Mancini:2002}
Mancini S, Giovannetti V, Vitali D and Tombesi P 2002 {\em Physical review
  letters\/} {\bf 88} 120401

\bibitem{Pirandola:2003b}
Pirandola S, Mancini S, Vitali D and Tombesi P 2003 {\em Phys. Rev. A\/} {\bf
  68}(6) 062317
  \urlprefix\url{http://link.aps.org/doi/10.1103/PhysRevA.68.062317}

\bibitem{Mancini:2003}
Mancini S, Vitali D and Tombesi P 2003 {\em Physical review letters\/} {\bf 90}
  137901

\bibitem{Zhang:2003}
Zhang J, Peng K and Braunstein S~L 2003 {\em Phys. Rev. A\/} {\bf 68}(1) 013808
  \urlprefix\url{http://link.aps.org/doi/10.1103/PhysRevA.68.013808}

\bibitem{Pinard:2007}
Pinard M, Dantan A, Vitali D, Arcizet O, Briant T and Heidmann A 2007 {\em EPL
  (Europhysics Letters)\/} {\bf 72} 747

\bibitem{Bhattacharya:2008c}
Bhattacharya M, Giscard P and Meystre P 2008 {\em Physical Review A\/} {\bf 77}
  030303

\bibitem{Rehbein:2007}
Rehbein H, M{\"u}ller-Ebhardt H, Somiya K, Li C, Schnabel R, Danzmann K and
  Chen Y 2007 {\em Physical Review D\/} {\bf 76} 062002

\bibitem{Ludwig:2010}
Ludwig M, Hammerer K and Marquardt F 2010 {\em Physical Review A\/} {\bf 82}
  012333

\bibitem{Loock:2000}
{van Loock} P and {Braunstein} S~L 2000 {\em Physical Review A\/} {\bf 61}
  010302 (\textit{Preprint} \eprint{arXiv:quant-ph/9907073})

\bibitem{Lee:2011}
Lee K~C, Sprague M~R, Sussman B~J, Nunn J, Langford N~K, Jin X~M, Champion T,
  Michelberger P, Reim K~F, England D, Jaksch D and Walmsley I~A 2011 {\em
  Science\/} {\bf 334} 1253--1256 (\textit{Preprint}
  \eprint{http://www.sciencemag.org/content/334/6060/1253.full.pdf})
  \urlprefix\url{http://www.sciencemag.org/content/334/6060/1253.abstract}

\bibitem{Lee:2011b}
Lee K, Sussman B, Sprague M, Michelberger P, Reim K, Nunn J, Langford N,
  Bustard P, Jaksch D and Walmsley I 2011 {\em Nature Photonics\/}

\bibitem{Duan:2001}
Duan L, Lukin M, Cirac I and Zoller P 2001 {\em arXiv preprint
  quant-ph/0105105\/}

\bibitem{Vanner:2013}
Vanner M~R, Aspelmeyer M and Kim M~S 2013 {\em Phys. Rev. Lett.\/} {\bf 110}(1)
  010504 \urlprefix\url{http://link.aps.org/doi/10.1103/PhysRevLett.110.010504}

\bibitem{khalili:2010}
Khalili F, Danilishin S, Miao H, M{\"u}ller-Ebhardt H, Yang H and Chen Y 2010
  {\em Physical review letters\/} {\bf 105} 70403

\bibitem{Misra:1977}
Misra B and Sudarshan E 1977 {\em Journal of Mathematical Physics\/} {\bf 18}
  756--763

\bibitem{Gupta:2007}
Gupta S, Moore K, Murch K and Stamper-Kurn D 2007 {\em Physical review
  letters\/} {\bf 99} 213601

\bibitem{Purdy:2010}
Purdy T, Brooks D, Botter T, Brahms N, Ma Z and Stamper-Kurn D 2010 {\em
  Physical review letters\/} {\bf 105} 133602

\bibitem{Painter:private}
Painter O 2012 Private communication.

\bibitem{Santamore:2004}
Santamore D~H, Doherty A~C and Cross M~C 2004 {\em Phys. Rev. B\/} {\bf 70}(14)
  144301 \urlprefix\url{http://link.aps.org/doi/10.1103/PhysRevB.70.144301}

\bibitem{Martin:2007}
Martin I and Zurek W~H 2007 {\em Phys. Rev. Lett.\/} {\bf 98}(12) 120401
  \urlprefix\url{http://link.aps.org/doi/10.1103/PhysRevLett.98.120401}

\bibitem{Gangat:2011}
Gangat A, Stace T and Milburn G 2011 {\em New Journal of Physics\/} {\bf 13}
  043024

\bibitem{Ludwig:2012}
Ludwig M, Safavi-Naeini A, Painter O and Marquardt F 2012 {\em arXiv preprint
  arXiv:1202.0532\/}

\bibitem{Thompson:2008}
Thompson J, Zwickl B, Jayich A, Marquardt F, Girvin S and Harris J 2008 {\em
  Nature\/} {\bf 452} 72--75

\bibitem{Jayich:2008}
Jayich A~M, Sankey J~C, Zwickl B~M, Yang C, Thompson J~D, Girvin S~M, Clerk
  A~A, Marquardt F and Harris J~G~E 2008 {\em New Journal of Physics\/} {\bf
  10} 095008 \urlprefix\url{http://stacks.iop.org/1367-2630/10/i=9/a=095008}

\bibitem{Romero-Isart:2011c}
Romero-Isart O, Pflanzer A, Blaser F, Kaltenbaek R, Kiesel N, Aspelmeyer M and
  Cirac J 2011 {\em Physical Review Letters\/} {\bf 107} 20405

\bibitem{Nunnenkamp:2010}
Nunnenkamp A, B\o{}rkje K, Harris J~G~E and Girvin S~M 2010 {\em Phys. Rev.
  A\/} {\bf 82}(2) 021806
  \urlprefix\url{http://link.aps.org/doi/10.1103/PhysRevA.82.021806}

\bibitem{Jacobs:2009}
Jacobs K, Tian L and Finn J 2009 {\em Phys. Rev. Lett.\/} {\bf 102}(5) 057208
  \urlprefix\url{http://link.aps.org/doi/10.1103/PhysRevLett.102.057208}

\bibitem{Miao:2009}
Miao H, Danilishin S, Corbitt T and Chen Y 2009 {\em Phys. Rev. Lett.\/} {\bf
  103}(10) 100402
  \urlprefix\url{http://link.aps.org/doi/10.1103/PhysRevLett.103.100402}

\bibitem{Clerk:2010}
Clerk A~A, Marquardt F and Harris J~G~E 2010 {\em Phys. Rev. Lett.\/} {\bf
  104}(21) 213603
  \urlprefix\url{http://link.aps.org/doi/10.1103/PhysRevLett.104.213603}

\bibitem{Clerk:2011}
Clerk A 2011 {\em Physical Review A\/} {\bf 84} 043824

\bibitem{Bose:1997}
Bose S, Jacobs K and Knight P~L 1997 {\em Phys. Rev. A\/} {\bf 56}(5)
  4175--4186 \urlprefix\url{http://link.aps.org/doi/10.1103/PhysRevA.56.4175}

\bibitem{Rabl:2011}
Rabl P 2011 {\em Phys. Rev. Lett.\/} {\bf 107}(6) 063601
  \urlprefix\url{http://link.aps.org/doi/10.1103/PhysRevLett.107.063601}

\bibitem{Nunnenkamp:2011}
Nunnenkamp A, B\o{}rkje K and Girvin S~M 2011 {\em Phys. Rev. Lett.\/} {\bf
  107}(6) 063602
  \urlprefix\url{http://link.aps.org/doi/10.1103/PhysRevLett.107.063602}

\bibitem{Qian:2012}
Qian J, Clerk A~A, Hammerer K and Marquardt F 2012 {\em Phys. Rev. Lett.\/}
  {\bf 109}(25) 253601
  \urlprefix\url{http://link.aps.org/doi/10.1103/PhysRevLett.109.253601}

\bibitem{Nunnenkamp:2012}
Nunnenkamp A, B{\o}rkje K and Girvin S 2012 {\em Physical Review A\/} {\bf 85}
  051803

\bibitem{hong:2011}
Hong T, Yang H, Miao H and Chen Y 2011 {\em arXiv preprint arXiv:1110.3348\/}

\bibitem{Liao:2012}
Liao J~Q, Cheung H~K and Law C~K 2012 {\em Phys. Rev. A\/} {\bf 85}(2) 025803
  \urlprefix\url{http://link.aps.org/doi/10.1103/PhysRevA.85.025803}

\bibitem{Marshall:2003}
Marshall W, Simon C, Penrose R and Bouwmeester D 2003 {\em Phys. Rev. Lett.\/}
  {\bf 91}(13) 130401
  \urlprefix\url{http://link.aps.org/doi/10.1103/PhysRevLett.91.130401}

\bibitem{Bassi:2005}
Bassi A, Ippoliti E and Adler S 2005 {\em Physical review letters\/} {\bf 94}
  30401

\bibitem{Liao:2012b}
Liao J and Law C 2012 {\em arXiv preprint arXiv:1206.3085\/}

\bibitem{Vanner:2011b}
Vanner M~R 2011 {\em Phys. Rev. X\/} {\bf 1}(2) 021011
  \urlprefix\url{http://link.aps.org/doi/10.1103/PhysRevX.1.021011}

\bibitem{Penrose:1998}
Penrose R 1998 {\em Philosophical Transactions of the Royal Society of London.
  Series A: Mathematical, Physical and Engineering Sciences\/} {\bf 356}
  1927--1939 (\textit{Preprint}
  \eprint{http://rsta.royalsocietypublishing.org/content/356/1743/1927.full.pdf+html})
  \urlprefix\url{http://rsta.royalsocietypublishing.org/content/356/1743/1927.abstract}

\bibitem{Penrose:2006}
Penrose R and Jorgensen P 2006 {\em The Mathematical Intelligencer\/} {\bf 28}
  59--61

\bibitem{Diosi:2007}
Di\'osi L 2007 {\em Journal of Physics A: Mathematical and Theoretical\/} {\bf
  40} 2989 \urlprefix\url{http://stacks.iop.org/1751-8121/40/i=12/a=S07}

\bibitem{vanWezel:2008}
van Wezel J, Oosterkamp T and Zaanen J 2008 {\em Philosophical Magazine\/} {\bf
  88} 1005--1026 (\textit{Preprint}
  \eprint{http://www.tandfonline.com/doi/pdf/10.1080/14786430801941824})
  \urlprefix\url{http://www.tandfonline.com/doi/abs/10.1080/14786430801941824}

\bibitem{vanWezel:2012}
van Wezel J and Oosterkamp T~H 2012 {\em Proceedings of the Royal Society A:
  Mathematical, Physical and Engineering Science\/} {\bf 468} 35--56
  (\textit{Preprint}
  \eprint{http://rspa.royalsocietypublishing.org/content/468/2137/35.full.pdf+html})
  \urlprefix\url{http://rspa.royalsocietypublishing.org/content/468/2137/35.abstract}

\bibitem{Housley:1966}
Housley R and Hess F 1966 {\em Physical Review\/} {\bf 146} 517

\bibitem{Yang:2012}
Yang H, Miao H, Lee D, Helou B and Chen Y 2012 {\em arXiv preprint
  arXiv:1210.0457\/}

\bibitem{Ando:2010}
Ando M, Ishidoshiro K, Yamamoto K, Yagi K, Kokuyama W, Tsubono K and Takamori A
  2010 {\em Physical review letters\/} {\bf 105} 161101

\bibitem{Romero-Isart:2011}
Romero-Isart O 2011 {\em Physical Review A\/} {\bf 84} 052121

\bibitem{Kaltenbaek:2012}
Kaltenbaek R, Hechenblaikner G, Kiesel N, Romero-Isart O, Schwab K, Johann U
  and Aspelmeyer M 2012 {\em Experimental Astronomy\/}  1--42

\bibitem{Feynman:1995}
Feynman R, Morinigo F and Wagner W 1995 {\em Reading, MA: Addison-Wesley,|
  c1995, edited by Hatfield, Brian\/} {\bf 1}

\bibitem{Carlip:2008}
Carlip S 2008 {\em Classical and Quantum Gravity\/} {\bf 25} 154010

\bibitem{moller:1962}
Moller C 1962 {\em CRNS, Paris\/}

\bibitem{Rosenfeld:1963}
Rosenfeld L 1963 {\em Nuclear Physics\/} {\bf 40} 353--356

\bibitem{Wald:1984}
Wald R 1984 {\em General relativity\/} (University of Chicago press)

\bibitem{Everett:1957}
Everett~III H 1957 {\em Reviews of modern physics\/} {\bf 29} 454

\bibitem{Page:1981}
Page D and Geilker C 1981 {\em Physical Review Letters\/} {\bf 47} 979--982

\bibitem{Moroz:1999}
Moroz I, Penrose R and Tod P 1999 {\em Classical and Quantum Gravity\/} {\bf
  15} 2733

\bibitem{Salzman:2006}
Salzman P and Carlip S 2006 {\em arXiv preprint gr-qc/0606120\/}

\bibitem{Maggiore:1994}
Maggiore M 1994 {\em Phys. Rev. D\/} {\bf 49}(10) 5182--5187
  \urlprefix\url{http://link.aps.org/doi/10.1103/PhysRevD.49.5182}

\bibitem{Das:2008}
Das S and Vagenas E~C 2008 {\em Phys. Rev. Lett.\/} {\bf 101}(22) 221301
  \urlprefix\url{http://link.aps.org/doi/10.1103/PhysRevLett.101.221301}

\bibitem{Ali:2011}
Ali A~F, Das S and Vagenas E~C 2011 {\em Phys. Rev. D\/} {\bf 84}(4) 044013
  \urlprefix\url{http://link.aps.org/doi/10.1103/PhysRevD.84.044013}

\bibitem{Pikovski:2012}
Pikovski I, Vanner M, Aspelmeyer M, Kim M and Brukner {\v{C}} 2012 {\em Nature
  Physics\/}

\bibitem{RYang:2010}
Yang R, Gong X, Pei S, Luo Z and Lau Y 2010 {\em Physical Review A\/} {\bf 82}
  032120

\bibitem{testqm:2012}
Adhikari R, Chen Y, Helou B, Kazarian P, Miao H, Price L, Smith N and Yang H
  2012 {\em Research in progress.\/}

\bibitem{tobar:2012}
Tobar M 2012 {\em Private Communications\/}

\bibitem{Will:2006}
Will C 2006 {\em Living Rev. Relativity\/} {\bf 9}

\bibitem{Weinberg:1989}
Weinberg S 1989 {\em Physical Review Letters\/} {\bf 62} 485--488

\bibitem{Weinberg:1989b}
Weinberg S 1989 {\em Annals of Physics\/} {\bf 194} 336--386

\bibitem{testqm2:2012}
Yang H {\em et~al.\/} 2012 {\em Research in progress.\/}

\bibitem{Wang:2012}
Wang Y and Clerk A 2012 {\em Physical Review Letters\/} {\bf 108} 153603

\bibitem{Verhagen:2012}
Verhagen E, Del{\'e}glise S, Weis S, Schliesser A and Kippenberg T 2012 {\em
  Nature\/} {\bf 482} 63--67

\bibitem{Palokaki:2012}
Palomaki T, Harlow J, Teufel J, Simmonds R and Lehnert K 2012 {\em arXiv
  preprint arXiv:1206.5562\/}

\bibitem{Aoki:2006}
Aoki T, Dayan B, Wilcut E, Bowen W, Parkins A, Kippenberg T, Vahala K and
  Kimble H 2006 {\em Nature\/} {\bf 443} 671--674

\bibitem{Hammerer:2009}
Hammerer K, Wallquist M, Genes C, Ludwig M, Marquardt F, Treutlein P, Zoller P,
  Ye J and Kimble H 2009 {\em Physical review letters\/} {\bf 103} 63005

\bibitem{Glaetzle:2010}
Glaetzle A, Hammerer K, Daley A, Blatt R and Zoller P 2010 {\em Optics
  Communications\/} {\bf 283} 758--765

\bibitem{Wallquist:2010}
Wallquist M, Hammerer K, Zoller P, Genes C, Ludwig M, Marquardt F, Treutlein P,
  Ye J and Kimble H 2010 {\em Physical Review A\/} {\bf 81} 023816

\bibitem{Singh:2008}
Singh S, Bhattacharya M, Dutta O and Meystre P 2008 {\em Physical Review
  Letters\/} {\bf 101} 263603

\bibitem{Hammerer:2008}
Hammerer K, Aspelmeyer M, Polzik E and Zoller P 2008 {\em arXiv preprint
  arXiv:0804.3005\/}

\bibitem{Muschik:2011}
Muschik C, Krauter H, Hammerer K and Polzik E 2011 {\em Quantum Information
  Processing\/}  1--25

\bibitem{Vasilyev:2012}
Vasilyev D, Hammerer K, Korolev N and S{\o}rensen A 2012 {\em Journal of
  Physics B: Atomic, Molecular and Optical Physics\/} {\bf 45} 124007

\bibitem{Hammerer:2010}
Hammerer K, S\o{}rensen A~S and Polzik E~S 2010 {\em Rev. Mod. Phys.\/} {\bf
  82}(2) 1041--1093
  \urlprefix\url{http://link.aps.org/doi/10.1103/RevModPhys.82.1041}

\bibitem{Steinke:2011}
Steinke S, Singh S, Tasgin M, Meystre P, Schwab K and Vengalattore M 2011 {\em
  Physical Review A\/} {\bf 84} 023841

\bibitem{Lin:2010}
Lin Q, Rosenberg J, Chang D, Camacho R, Eichenfield M, Vahala K and Painter O
  2010 {\em Nature Photonics\/} {\bf 4} 236--242

\bibitem{Stannigel:2010}
Stannigel K, Rabl P, S{\o}rensen A, Zoller P and Lukin M 2010 {\em Physical
  review letters\/} {\bf 105} 220501

\bibitem{Stannigel:2011}
Stannigel K, Rabl P, S{\o}rensen A, Lukin M and Zoller P 2011 {\em Physical
  Review A\/} {\bf 84} 042341

\bibitem{Hill:2012}
Hill J, Safavi-Naeini A, Chan J and Painter O 2012 {\em arXiv preprint
  arXiv:1206.0704\/}

\bibitem{Winger:2011}
Winger M, Blasius T, Mayer~Alegre T, Safavi-Naeini A, Meenehan S, Cohen J,
  Stobbe S and Painter O 2011 {\em Optics Express\/} {\bf 19} 24905--24921

\bibitem{Hafezi:2012}
Hafezi M and Rabl P 2012 {\em Optics Express\/} {\bf 20} 7672--7684

\bibitem{Dong:2012}
Dong C, Fiore V, Kuzyk M and Wang H 2012 {\em Science\/} {\bf 338} 1609--1613

\bibitem{Hammerer:2010b}
Hammerer K, Stannigel K, Genes C, Zoller P, Treutlein P, Camerer S, Hunger D
  and H{\"a}nsch T 2010 {\em Physical Review A\/} {\bf 82} 021803

\bibitem{Bhattacharya:2010}
Bhattacharya M, Singh S, Giscard P and Meystre P 2010 {\em Laser Physics\/}
  {\bf 20} 57--67

\bibitem{Schmidt:2012}
Schmidt M, Ludwig M and Marquardt F 2012 {\em arXiv preprint arXiv:1202.3659\/}

\bibitem{Vanner:2012}
Vanner M, Aspelmeyer M and Kim M 2012 {\em arXiv preprint arXiv:1203.4525\/}

\bibitem{Habraken:2012}
Habraken S, Stannigel K, Lukin M, Zoller P and Rabl P 2012 {\em arXiv preprint
  arXiv:1205.7008\/}

\bibitem{Stannigel:2012}
Stannigel K, Komar P, Habraken S, Bennett S, Lukin M, Zoller P and Rabl P 2012
  {\em Physical Review Letters\/} {\bf 109} 13603

\bibitem{Seok:2012}
Seok H, Buchmann L, Singh S and Meystre P 2012 {\em arXiv preprint
  arXiv:1208.5821\/}

\bibitem{Buchmann:2012}
Buchmann L, Zhang L, Chiruvelli A and Meystre P 2012 {\em Physical Review
  Letters\/} {\bf 108} 210403

\bibitem{Bhattacharya:2008b}
Bhattacharya M and Meystre P 2008 {\em Physical Review A\/} {\bf 78} 041801

\bibitem{Ludwig:2012b}
Ludwig M and Marquardt F 2012 {\em arXiv preprint arXiv:1208.0327\/}

\bibitem{Jacobs:2012b}
Jacobs K 2012 {\em arXiv preprint arXiv:1209.2499\/}

\bibitem{Holmes:2012}
Holmes C, Meaney C and Milburn G 2012 {\em Physical Review E\/} {\bf 85} 066203

\bibitem{ChenW:2009}
Chen W, Zhang K, Goldbaum D, Bhattacharya M and Meystre P 2009 {\em Physical
  Review A\/} {\bf 80} 011801

\bibitem{Zhang:2010}
Zhang K, Chen W, Bhattacharya M and Meystre P 2010 {\em Physical Review A\/}
  {\bf 81} 013802

\bibitem{Kippenberg:2008}
Kippenberg T and Vahala K 2008 {\em science\/} {\bf 321} 1172--1176

\bibitem{Marquardt:2009}
Marquardt F and Girvin S 2009 {\em arXiv preprint arXiv:0905.0566\/}

\bibitem{Wallquist:2009}
Wallquist M, Hammerer K, Rabl P, Lukin M and Zoller P 2009 {\em Physica
  Scripta\/} {\bf 2009} 014001

\bibitem{Aspelmeyer:2010}
Aspelmeyer M, Groeblacher S, Hammerer K and Kiesel N 2010 {\em JOSA B\/} {\bf
  27} A189--A197

\bibitem{Meystre:2012}
Meystre P 2012 {\em Annalen der Physik\/}

\end{thebibliography}

\end{document}